\documentclass{article}

\usepackage[preprint,nonatbib]{neurips_2026}

\usepackage{amsmath,amssymb,amsthm,bm,bbm,mathrsfs,amscd}
\usepackage{braket}
\usepackage{cancel}

\usepackage{algorithm}
\usepackage{algorithmicx}
\usepackage{algpseudocode}

\usepackage{booktabs}

\usepackage[hidelinks]{hyperref}

\usepackage[
    backend=biber,
    sorting=none,
    maxbibnames=10,
    style=numeric-comp
]{biblatex}

\usepackage{graphicx}
\usepackage{xcolor}

\newcommand{\dd}{\!\mathrm{d}}

\newcommand{\inv}{^{-1}}

\newcommand{\R}{\mathbb{R}}

\newcommand{\E}{\mathbb{E}}
\newcommand{\KL}{\mathrm{KL}}
\newcommand{\Var}{\mathrm{Var}}
\newcommand{\Deff}{D_{\mathrm{eff}}}

\title{Generative Nested Sampling of Atomistic Thermodynamic Landscapes}

\author{Alessandro Coretti\thanks{These authors contributed equally to this work.}{\ \ }\thanks{\texttt{alessandro.coretti@univie.ac.at}}\\
Faculty of Physics,\\
University of Vienna,\\
1090 Vienna, Austria\\
\And
Nico Unglert\footnotemark[1]{\ \ }\thanks{\texttt{nico.unglert@tuwien.ac.at}}\\
Institute of Materials Chemistry,\\
TU Wien,\\
1060 Vienna, Austria\\
\And
Sebastian Falkner\\
Institute of Physics,\\
University of Augsburg,\\
86159 Augsburg, Germany
\AND
Georg K. H. Madsen\\
Institute of Materials Chemistry,\\
TU Wien,\\
1060 Vienna, Austria
\And
Christoph Dellago\\
Faculty of Physics,\\
University of Vienna,\\
1090 Vienna, Austria
}

\newcommand{\Umax}{U_{\max}}
\newcommand{\Utrain}{U_{\max}^{\mathrm{train}}}

\usepackage{tabularx}
\usepackage{multirow}

\begin{document}

\begin{refsection}
\maketitle

\begin{abstract}
Nested sampling (NS) resolves the thermodynamics of an atomistic system from a single simulation, but its practical reach is limited by the Markov-chain updates needed to decorrelate walkers within each likelihood-constrained ensemble. Flow-based NS has reduced this bottleneck for gravitational-wave (GW) inference, yet its transfer to atomistic systems is not merely a change of application. Comparing a GW150914-like binary-black-hole likelihood with an eight-particle two-dimensional Lennard-Jones (LJ) system of comparable dimensionality, we show that the two landscapes differ fundamentally: atomistic multimodality is discrete and combinatorial, generated by particle permutations separated by hard collision walls, and its coordinate coupling is dense and collective, whereas the GW posterior exhibits smooth degeneracies and localized parameter coupling. Guided by this diagnosis, we introduce NS-Flows: a single conditional normalizing flow, conditioned on the NS energy bound and trained on a sliding window of recent live sets, that replaces MCMC by direct parallel draws corrected by importance-weighted rejection resampling. Live sets supply the data self-consistently, allowing flow training without structured priors or a pre-existing dataset. For LJ disks under periodic boundary conditions, the algorithm can reduce energy evaluations by two orders of magnitude and wall-clock time by about $30\%$, an advantage that becomes more favourable as the cost of the potential grows. The flow's generative efficiency further acts as a physical diagnostic: it varies non-monotonically along the annealing trajectory, is lowest in the dense disordered regime, and is quantitatively captured by the constrained ensemble's internal mode complexity together with target drift across the training window, identifying liquid-like ensembles, not prior-target separation, as the hard case for current flow architectures.
\end{abstract}

\section{Introduction}

The prediction of thermodynamic phase diagrams remains one of the central challenges in computational materials science. Although atomistic simulation techniques such as molecular dynamics and Monte Carlo have become highly sophisticated, their exploration of configuration space is fundamentally based on local updates, making the sampling of systems with large free-energy barriers and long-lived metastable states prohibitively slow. Enhanced-sampling techniques can accelerate exploration in specific settings, but they typically rely on carefully designed collective variables or bias potentials whose construction is highly system dependent.

Nested Sampling (NS) provides a fundamentally different approach to thermodynamic sampling. Originally developed for Bayesian evidence estimation~\cite{skilling_nested_2004}, it was later adapted to statistical mechanics~\cite{partay_efficient_2010}, with the potential energy taking the role of the likelihood as the scalar quantity used to define successive nested constraints. In this formulation, it transforms the partition function into a one-dimensional integral over nested regions of configuration space with progressively decreasing energy bounds. A single NS simulation therefore provides direct access to the partition function, free energies and thermodynamic observables over a broad temperature range, making the method particularly attractive for automated phase-diagram prediction.

In practice, however, the efficiency of NS is limited by the need to repeatedly generate independent samples from likelihood-constrained ensembles. Existing implementations for atomistic systems rely almost exclusively on Markov-chain Monte Carlo methods (MCMC), for which decorrelation times can grow rapidly and severe mixing pathologies can arise on the rugged, high-dimensional energy landscapes characteristic of atomistic systems~\cite{baldock_constant-pressure_2017,partay_nested_2014,dorrell_thermodynamics_2019,marchant_exploring_2023,unglert_neural-network_2023,unglert_active_2026,kroupa_resonances_2025}. Even with recent developments such as superposition-enhanced~\cite{martiniani_superposition_2014} and replica-exchange NS~\cite{unglert_replica_2025}, MCMC-based constrained sampling remains the dominant computational bottleneck. 

Recent progress in generative machine learning offers an attractive alternative for generating the constrained samples required by NS. Williams \emph{et al.}~\cite{williams_nested_2021} first combined normalizing flows (NFs) with NS in the context of gravitational-wave (GW) inference, demonstrating that learned proposals can substantially reduce the number of likelihood evaluations required for convergence. This pioneering work has since inspired a growing body of research extending and refining the interplay between NS and generative models, with applications ranging from astrophysics to high-energy particle physics~\cite{williams_importance_2023,lemos_improving_2023,prathaban_accelerated_2025,baruah_normalizing_2025,villa_improving_2026}. In general, NFs learn invertible transformations between a simple latent distribution and a complex target distribution while retaining exact likelihood evaluation through the change-of-variables formula~\cite{rezende_variational_2016,papamakarios_normalizing_2021}. Their ability to generate independent samples together with exact density evaluation has led to numerous successful applications in statistical mechanics, including equilibrium sampling, free-energy estimation and rare-event simulations~\cite{noe_boltzmann_2019,wirnsberger_targeted_2020,gabrie_adaptive_2021,wirnsberger_normalizing_2022,schebek_efficient_2024,falkner_conditioning_2023,asghar_efficient_2024}. In most such applications, however, the flow is required to learn only a comparatively modest deformation of a well-chosen reference distribution, e.g. a harmonic reference for a crystalline solid, or a simple thermodynamic path along which the target evolves smoothly. The GW case shares this favorable structure: a single, sharply peaked mode with smooth degeneracy ridges and coupling concentrated in a few physically identifiable coordinate blocks.

Extending learned NFs to generate independent samples from likelihood-constrained ensembles in atomistic NS is, however, not merely a change of application domain. Even for systems with short-ranged interactions, the energy constraint can induce strong collective correlations among particle coordinates. Atomistic systems further obey characteristic symmetries, including permutations of identical particles, rotations and translations together with the point group of the simulation cell, which flow architectures must respect, motivating a growing body of equivariant NF designs~\cite{noe_boltzmann_2019,wirnsberger_targeted_2020,kohler_equivariant_2020}.
Learning such targets poses a qualitatively different challenge for NF-based methods than the applications in which they have so far succeeded.

In this work, we show quantitatively that the atomistic landscape is combinatorially multimodal in a way the GW posterior is not, and no comparably simple reference distribution is available to absorb this structure. We then develop a conditional NF framework for NS of atomistic systems that addresses this challenge directly. Rather than learning each constrained distribution independently or relying on a fixed reference close to the target, we represent the entire sequence of likelihood-constrained ensembles by a single flow conditioned on the NS energy bound, thereby representing the full sequence of constrained ensembles. The method preserves the statistical rigor of NS while replacing expensive MCMC exploration by direct generative proposals combined with exact rejection correction. We further show how the generative efficiency of the trained NF varies nonmonotonically along the NS trajectory, falling sharply in the dense-fluid regime before recovering as the system orders. To explain this behavior, we introduce the internal mode complexity, a diagnostic computed directly from the geometry of the live-point ensemble. It identifies combinatorial multimodality, rather than geometric distance from a reference distribution, as the principal source of difficulty. We present the probabilistic formulation and algorithmic implementation of the method, and demonstrate both its performance and the diagnostic for a two-dimensional Lennard-Jones (LJ) system.

\section{Atomistic posterior landscapes}
\label{sec:landscapes}

Before presenting the methodological developments required for applying flow-based nested sampling to atomistic systems, we first compare representative posterior landscapes from atomistic and GW inference. As we show below, the distinction between the two is not primarily one of dimensionality, but of geometric and topological organization. It is this qualitative difference that ultimately determines the difficulty of learning the target distribution with a NF.
The GW posteriors for which flow-based nested sampling was originally developed~\cite{williams_nested_2021} and the atomistic posteriors we target here differ sharply in exactly this respect. 

To make the distinction concrete and quantitative, we probe two benchmark posteriors of comparable dimensionality. 
The first is an atomistic eight-particle two-dimensional LJ system with periodic boundary conditions, $x\in[0,L)^{16}$, where, in reduced units, $L=2.9$ and the density $\rho\approx0.95$ places the system in the high-density region of the phase diagram~\cite{li_phase_2020}.
The pair interaction is described by the standard LJ potential,
\[
U(r_{ij}) = 4\left[r_{ij}^{-12} - r_{ij}^{-6}\right],
\]
where $r_{ij}$ is the distance between particles $i$ and $j$. To enforce the minimum-image convention under periodic boundary conditions, the interaction is cut and shifted at a reduced cutoff radius $R_c = 0.49L$. In addition, the strongly repulsive short-range part of the interaction is linearized below $R_{\mathrm{cutin}} = 0.8$ to avoid numerical instabilities arising from nearly overlapping configurations during generation. Long-range corrections to the potential energy are included consistently throughout the simulation.

The second benchmark models a GW150914-like binary-black-hole merger~\cite{gw150914} with parameters $\vartheta\in\mathbb{R}^{15}$. Its likelihood $\mathcal{L}(\vartheta)$ measures the agreement between simulated detector data and the gravitational waveform predicted for a given set of source parameters, assuming stationary Gaussian detector noise (see Appendix~\ref{app:gw}).

\begin{figure}[htbp]
  \centering
  \includegraphics[width=\linewidth]{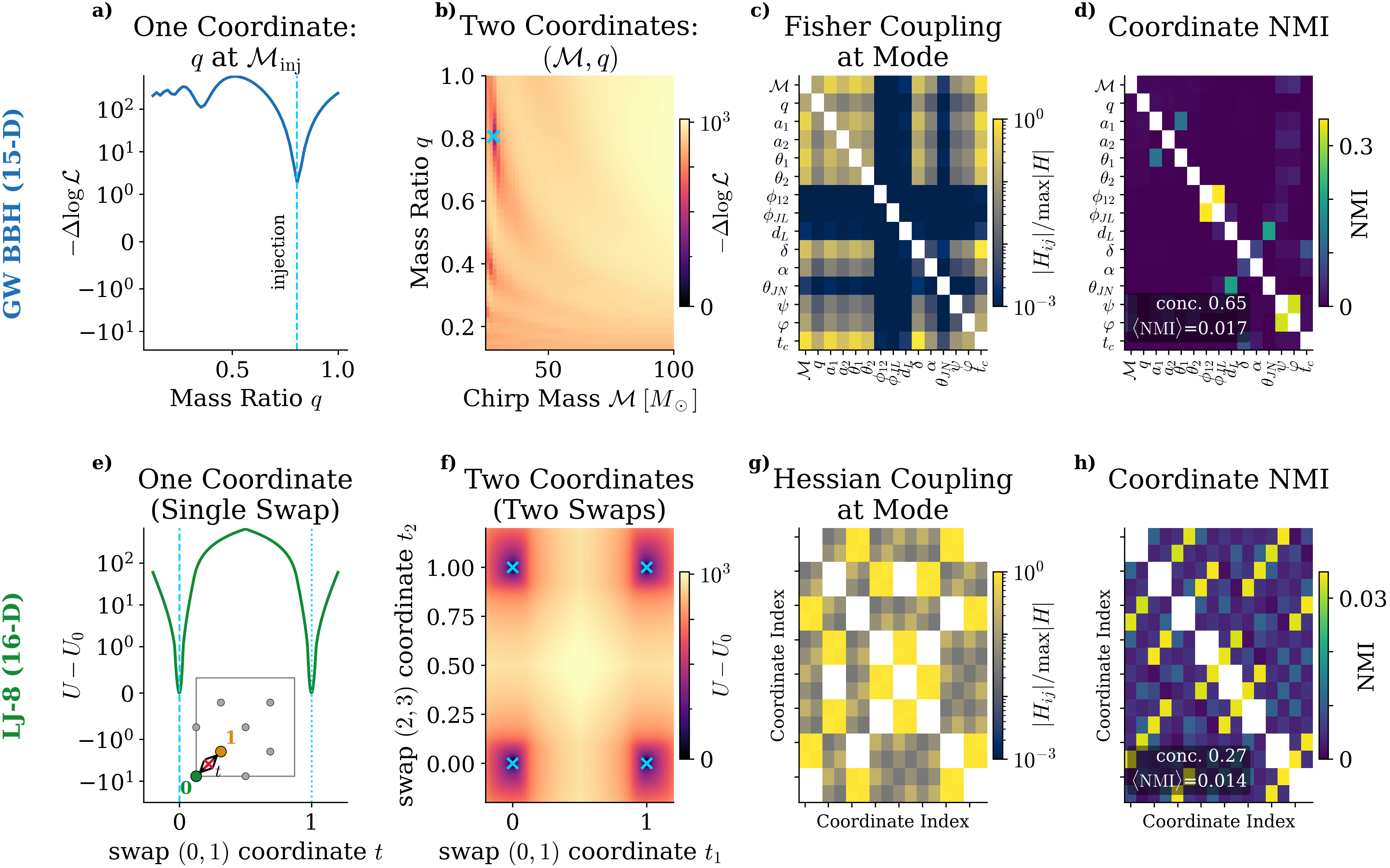}
  \caption{\textbf{GW binary-black-hole (top) vs.\ atomistic (LJ-8, bottom) posterior
    landscapes.} \emph{Column 1:} a one-coordinate cut --- GW
    log-likelihood along mass ratio $q$ at the injected chirp mass a),
    LJ energy along a single particle-swap interpolation e). \emph{Column 2:} the two-coordinate extension --- the GW
    chirp-mass / mass-ratio ``banana'', a single continuous degeneracy
    ridge b), vs.\ the LJ two-swap grid with four permutation-equivalent
    minima walled off by collision barriers f). \emph{Column 3:}
    normalized coupling magnitude of the Fisher / Hessian matrix at the
    mode --- sparse, block-structured for GW c); dense, all-to-all for
    LJ g). \emph{Column 4:} pairwise normalized mutual information of
    Laplace samples at the mode --- concentrated in a few blocks for GW
    d), a diffuse low-amplitude haze for LJ h). Note the differing NMI
    color scales: GW couples a few coordinates strongly, LJ couples all
    coordinates at a similar scale. The leading diagonal is masked (white)
    in all four matrix panels; in the LJ panels g) and h) the $2\times2$
    on-diagonal blocks --- each particle's Cartesian pair $(x_p, y_p)$ ---
    are additionally masked, since the intra-particle $x$--$y$ coupling is
    trivially strong and structurally uninformative, leaving only the
    inter-particle coupling on display (which also sets the colour-scale
    normalization in g)). The quoted $\langle\mathrm{NMI}\rangle$ and
    concentration are computed over all coordinate pairs.}
  \label{fig:landscapes}
\end{figure}

The two posterior landscapes are thus defined by (or proportional to) negative log-likelihoods of different physical origin but similar dimensionality: the potential energy $U(x)$ for the LJ system and the noise-weighted waveform mismatch $-\log\mathcal{L}(\vartheta)$ for the GW model. Figure~\ref{fig:landscapes} contrasts the two posteriors on equal footing: the two left-hand columns probe the global landscape along interpretable directions, while the two right-hand columns characterize the local geometry around the dominant mode.

We first consider a single representative coordinate. Along the mass-ratio direction at the injected chirp mass, the negative GW log-likelihood is a single smooth basin (Fig.~\ref{fig:landscapes}a). 
An atomistic counterpart is e.g. the energy along a single particle-swap interpolation $x(t)=(1-t)\,x_\star + t\,P_{(ij)}x_\star$ between a configuration and its label-swapped copy. Here, $x_\star$ is a local minimum configuration and $P_{(ij)}$ is the permutation operator swapping particle $i$ and $j$ to yield the symmetry equivalent configuration $P_{(ij)}x_\star$. 
This choice of coordinate already reveals the strong multimodality of this simple atomistic system (Fig.~\ref{fig:landscapes}e): two equivalent minima at $t=0$ and $t=1$ separated by a steep divergent collision barrier near $t=\tfrac12$. For the unmodified LJ potential, the energy would diverge at the coincidence point \(t=\tfrac12\); in our simulations, this divergence is regularized by the linearization of the short-range interaction introduced above.
Discrete multimodality appears in the atomistic system at the level of a \emph{single} coordinate, where the GW likelihood is still effectively unimodal.

Extending each cut to a second coordinate sharpens the contrast. 
The GW chirp-mass--mass-ratio plane reveals the canonical ``banana'' shape (Fig.~\ref{fig:landscapes}b): a single, smooth, connected ridge of near-maximal likelihood, reflecting a continuous degeneracy. 
The atomistic two-swap grid (Fig.~\ref{fig:landscapes}f) instead shows four permutation-equivalent minima at the corners, mutually walled off by collision barriers and thus forming a discrete, combinatorial multimodality. 
The same dimensional extension exposes qualitatively different structure in the two systems.

The local geometry (see also Appendix~\ref{app:diagnostics}) at the mode tells the same story. 
The Hessian of $U$ at the LJ inherent structure and the Fisher information matrix~\cite{vallisneri_use_2008}
evaluated at the GW injection encode the local curvature and coordinate coupling. 
Displayed as normalized coupling magnitude $|H_{ij}|/\max|H|$ (Fig.~\ref{fig:landscapes}c,g), the GW Fisher matrix is sparse, with coupling confined to a few named blocks, whereas the LJ Hessian is dense: every coordinate couples appreciably to every other. 
An independent measurement, the pairwise normalized mutual information (NMI)~\cite{vinh_information_2010} of Laplace samples drawn at each mode (Fig.~\ref{fig:landscapes}d,h), confirms this from the marginal side.
The GW NMI is concentrated (the top $5\%$ of coordinate pairs carry
$65\%$ of the total mutual-information mass), while the LJ NMI is a
diffuse, low-amplitude haze spread over all pairs (concentration $27\%$).
Notably the \emph{mean} off-diagonal NMI is larger for GW ($0.017$) than
for LJ-8 ($0.014$): the atomistic disadvantage for a coupling-layer flow of the kind described below is
not that its coordinates are more strongly correlated, but that the
correlation is collective and admits no sparse, axis-aligned block
structure that a coordinate split could exploit.

Across all four diagnostics, the same dichotomy recurs. The GW posterior
is a single concentrated mode dressed with smooth, continuous degeneracy
ridges and a sparse, block-structured coupling that decades of
reparameterization \cite{williams_nested_2021,veitch_parameter_2015} have isolated into named groups. 
In contrast, the atomistic posterior is a combinatorial proliferation of discrete, hard-walled permutation copies with dense, collective coordinate coupling. 
These landscapes differ not merely in degree along a single axis of ``difficulty'' but in their global organization. Indeed by the same local measures, such as anisotropy and mean coupling, the GW likelihood is the more challenging of the two. It is this difference in kind that a normalizing flow must contend with.

\section{Methods}
\label{sec:methods}

We present the two ingredients of our method in a notation chosen to make their composition transparent.
Section~\ref{subsec:NS} reframes the likelihood-constrained priors that drive nested sampling as a one-parameter family of conditional distributions indexed by the energy bound.
Section~\ref{subsec:CNF} introduces a conditional normalizing flow $F_\theta$ that learns a parameterized approximation of exactly such a family.
Section~\ref{subsec:NSF} combines the two ingredients by identifying the flow conditioning parameter with the energy bound, $c\equiv U_\mathrm{max}$. The flow thereby becomes a direct generator for the constrained prior at each NS iteration, replacing the costly MCMC decorrelation procedure of classical NS by parallel, independent draws from a differentiable flow model.

\subsection{Nested Sampling}
\label{subsec:NS}
Nested sampling is widely employed as a tool for Bayesian inference \cite{skilling_nested_2006,ashton_nested_2022}, as it provides an approximation of the evidence, which appears in the well-known Bayes' theorem
\begin{align}
    \pi(x) = \frac{\mathcal{L}(x) \; \eta(x)}{Z},
    \label{eq:bayes_general}
\end{align}
where \( \pi \) represents the posterior, \( \mathcal{L} \) the likelihood, \( \eta \) the prior, and \( Z \) the evidence for a parameter $x$. 

In the context of atomistic systems at constant volume, this framework is analogous to
\begin{align}
    \pi(x) = \frac{\exp[-\beta U(x)] \; \eta(x)}{Z_{NVT}},
    \label{eq:bayes_npt}
\end{align}
where the posterior corresponds to the distribution of the canonical ensemble, normalized by the canonical partition function \( Z_{NVT} \), and $N$, $V$ and $T$ are the number of particles, volume and temperature, respectively. The likelihood is expressed as the Boltzmann factor, which depends on the inverse temperature $\beta=(k_\mathrm{B} T)^{-1}$ and the potential energy \( U \) as a function of the configuration \( x \). Here, $x$ contains the complete structural information of the system, which, in the case of periodic boundary conditions, can be decomposed into the Cartesian coordinates of a repeating unit and the simulation cell.
In this framework, the prior distribution, $\eta$, is uniform, reflecting the principle of equal {\it a priori} probability. For simplicity, we consider only the configurational part of the distribution, as it can be straightforwardly decoupled from the momentum-dependent part, as follows from classical statistical mechanics.

\begin{algorithm}[htbp]
\caption{Nested sampling}
\label{alg:nested-sampling}
\begin{algorithmic}[1]
\Require Live-set size $K$, termination parameter $\varepsilon$
\State $\alpha_0 \gets 1$
\For{$k=1$ \textbf{to} $K$}
  \State draw $X_0^{(k)} \sim \eta$
\EndFor
\For{$t=1,2,\dots$}
    \State $m \gets \arg\max_{1\le k\le K} U\big(X_{t-1}^{(k)}\big)$ \Comment{worst-performing walker}
    \State $U_\mathrm{max}^t \gets U\big(X_{t-1}^{(m)}\big)$
    \State $\tilde X_{t-1} \gets X_{t-1}^{(m)}$ \Comment{store sample for inference}
    \State $\alpha_t \gets \alpha_{t-1}\, K/(K+1)$
    \State $X_t^{(k)} \gets X_{t-1}^{(k)}$ for $k\neq m$
    \State $X_t^{(m)} \gets$ independent draw from $\eta(\,\cdot\,\mid U_\mathrm{max}^t)$
    \Comment{replace via constrained prior}
    \If{stopping condition based on $\varepsilon$ is satisfied}
      \State $T \gets t$; \textbf{break}
    \EndIf
\EndFor
\State $Z \gets \displaystyle \sum_{t=0}^{T-1} (\alpha_{t-1}-\alpha_t) \exp \{ -\beta U_\mathrm{max}^t \}$
\end{algorithmic}
\end{algorithm}

Algorithm~\ref{alg:nested-sampling} outlines the basic NS algorithm. An initial set of $K$ walkers $\{X^{(k)}_0\}_{k=1}^K$ is drawn from the prior $\eta$, typically chosen uniform. At each iteration, the lowest-likelihood (highest-energy) walker $X_{t-1}^{(m)}$ is identified, fixing the threshold $\mathcal{L}_t = \mathcal{L}(X_{t-1}^{(m)})$ or, equivalently, the energy bound $U_\mathrm{max}^t = U(X_{t-1}^{(m)})$. The walker is removed and stored as $\tilde X_{t-1}$, then replaced by an independent draw from the \emph{constrained prior}
\begin{equation}
\label{eq:constrained_prior}
  \eta(x\mid U_\mathrm{max}^t)
    = \frac{\eta(x)\,\mathbf{1}\{U(x)<U_\mathrm{max}^t\}}{P_t}
    = \begin{cases}
        \eta(x)\,P_t^{-1}, & U(x)<U_\mathrm{max}^t,\\[2pt]
        0, & \text{otherwise,}
      \end{cases}
\end{equation}
where $P_t=\int\eta(x)\,\mathbf{1}\{U(x)<U_\mathrm{max}^t\}\, \dd x$ is the prior mass below the bound. Viewed as a function of $U_\mathrm{max}^t$, Eq.~\eqref{eq:constrained_prior} defines a one-parameter \emph{family} of conditional distributions indexed by the energy bound---a fact we exploit in Sec.~\ref{subsec:NSF} by training a single conditional normalizing flow to model the entire family. Assuming a uniform prior $\eta$ and ignoring normalization,
\[
  \eta(x\mid U_\mathrm{max}^t)
    \propto \mathbf{1}\{U(x)<U_\mathrm{max}^t\}
    \;=\; \mathbf{1}\{\mathcal{L}(x)>\mathcal{L}_t\},
\]
which highlights the invariance of NS with respect to temperature.

In practice, classical NS samples the distribution of Eq.~\eqref{eq:constrained_prior} by cloning a uniformly chosen remaining walker (with index $A^\mathrm{clone}_t$) and running MCMC trajectory of length $\mathscr{L}$ whose Markov kernel $\kappa_t(x,\dd x)$ has $\eta(\,\cdot\,\mid U_\mathrm{max}^t)$ as its stationary distribution (line 11 in Alg. \ref{alg:nested-sampling}). The cost of running these chains long enough to obtain effectively independent samples is the dominant bottleneck of NS in atomistic settings, and motivates the flow-based replacement of Sec.~\ref{subsec:NSF}.

The enclosed prior mass at each iteration can be obtained by
\begin{equation}
    \label{eq:alphat}
    \alpha_t = \left( \frac{K}{K+1} \right)^t,
\end{equation}
and is directly proportional to the enclosed configuration space volume. From $\alpha_t$, the partition function can be estimated according to Alg.~\ref{alg:nested-sampling}.

\subsection{Conditional Normalizing Flows}
\label{subsec:CNF}

Let $\mathcal{Z}$ be a latent space whose elements $z\in\mathcal{Z}$ follow an easy-to-sample density $\rho_Z$ (e.g.\ standard Gaussian or uniform), and let $\mathcal{X}$ be a data space whose elements $x\in\mathcal{X}$ follow a target density $\rho_X$. Throughout the remainder of this paper, we use the term base distribution to refer to $\rho_Z(z)$ and the term target distribution to refer to $\rho_X(x)$. We deliberately avoid the terms prior and posterior, which are also common in the NF literature, in order to avoid confusion with the Bayesian prior $\eta(x)$ and posterior $\pi(x)$ introduced in the previous section. Throughout this work, the latter terminology is reserved exclusively for NS quantities. A NF is an invertible, differentiable map $F:\mathcal{Z}\to\mathcal{X}$,
\[
  x = F(z), \qquad z = F\inv(x),
\]
whose Jacobian and inverse Jacobian
\[
  J_F(z) = \frac{\dd F(z)}{\dd z}, \qquad J_{F\inv}(x) = \frac{\dd F\inv(x)}{\dd x}
\]
satisfy
\begin{equation}
\label{eq:inverse_jacobians}
  J_F(z) \;=\; \bigl[J_{F\inv}(x)\bigr]\inv,
\end{equation}
so that the change-of-variables formula relates the two densities by
\begin{equation}
\label{eq:change_of_vars}
  \rho_X(x) = \rho_Z(z)\,\lvert\det J_F(z)\rvert\inv,
  \qquad
  \rho_Z(z) = \rho_X(x)\,\lvert\det J_{F\inv}(x)\rvert\inv.
\end{equation}
Note that in these expressions $x$ and $z$ are related by $x=F(z)$ and $z = F\inv(x)$. 

In what follows, the flow is realized by a neural network with parameters $\theta$ and denoted $F_\theta$. We write $J_\theta(z)\equiv J_{F_\theta}(z)$ and $J_\theta\inv(x)\equiv J_{F_\theta\inv}(x)$, so that Eq.~\eqref{eq:inverse_jacobians} becomes $J_\theta(z)=[J_\theta\inv(x)]\inv$. Pushing $\rho_Z$ through $F_\theta$ and pulling $\rho_X$ back through $F_\theta\inv$ yields the model densities
\begin{equation}
\label{eq:model_densities}
  \rho_X^\theta(x) = \rho_Z(z)\,\lvert\det J_\theta(z)\rvert\inv,
  \qquad
  \rho_Z^\theta(z) = \rho_X(x)\,\lvert\det J_\theta\inv(x)\rvert\inv,
\end{equation}
which approximate $\rho_X$ and $\rho_Z$ respectively.

We now extend this construction to a family of target distributions indexed by a condition drawn from $p_C$. For each $c\in\mathcal{C}$ we want to sample the conditional target density $\rho_{X|C}(x\mid c)$. We condition the transformation directly without affecting the base distribution, so that $\rho_{Z|C}(z\mid c)=\rho_Z(z)$ by design, and
\[
  x = F_\theta(z\mid c), \qquad z = F_\theta\inv(x\mid c).
\]
The change of variables in Eq.~\eqref{eq:change_of_vars} then reads
\begin{equation}
\label{eq:cond_change_of_vars}
  \rho_{X|C}^\theta(x\mid c) = \rho_Z(z)\,\lvert\det J_\theta(z\mid c)\rvert\inv,
  \qquad
  \rho_{Z|C}^\theta(z\mid c) = \rho_{X|C}(x\mid c)\,\lvert\det J_\theta\inv(x\mid c)\rvert\inv.
\end{equation}

An important consequence of the exact density evaluation provided by normalizing flows is that any residual bias arising from imperfect training can be corrected through reweighting: each generated sample $x = F_\theta(z\mid c)$ carries an importance weight
\begin{equation}
\label{eq:flow_weight}
  w\bigl(F_\theta(z\mid c)\bigr) \;=\; \frac{\rho_{X|C}\bigl[F_\theta(z\mid c)\bigr]}{\rho_{X|C}^\theta\bigl[F_\theta(z\mid c)\bigr]}
              \;=\; \frac{\rho_{X|C}\bigl[F_\theta(z\mid c)\bigr]}{\rho_Z(z)\,\lvert\det J_\theta(z\mid c)\rvert\inv}.
\end{equation}
These weights can be used to reweight observables or to resample configurations, provided there is sufficient overlap between the true target distribution $\rho_{X|C}$ and the distribution $\rho_{X|C}^\theta$ generated by the flow. A convenient quantitative measure of this overlap is the Kish (relative) effective sample size (RESS)~\cite{Kish1966-hz}, defined as
\begin{equation}
\label{eq:ress}
\mathrm{RESS}
=
\frac{1}{N}\frac{\left(\sum_{i=1}^{N} w_\theta(z_i\mid c)\right)^2}
{\sum_{i=1}^{N} w^2_\theta(z_i\mid c)},
\end{equation}
where $w_\theta(z_i\mid c) \equiv w\bigl(F_\theta(z_i\mid c)\bigr)$ denotes the importance weight of the $i$-th generated sample. The RESS takes values between $0$ and $1$, with unity corresponding to perfectly uniform weights and therefore an exact representation of the target distribution by the flow.

To train the conditional normalizing flow, we construct loss functions from the conditional Kullback--Leibler divergence
\begin{equation}
\label{eq:cond_KLD}
  \mathrm{KL}\bigl[p(\,\cdot\,\mid c)\,\|\,q(\,\cdot\,\mid c)\bigr]
  \;=\; \mathbb{E}_{c\sim p_C}\!\Bigl[-H_p(c) \;-\; \mathbb{E}_{x\sim p(\,\cdot\,\mid c)}\!\bigl[\ln q(x\mid c)\bigr]\Bigr],
\end{equation}
where $H_p(c)$ is the Shannon entropy of $p(\,\cdot\,\mid c)$. When applied to the flow, this divergence leads to two complementary training objectives.

\paragraph{Forward loss -- training by energy.}
Setting $p=\rho_{Z|C}$ and $q=\rho_{Z|C}^\theta$ in Eq.~\eqref{eq:cond_KLD}, substituting the second relation of Eq.~\eqref{eq:cond_change_of_vars}, and dropping the $\theta$-independent entropy yields
\begin{equation}
\label{eq:loss_fwd_generic}
  \mathcal{L}_\mathrm{fwd}(\theta)
    \;=\; -\,\mathbb{E}_{c\sim p_C}\Bigl[\,\mathbb{E}_{z\sim\rho_Z}\!\bigl[\,\ln\rho_{X|C}\!\bigl(F_\theta(z\mid c)\bigr) \;+\; \ln\bigl|\det J_\theta(z\mid c)\bigr|\,\bigr]\Bigr].
\end{equation}
Minimizing $\mathcal{L}_\mathrm{fwd}$ requires only the ability to \emph{evaluate} the target density $\rho_{X|C}$ pointwise --- no target-distributed data are needed. The training loop draws $c\sim p_C$ and $z\sim\rho_Z$, evaluates $F_\theta$ and its Jacobian, and minimizes Eq.~\eqref{eq:loss_fwd_generic} with respect to $\theta$. Since the normalization constant of $\rho_{X|C}$ is independent of the network parameters $\theta$, it contributes only an additive constant to the loss and therefore does not need to be known during training.

\paragraph{Backward loss -- training by example.}
Setting $p=\rho_{X|C}$ and $q=\rho_{X|C}^\theta$ in Eq.~\eqref{eq:cond_KLD}, substituting the first relation of Eq.~\eqref{eq:cond_change_of_vars}, and again dropping the $\theta$-independent entropy gives
\begin{equation}
\label{eq:loss_bwd_generic}
  \mathcal{L}_\mathrm{bwd}(\theta)
    \;=\; -\,\mathbb{E}_{c\sim p_C}\Bigl[\,\mathbb{E}_{x\sim\rho_{X|C}}\!\bigl[\,\ln\rho_Z\!\bigl(F_\theta\inv(x\mid c)\bigr) \;+\; \ln\bigl|\det J_\theta\inv(x\mid c)\bigr|\,\bigr]\Bigr].
\end{equation}
The backward loss requires no analytic form for $\rho_{X|C}$; only samples are needed. The same is also true for $p_C$ for which an analytical expression is not required, and it is inferred from data obtained from certain values of $c$. In the NS-Flows algorithm of Sec.~\ref{subsec:NSF}, the live sets accumulated by classical NS serve precisely as such samples, so both losses become available simultaneously.

\subsection{Nested Sampling with Conditional Normalizing Flows}
\label{subsec:NSF}

We can now compose the two ingredients. Setting $c\equiv U_\mathrm{max}$, the likelihood-constrained prior of Eq.~\eqref{eq:constrained_prior} is the conditional target $\rho_{X|C}$ of Sec.~\ref{subsec:CNF}: up to a normalizing constant,
\begin{equation}
\label{eq:target_rho}
  \rho_{X|C}(x\mid U_\mathrm{max}) \;\propto\; \mathbf{1}\{U(x)<U_\mathrm{max}\}.
\end{equation}

The indicator function in Eq.~\eqref{eq:target_rho} is non-differentiable, which precludes its direct use in the gradient-based loss of Eq.~\eqref{eq:loss_fwd_generic}. A common approach is to replace the discontinuous energy constraint with a continuous surrogate potential that penalizes configurations with energies $U(x)>U_{\mathrm{max}}$, thereby replacing Eq.~\eqref{eq:target_rho} with
\begin{equation}
\label{eq:target_rho_wall}
\rho_{X|C}^{W}(x\mid U_\mathrm{max}) \propto \exp\bigl[-W(x\mid U_\mathrm{max})\bigr].
\end{equation}
Different choices for the penalty function $W$ are possible depending on the specific application; common examples include linear, quadratic, or smooth logarithmic penalties that vanish (or become negligible) below the energy threshold and increase progressively once the constraint $U(x)>U_{\mathrm{max}}$ is violated. Substituting Eq.~\eqref{eq:target_rho_wall} into the generic loss of Eq.~\eqref{eq:loss_fwd_generic}, the term $-\ln\rho_{X|C}(x\mid c)$ reduces to $W(x\mid U_\mathrm{max})$ up to an additive normalization constant, yielding the NS-Flows forward loss
\begin{equation}
\label{eq:loss_fwd_NSF}
  \mathcal{L}_\mathrm{fwd}(\theta)
    \;=\; \mathbb{E}_{U_\mathrm{max}\sim p_C}\!\Bigl[\,\mathbb{E}_{z\sim\rho_Z}\!\bigl[\,W\!\bigl(F_\theta(z\mid U_\mathrm{max})\bigr) \;-\; \ln\bigl|\det J_\theta(z\mid U_\mathrm{max})\bigr|\,\bigr]\Bigr].
\end{equation}
The conditioning variable can be sampled from any distribution $p_C$ with support restricted to the interval $[C_{\min}, C_{\max}]$, where $C_{\min}$ denotes the energy bound reached at the current stage of the NS simulation. Typical choices include a uniform distribution over this interval or distributions biased toward lower energy bounds, which more closely reproduce the distribution of thresholds generated during the nested-sampling trajectory.

Specializing Eq.~\eqref{eq:loss_bwd_generic} likewise gives the backward loss
\begin{equation}
\label{eq:loss_bwd_NSF}
  \mathcal{L}_\mathrm{bwd}(\theta)
    \;=\; -\,\mathbb{E}_{U_\mathrm{max}\sim p_C}\!\Bigl[\,\mathbb{E}_{x\sim\eta(\,\cdot\,\mid U_\mathrm{max})}\!\bigl[\,\ln\rho_Z\!\bigl(F_\theta\inv(x\mid U_\mathrm{max})\bigr) \;+\; \ln\bigl|\det J_\theta\inv(x\mid U_\mathrm{max})\bigr|\,\bigr]\Bigr],
\end{equation}
with the outer expectation realized over the collection of live sets accumulated during the nested-sampling trajectory, each associated with the energy bound at which it was generated. We adopt either a standard Gaussian $\rho_Z(z)\propto\exp(-\tfrac{1}{2}\|z\|^2)$ on $\mathbb{R}^n$ or a uniform $\rho_Z$ on a finite domain $\Omega$; for the uniform choice the term $\ln\rho_Z$ in Eq.~\eqref{eq:loss_bwd_NSF} is constant and drops, while for the Gaussian it becomes $-\tfrac{1}{2}\|F_\theta\inv(x\mid U_\mathrm{max})\|^2$ up to a constant. 

In practice, the complementary information provided by the forward and backward objectives can be combined during training. Training NS-Flows then minimizes a weighted sum for forward and backward losses
\begin{equation}
\label{eq:total_loss}
\mathcal{L}_{\mathrm{tot}} = w_{\mathrm{fwd}}\mathcal{L}_\mathrm{fwd}+w_{\mathrm{bwd}}\mathcal{L}_\mathrm{bwd},
\end{equation}
where $w_{\mathrm{fwd}}$ and $w_{\mathrm{bwd}}$ are hyperparameters tuned in the validation stage.

\subsection{A flow-based nested sampling algorithm for atomistic systems}
\label{subsec:nsf_alg}

Following the backbone of Williams \emph{et al.}~\cite{williams_nested_2021}, samples generated by the trained flow populate a pool from which NS draws replacements for its worst-performing walker, removing the MCMC bottleneck. We extend their approach by conditioning the flow on the energy bound and training it on a sliding window of the most recent $h$ live sets, so that a single network represents the family $\{\eta(\,\cdot\,\mid U_\mathrm{max}^t)\}_t$ along the NS trajectory~\cite{falkner_conditioning_2023}. Since consecutive constrained ensembles differ only incrementally along the NS trajectory, conditioning on previous live sets allows the flow to exploit continuity in the annealing process and progressively adapt to the evolving constrained distribution. This temporal continuity also enables different flow training strategies: each retraining stage may either start from randomly initialized network parameters or continue from the parameters obtained in the previous optimization stage, effectively resulting in an online fine-tuning procedure. The former strategy may help avoid poor local minima inherited from earlier training cycles, whereas the latter exploits the strong similarity between consecutive constrained distributions to progressively refine a single model throughout the NS trajectory. An intermediate strategy is also possible, where the network is periodically reinitialized after a prescribed number of retraining stages while continuing to fine-tune the previously learned parameters in the intervening updates.

\begin{algorithm}[htbp]
\caption{Nested Sampling with Conditional Normalizing Flows (NS-Flows). }
\label{alg:nsf}
\begin{algorithmic}[1]
\Require Live-set size $K$, pool size $N$, warm-up length $\mathscr{M}'$, dilution length $\mathscr{M}$, retention horizon $h$, termination parameter $\varepsilon$
\For{$k=1$ \textbf{to} $K$}
  \State draw $X_0^{(k)}\sim\eta$
\EndFor
\State run $\mathscr{M}'$ steps of Alg.~\ref{alg:nested-sampling}
\State $\mathcal{H}\gets\emptyset$ \Comment{sliding window of $\bigl(\text{live set},\,U_\mathrm{max}\bigr)$ snapshots}
\State $\mathcal{P}\gets\emptyset$ \Comment{empty pool}
\For{$t=\mathscr{M}'{+}1, \mathscr{M}'{+}2,\dots$}
    \State $m\gets\arg\max_{k} U\!\big(X_{t-1}^{(k)}\big)$;\quad $U_\mathrm{max}^t\gets U\!\big(X_{t-1}^{(m)}\big)$
    \State
    \If{$\mathcal{P}=\emptyset$}
        \State run $\mathscr{M}$ steps of Alg.~\ref{alg:nested-sampling} \Comment{dilute synthetic samples before next retraining}
        \State append $\bigl(\{X_{t-1}^{(k)}\}_{k=1}^{K},\,U_\mathrm{max}^t\bigr)$ to $\mathcal{H}$; truncate $\mathcal{H}$ to the last $h$ entries
        \State train $F_\theta$ on $\mathcal{H}$ by minimizing $\mathcal{L}_{\mathrm{tot}}$ (Eq.~\eqref{eq:total_loss})
        \State $U_\mathrm{max}^\mathrm{train}\gets U_\mathrm{max}^t$
        \State draw $\{z^{(j)}\}\sim\rho_Z$; compute $x^{(j)}=F_\theta(z^{(j)}\mid U_\mathrm{max}^\mathrm{train})$ and weights $w_\theta(z^{(j)}\mid U_\mathrm{max}^\mathrm{train})$
        \State $\mathcal{P}\gets$ accept/reject the $x^{(j)}$ according to $w_\theta$ until $|\mathcal{P}|=N$ \Comment{Eq.~\eqref{eq:flow_weight}}
        \State \textbf{continue}
    \EndIf
    \State
    \State draw $X'$ uniformly from $\mathcal{P}$;\quad $\mathcal{P}\gets\mathcal{P}\setminus\{X'\}$
    \If{$U(X')\ge U_\mathrm{max}^t$} \Comment{$U_\mathrm{max}^t<U_\mathrm{max}^\mathrm{train}$ in general}
        \State \textbf{continue}
    \EndIf
    \State $\tilde X_{t-1}\gets X_{t-1}^{(m)}$;\quad $\tilde w_t\gets(\alpha_{t-1}-\alpha_t) \exp \{ -\beta U_\mathrm{max}^t \}$
    \State $X_t^{(k)}\gets X_{t-1}^{(k)}$ for $k\neq m$;\quad $X_t^{(m)}\gets X'$
    \If{stopping condition based on $\varepsilon$ is satisfied}
        \State $T\gets t$;\quad\textbf{break}
    \EndIf
\EndFor
\State \Return evidence estimator $Z=\sum_t\tilde w_t$ and weighted samples $\bigl\{\tilde X_t,\;\tilde w_t/\!\sum_s\tilde w_s\bigr\}_t$
\end{algorithmic}
\end{algorithm}

The entire flow-based NS procedure is summarized in Alg.~\ref{alg:nsf}.
After generation, an acceptance-rejection step uses the flow-derived weight $w_\theta(z\mid c)$ of Eq.~\eqref{eq:flow_weight} to correct for residual mismatch between $\rho_{X|C}^\theta$ and the indicator target~\cite{williams_nested_2021}; unlike a multinomial bootstrap, this avoids repeating configurations in the pool, which remain unique instead. Once generated, configurations can be drawn efficiently from the pool only while $U_\mathrm{max}^t$ does not drift far below $U_\mathrm{max}^\mathrm{train}$: each NS iteration tightens the bound, so efficiency degrades monotonically and triggers a retraining once exhausted. A number $\mathscr{M}$ of MCMC steps interleaved between pool exhaustion and retraining serves to dilute synthetic samples in the live set before the next training cycle, mitigating the accumulation of distributional bias and correlations associated with repeatedly retraining generative models on self-generated samples~\cite{shumailov_ai_2024}.

\subsection{Flow architecture and symmetries}
\label{subsec:arch}

The performance of the proposed approach depends crucially on the ability of the NF to represent the highly structured constrained distributions encountered along the NS trajectory. In atomistic systems, this capability is strongly influenced by how physical symmetries are incorporated into the architecture. We therefore employ an equivariant transformer-based flow architecture specifically designed to encode the permutation, translational, and periodic symmetries characteristic of condensed-matter systems~\cite{wirnsberger_targeted_2020,wirnsberger_normalizing_2022}. In practice, the invertible transformation defining the flow is constructed from rational quadratic spline bijectors~\cite{durkan_neural_2019}, whose parameters are predicted by an equivariant transformer network~\cite{vaswani_attention_2017}.

Translational invariance is treated explicitly by expressing each configuration in internal coordinates. Rather than using absolute Cartesian positions, all particle coordinates are represented relative to a reference particle, chosen initially as the particle with index $0$. In addition, the training dataset is continuously augmented during optimization. Whenever a configuration is loaded by the data loader, the reference particle used to define the internal coordinates is selected randomly and a random symmetry operation of the point group of the simulation cell ($D_4$ in the two-dimensional square box) is applied to the configuration. This augmentation procedure strongly increases the effective diversity of the training data and, in practice, largely suppresses overfitting throughout the simulation.

The choice of a square simulation cell is also motivated by this augmentation strategy. Although the minimum-energy configuration of the two-dimensional LJ system is a hexagonal lattice, which is naturally accommodated by a rectangular simulation cell with aspect ratio $L_x/L_y=\sqrt{3}/2$, such a geometry preserves only a subset of the symmetry operations available in the square box. Consequently, the number of transformations that can be exploited for data augmentation is significantly reduced, leading in practice to less effective training. For this reason, all results presented in the main text are obtained using square simulation cells. Results for a nine-particle system in a rectangular simulation cell with aspect ratio $L_x/L_y=\sqrt{3}/2$ are reported in the Supplementary Material.

\subsection{Simulation and training details}
\label{subsec:simdetails}

We now specify the parameters of the NS and training protocol used for the LJ simulations of Sec.~\ref{sec:results}.

The NS simulations are initialized from a common live set of $K=10^4$ walkers, obtained by partially annealing for $10^3$ standard NS steps an initially uniform population using conventional NS. Starting all conventional and flow-assisted simulations from this same live set ensures a fair comparison, and the cost of generating it is excluded from the reported energy-evaluation counts and wall-clock times. Before activating the normalizing-flow component, the system is annealed for a further $\mathscr{M}'=100$ conventional NS iterations. During these standard NS stages, batches of $10^3$ walkers are evolved in parallel through rejection-based MCMC updates consisting of $10^2$ trial particle displacements per walker. To maintain a reasonable acceptance probability, the maximum displacement employed in the MCMC proposals is reduced by a factor of two whenever the acceptance ratio falls below $0.5$. Unless otherwise specified, the same standard NS protocol and MCMC parameters are employed throughout all subsequent refresh stages where the algorithm temporarily switches back from flow-based generation to conventional NS (in particular, in Alg.~\ref{alg:nsf} we set $\mathscr{M} = \mathscr{M}'$).

Following the sliding-window scheme of Sec.~\ref{subsec:nsf_alg}, each live set is stored together with its associated NS energy threshold and used to train the conditional flow, with the energy bound acting as the conditioning variable. To limit memory growth and maintain relevance to the current stage of the simulation, only the three most recent live sets are retained during training, the highest-energy dataset being discarded whenever a new one is added. In the present implementation the flow is trained exclusively in the backward direction (i.e. \(x\rightarrow z\)) through the transformation $F\inv_{\theta}$ (namely $w_{\text{fwd}} =0$ and $w_{\text{bwd}}=1$ in Eq.~\eqref{eq:total_loss}), using only configurations sampled from the constrained target distribution, so that no target-system energy evaluations enter the optimization. This choice is motivated by the observation that the forward objective derived from Eq.~\eqref{eq:target_rho} (and analogously from its differentiable counterpart, Eq.~\eqref{eq:target_rho_wall}) provides only a weak learning signal about the internal organization of the constrained ensemble. While the target density specifies the constrained region, it provides comparatively little information about how probability mass should be transported from the base distribution into that region. This information is instead conveyed directly by the sampled configurations, which provide abundant examples of the target distribution through the current and previous live sets accumulated during the nested-sampling trajectory. Furthermore, avoiding repeated energy evaluations during training is particularly advantageous for atomistic systems, where energy calculations often constitute the dominant computational cost. Training is performed with the Adam optimizer together with a learning-rate scheduling strategy; the precise choice of scheduler, learning-rate protocol, and number of optimization steps constitutes an important set of hyperparameters of the method and is examined comparatively in the results (Sec.~\ref{sec:results}). The different strategies introduced in Sec.~\ref{subsec:nsf_alg} are used in what follows: the cosine-annealing runs carry the parameters over from one training stage to the next, effectively resulting in an online fine-tuning procedure; the 1C+CA/CA(5) runs re-initialize them whenever the full schedule is applied and fine-tune in between; the 1C+CA protocol re-initializes the parameters of the network before every training stage for the larger pool case (Fig.~\ref{fig:timings}a), while the previous network parameters are used for the smaller pool size (Fig.~\ref{fig:timings}b), where the live sets are closer to each other. The choice made in each run is listed in the Supplementary Material.

After each training stage, the conditioned network is used to generate a pool of configurations through the rejection-resampling procedure of Sec.~\ref{subsec:nsf_alg}. Generation begins by drawing configurations from an ideal-gas base distribution, corresponding to particles distributed uniformly within a simulation box of the same size as the target system. Apart from the thermodynamics constraints imposed by the number of particles and the simulation volume, this base distribution contains no information about the interactions, thermodynamic state, or structural organization of the system. These configurations are transformed by the conditioned normalizing flow, where the current training energy threshold $U^{\mathrm{train}}_{\max}$ is provided as the conditioning variable, and are subsequently subjected to the importance-weight-based rejection-resampling procedure in order to recover the desired constrained distribution and construct the pool used during the subsequent NS stage. The pool size constitutes an important hyperparameter controlling the trade-off between retraining frequency, generative efficiency, and computational cost, and its effect on the overall performance of the algorithm is examined in the results (Sec.~\ref{sec:results}). Although the target pool size is chosen a priori, the rejection-resampling procedure accepts only a fraction of the generated configurations. As a result, the total number of generated samples is not known in advance, and additional batches are produced until the prescribed pool size is reached. Since the rejection-resampling procedure employs relative importance weights normalized over finite generated batches, the effective normalization constant may differ slightly between independently generated candidate sets. Consequently, combining configurations originating from multiple independently normalized batches introduces a small finite-sample inconsistency analogous to that encountered in self-normalized importance sampling~\cite{chopin_introduction_2020}. In practice, this effect is expected to become negligible for sufficiently large generated batches and sufficiently accurate flow models, where the relative-weight distribution becomes increasingly stable across successive generations.

\section{Results and Discussion}\label{sec:results}

We now turn to a concrete demonstration of flow-assisted nested sampling. As a testbed, we consider the two-dimensional LJ system of eight disks already introduced in Sec.~\ref{sec:landscapes}. 
Although modest in size, this benchmark remains representative of the challenges posed by atomistic posterior landscapes. Recent advances based on graph neural networks~\cite{schebek_scalable_2026} and equivariant architectures~\cite{midgley_flow_2023} have enabled normalizing flows to scale to substantially larger crystalline systems, whereas extending these methods to structurally heterogeneous liquid or disordered phases remains an active area of research~\cite{jung_normalizing_2024,coretti_learning_2025,de_santis_sampling_2026}. At the same time, the present system is sufficiently small to permit a detailed analysis and a direct comparison with the original flow-assisted nested-sampling framework of Williams et al.~\cite{williams_nested_2021}. Despite its modest size, it already exhibits the dense, collective coupling and combinatorial, permutation-induced multimodality that characterize realistic atomistic posterior landscapes. We study it under periodic boundary conditions at a constant reduced density $\rho = 0.95$, which places the system in a relatively dense fluid/solid regime characterized by strong excluded-volume effects and multiple metastable structural motifs~\cite{li_phase_2020}.

\begin{figure}[htbp]
    \centering
    \includegraphics[width=\linewidth]{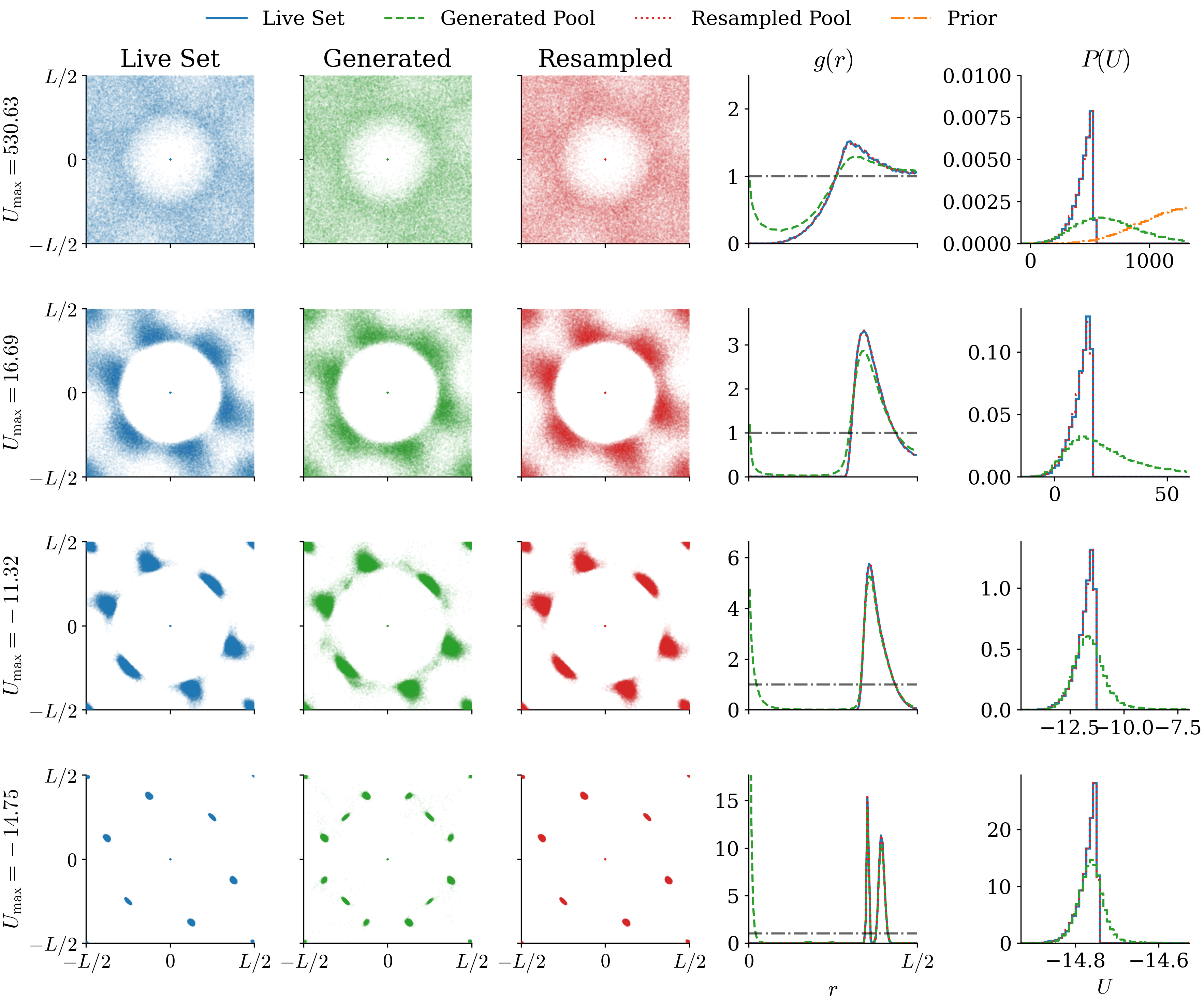}
    \caption{Representative configurations and corresponding structural and energetic distributions obtained at different stages of the nested-sampling trajectory for the two-dimensional Lennard--Jones system at $\rho=0.95$. Each row corresponds to a different nested-sampling energy threshold $U$, decreasing from top to bottom. The first three columns show ensembles of configurations drawn from the live set used for training, the generated pool obtained directly from the normalizing flow, and the final pool after rejection-resampling, respectively. In each panel, all configurations are superimposed after transformation to internal coordinates and alignment with respect to a common reference particle, so that the resulting image represents the positional probability density of the remaining particles relative to the reference. The alignment procedure, and the two-fold orientational degeneracy of the low-energy crystal that it removes, are described in the Supplementary Material. The fourth column reports the corresponding radial distribution functions $g(r)$, with the dash-dotted horizontal line indicating the ideal-gas limit $g(r)=1$ (the base distribution). The fifth column shows the corresponding energy probability distributions $P(U)$ for the live set, generated pool, and resampled pool. In the top row, the energy distribution associated with the ideal-gas prior is also reported for comparison. For the lower-energy rows this distribution is omitted, as the corresponding energy range becomes increasingly separated from that of the constrained target ensembles and would not be visible on the scale of the figure.}
    \label{fig:pools_aligned}
\end{figure}

\subsection{Simulation setup}

The complete simulation is performed for a total of $5\times10^5$ NS iterations, following the initialization, training, and generation protocol detailed in Sec.~\ref{subsec:simdetails}. During this process, the system evolves from an initial reduced energy of approximately $U = 2540.9$ down to a final value of approximately $U = -14.8$, spanning both highly disordered and strongly ordered regions of configuration space (see Fig.~\ref{fig:pools_aligned}). During the full annealing trajectory, the simulation alternates repeatedly between standard NS stages, normalizing-flow training phases, and pool-generation steps, resulting in multiple retraining cycles as the constrained distribution progressively evolves toward lower energies.

\subsection{Validation against standard nested sampling}
\begin{figure}[htbp]
    \centering
    \includegraphics[width=\linewidth]{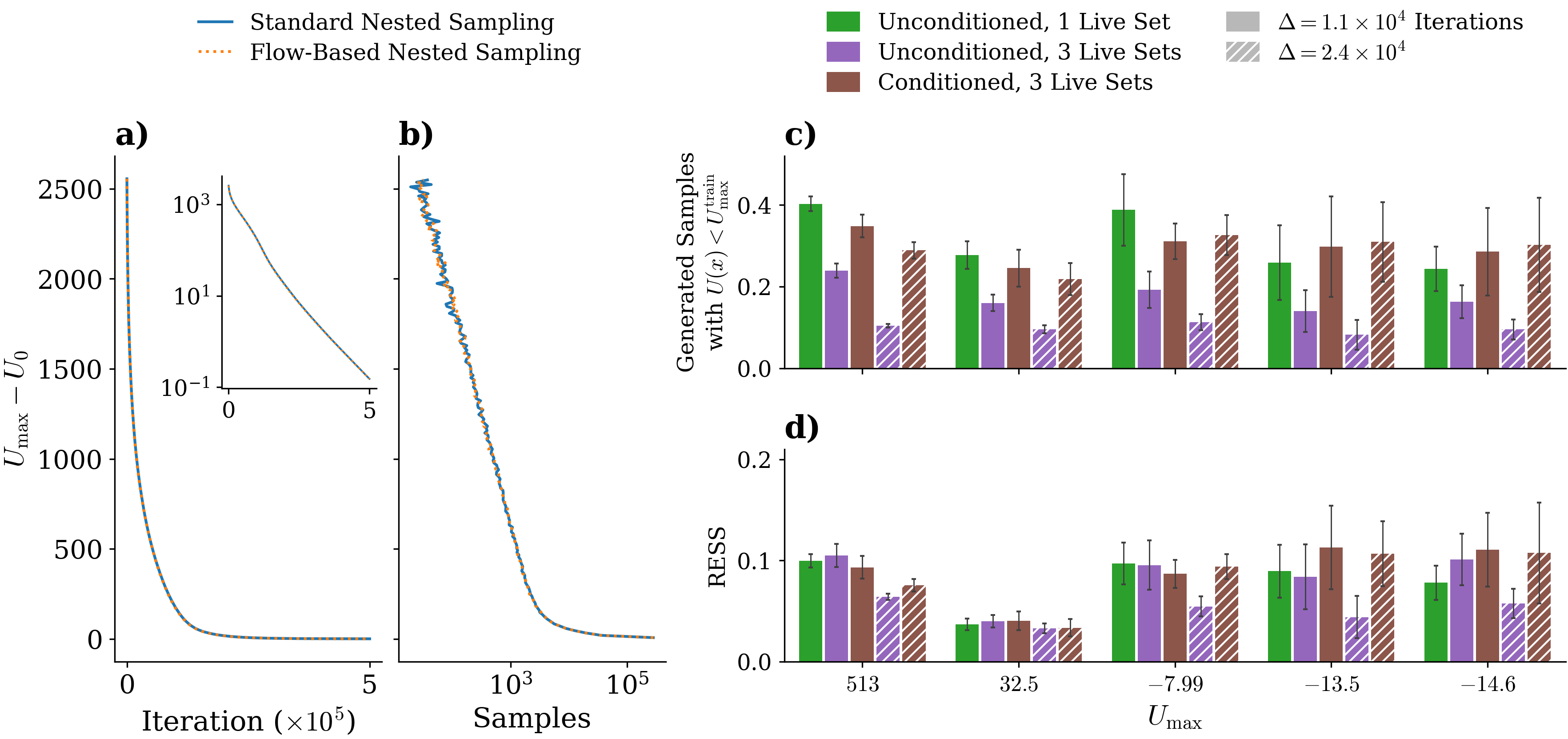}
\caption{Validation of the flow-assisted algorithm, and the effect of conditioning on the energy threshold. \textbf{a)} Nested-sampling energy bound $U_{\max}$ against iteration for conventional rejection Monte Carlo (solid blue) and the flow-assisted scheme (dotted orange), measured from the relaxed minimum energy $U_0$; the two trajectories are statistically indistinguishable. The inset repeats the comparison on a logarithmic scale, covering the whole range from the initial high-energy configurations to the lowest-energy states explored. \textbf{b)} Distribution of the sampled energy bounds; the overlap confirms that replacing the Monte Carlo proposal with the conditional normalizing flow preserves the statistical properties of the nested-sampling process. \textbf{c, d)} Three training strategies compared at five live sets approximately equally spaced along a reference run, the two that use a window of live sets at two window widths: an unconditioned flow trained on the target live set alone, an unconditioned flow trained on it together with the two preceding live sets corresponding to higher energy thresholds, and the proposed conditional flow trained on those same three live sets with each configuration labeled by its energy threshold. Color identifies the strategy and hatching the separation $\Delta$ between consecutive live sets of the window, $\Delta=1.1\times10^4$ and $\Delta=2.4\times10^4$ NS iterations, close to the retraining frequencies of the small- and large-pool production runs of Fig.~\ref{fig:timings}; the single-live-set strategy has no window and therefore appears once. \textbf{c)} Fraction of generated configurations satisfying the target energy constraint and \textbf{d)} relative effective sample size (RESS) of the generated distribution. Every bar is the mean over ten independently trained networks, each generating ten pools of $10^4$ configurations; the error bars are the standard deviation over networks, with the contribution of the generation estimated from the ten pools and removed, so that they measure the variability of the training alone.}
    \label{fig:validation_conditioning}
\end{figure}

Before addressing the performance of the proposed algorithm, we first verify that replacing the MCMC proposal with the flow-based generator preserves the statistical behavior of the nested-sampling procedure. Figure~\ref{fig:validation_conditioning} (panels a and b) compares representative NS trajectories obtained using the standard rejection Monte Carlo approach and the proposed flow-based algorithm (best performing run, see Fig.~\ref{fig:timings}f). The left panel reports the evolution of the energy threshold $U_{\max}$ as a function of the NS iteration. The two trajectories are statistically indistinguishable and exhibit the same progressive compression of the accessible configuration space throughout the simulation. The right panel compares the corresponding distributions of sampled energy thresholds, showing excellent agreement over the entire range explored during the run. These results demonstrate that the proposed replacement of the MCMC proposal with a conditional normalizing flow leaves the NS dynamics unchanged while providing an alternative mechanism for generating configurations below the current energy bound.

\subsection{Effect of conditioning on the energy threshold}
Having established that the flow-assisted algorithm faithfully reproduces the statistical behavior of standard NS, we next assess the role of conditioning on the NS energy threshold. The primary motivation for conditioning is to enable the effective reuse of configurations accumulated during previous stages of the NS trajectory. Without conditioning, two natural training strategies are possible. The first is to train an unconditioned flow using only the current live set, thereby matching the desired constrained distribution but discarding all information collected during previous retraining stages. The second is to increase the amount of training data by combining configurations from multiple live sets into a single dataset. While this provides the network with substantially more examples, the samples originate from different constrained distributions and are presented without any information identifying the energy thres\-hold at which they were generated. Consequently, the network is required to approximate several distinct target distributions simultaneously, despite having no mechanism to distinguish between them.

Panels c and d of Fig.~\ref{fig:validation_conditioning} compare these alternatives at five representative stages of a standard NS simulation. For each energy threshold, three flows are trained with the same optimization protocol: an unconditioned flow trained on the corresponding live set alone, an unconditioned flow trained on that live set together with the two preceding ones, and the proposed conditional flow trained on the identical three-live-set dataset with each configuration labeled by its energy threshold. The two multi-set strategies are compared at two window widths, $\Delta=1.1\times10^4$ and $\Delta=2.4\times10^4$ NS iterations between consecutive live sets, which are close to the retraining frequencies of the small- and large-pool production runs of Fig.~\ref{fig:timings}. Ten networks are trained independently in each case, and each generates ten pools of $10^4$ configurations; the error bars are the standard deviation over networks, with the contribution of the generation estimated from the ten pools and removed.

Conditioning recovers almost all of the cost of training on a window of live sets. At the narrower window the conditional flow generates acceptable configurations at $1.45$ to $2.12$ times the rate of the unconditioned flow trained on the same data, at every threshold, and at the wider window the factor rises to between $2.29$ and $3.76$. What remains to be investigated in more detail is whether conditioning beats training on the current live set alone: at the two highest thresholds training on the single, most recent live set seems to remain the most efficient of the three strategies (efficiency of $0.40$ against $0.35$ at $U_{\max}=513$), and at the two lowest the conditional flow is ahead only by $15$ and $18\%$, within the spread over networks. Nonetheless, at the two lowest thresholds the conditional flow is ahead in both quantities, the fraction of generated configurations satisfying the energy bound and the relative effective sample size. A clear benefit of conditioning is that it makes a window of live sets usable at all, rather than that a window is better than the current live set alone --- which raises the question of how wide that window may be.

The width of the window, and not the energy separation it spans, is what the conditioning repairs. Because NS compresses the prior mass at a fixed rate, live sets separated by $\Delta$ iterations always differ by the same factor $e^{-\Delta/K}$ in enclosed volume, wherever they sit on the trajectory, while the energy difference between them falls from $\Delta U\approx3\times10^2$ at the highest threshold considered here to $10^{-1}$ at the lowest. Widening the window from $\Delta=1.1\times10^4$ to $2.4\times10^4$ costs the unconditioned flow trained on the three live sets $40$ to $57\%$ of its efficiency at every threshold, and leaves the conditional flow unchanged within error. The energy spread is not predictive: it varies by three orders of magnitude between the thresholds considered while the penalty does not.

The relative effective sample size is a noisier estimator than the acceptance fraction, and at the narrower window it barely separates the three architectures at any threshold. At the wider window it does: the conditional flow reaches $1.72$, $2.42$ and $1.87$ times the RESS of the unconditioned flow trained on the same data at the three lowest thresholds. Pooling live sets without conditioning therefore costs acceptance everywhere, but degrades the weight distribution only once the window spans an appreciable range of the prior mass.

\subsection{Computational efficiency}
Assessing the performance of generative models in molecular simulation is not straightforward~\cite{jung_normalizing_2024,coretti_boltzmann_2024,john_comparison_2025}. Unlike conventional simulation methods, where the dominant computational bottleneck is usually well defined, flow-based approaches require balancing the cost of training the generative model against the quality of the generated samples. In this specific application, longer and more accurate training generally improves the approximation of the constrained target distribution and increases the acceptance of generated configurations, but at the expense of additional optimization time. Conversely, shorter training reduces the computational overhead while producing lower-quality proposals that require more rejection during pool construction.

In standard NS, the dominant computational cost arises from the repeated evaluation of the potential energy during the MCMC procedure used to generate approximately independent samples from the constrained prior. In the present implementation, each NS iteration evolves a batch of $10^3$ walkers in parallel through rejection-based MCMC updates, with $10^2$ trial particle displacements attempted for every walker. This results in approximately $10^5$ energy evaluations per iteration, or roughly $5\times10^{10}$ evaluations over the full annealing trajectory of $5\times10^5$ NS steps. Although the LJ system considered here is relatively inexpensive in terms of individual energy evaluations, the cumulative cost of this decorrelation procedure remains substantial, requiring approximately 12 hours of wall-clock time on the GPU architecture employed in the present work~\cite{turisini_leonardo_2024}.

The flow-assisted approach replaces most of these energy evaluations with the cost of training a generative model. In the present implementation, where the flow is trained exclusively ``by example'' through the backward loss, no target-system energy evaluations are required during training, since the optimization depends only on configurations already contained in the live sets. This contrasts with forward or energy-based training schemes, where energy evaluations explicitly enter the loss function. Likewise, generating configurations from the trained flow requires no energy computations. Energy evaluations re-enter only during the rejection-resampling stage used to construct the pool, where each generated configuration is evaluated once. Since the resulting energies can be retained together with the generated configurations, subsequent NS iterations can determine whether a sampled configuration satisfies the current threshold $U_{\max}$ without reevaluating the potential.

This reduction in energy evaluations is achieved at the expense of gradient-based optimization and backpropagation through a relatively large neural network. The architecture employed in the present work consists of 28 coupling blocks and approximately $2.3\times10^7$ trainable parameters. For the LJ system, where individual energy evaluations are comparatively inexpensive, the cost of training such a model can become comparable to, or even exceed, the cost of evaluating the potential itself. Consequently, the overall efficiency depends on the balance between neural-network training and the reduction in energy evaluations. Complete details of the network architecture and training protocol are provided in the Supplementary Material.

The introduction of a trainable generative model into the NS workflow creates a considerably richer hyperparameter landscape than that of conventional NS. Besides the optimization protocol itself, parameters such as the retraining frequency, the pool size, and the amount of accumulated training data all influence the overall efficiency of the method. In practice, however, the dominant trade-off arises from the interplay between the amount of optimization performed during training and the number of configurations generated and stored in the pool. If the computational bottleneck lies in expensive energy evaluations, longer training stages can be justified to obtain more accurate generative models, reducing rejection during pool construction and allowing larger pools to sustain many NS iterations before retraining becomes necessary (see SM for details). Conversely, when gradient evaluations dominate the computational cost, shorter optimization protocols may provide a better overall compromise despite increasing the rejection rate during pool generation. Unlike conventional NS, where efficiency is controlled primarily by the degree of MCMC decorrelation, this flexibility allows the flow-assisted algorithm to be adapted to systems with different computational bottlenecks.

\begin{figure}[htbp]
    \centering
    \includegraphics[width=0.95\linewidth]{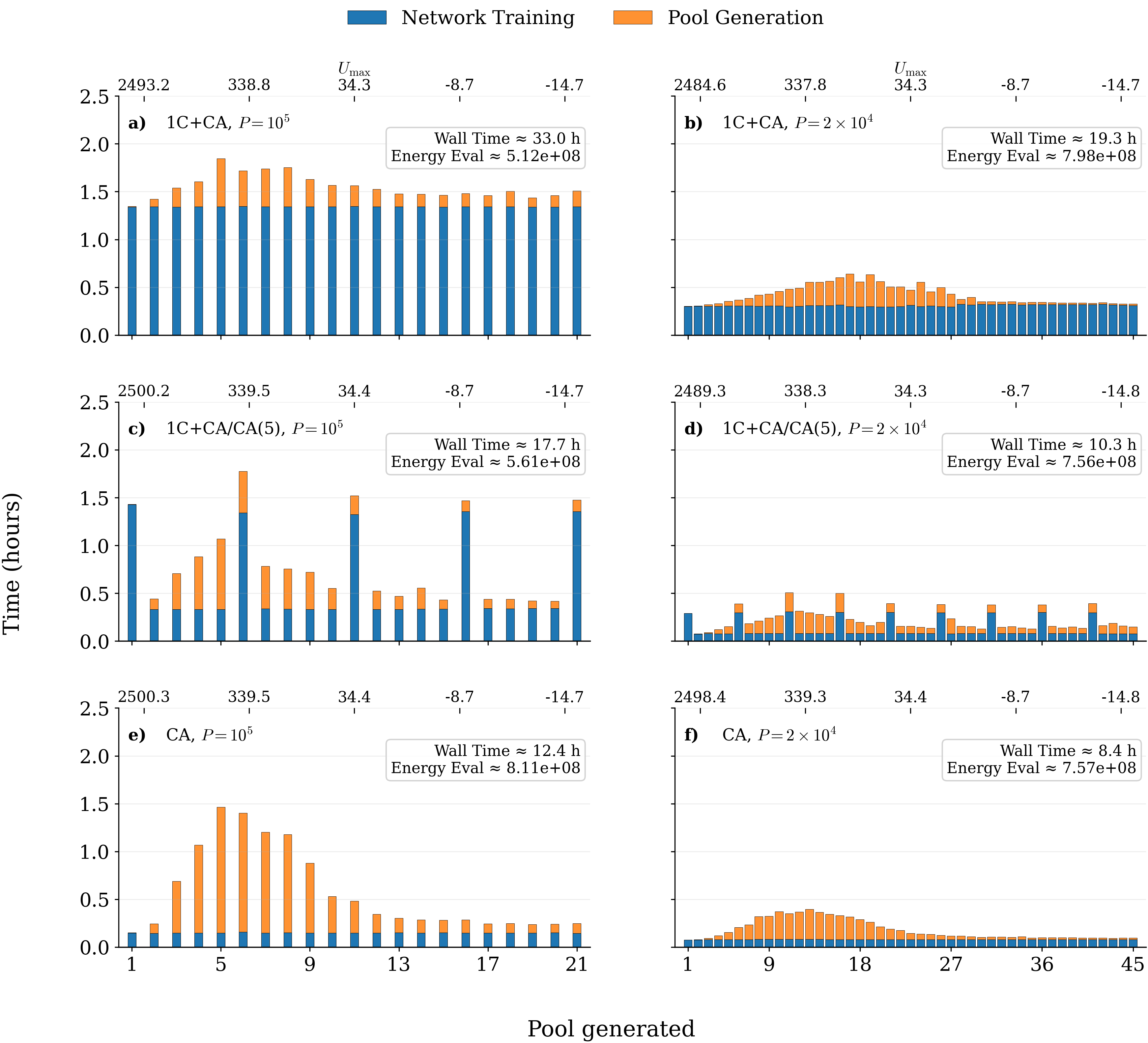}
    \caption{Time decomposition associated with the generation of successive configuration pools during nested-sampling simulations accelerated by conditional normalizing flows. Each bar corresponds to a newly generated pool and is partitioned into the time spent in the different stages of the algorithm, including network training (blue bars) and pool generation together with the rejection-resampling procedure (orange bars). All simulations start from the same initial live set of walkers and perform the same total number of nested-sampling iterations. Pool size of \textbf{Left column:} $10^5$ configurations and \textbf{Right column:} $2\times10^4$ configurations. \textbf{Top row:} One Cycle followed by Cosine Annealing at every training stage (1C+CA), with \textbf{a)} 3375 + 1125 optimization steps (OSs) and \textbf{b)} 750 + 250 OSs. \textbf{Middle row:} the same pair applied only at every fifth training stage, with CA alone in between (1C+CA/CA(5)), with \textbf{c)} 3375 + 1125 OSs and 1125 CA OSs in between, and \textbf{d)} 750 + 250 OSs and 250 CA OSs in between. \textbf{Bottom row:} CA at every training stage, with \textbf{e)} 500 OSs and \textbf{f)} 250 OSs. The three protocols are compared within a single training stage in the Supplementary Material. Wall time and number of energy evaluations for the whole nested sampling run are reported in the box on the upper right of each panel.}
    \label{fig:timings}
\end{figure}

This trade-off is illustrated in Fig.~\ref{fig:timings}, where different combinations of learning-rate schedules, optimization lengths, and pool sizes are compared through a decomposition of the computational time spent on training and pool generation throughout the nested-sampling trajectory. Two target pool sizes are considered, namely $10^5$ configurations (left column) and $2\times10^4$ configurations (right column). For each pool size, three learning-rate protocols are investigated, all built from the One Cycle (1C) and Cosine Annealing (CA) schedulers: 1C followed by CA at every training stage (1C+CA, top row); the same pair applied only at every fifth training stage, with CA alone in between (1C+CA/CA(5), middle row); and CA alone at every stage (CA, bottom row). Within each protocol, two optimization lengths are compared, corresponding to a longer and a shorter training schedule.
For the larger pool size of $10^5$ configurations (left column), only 21 retraining stages are required during the entire nested-sampling trajectory, allowing the training cost to be amortized over many subsequent NS iterations. In this regime, longer training schedules reduce the rejection rate during pool generation and therefore the number of energy evaluations, but at the expense of substantially longer optimization times. This trade-off is evident in the left column of Fig.~\ref{fig:timings}: moving from panel a) to panel e), the total number of energy evaluations increases from $5.12\times10^8$ to $8.11\times10^8$, while the wall-clock time decreases from approximately 33 hours to less than 12.5 hours. The full breakdown of the energy evaluations, separated into pool generation, standard NS stages and training, is given for every run in the Supplementary Material. For the smaller pool size of $2\times10^4$ configurations (right column), retraining occurs more frequently, but each pool is considerably cheaper to generate. All three simulations complete the trajectory in roughly $60\%$ of the wall-clock time of their large-pool counterparts. In terms of energy evaluations, however, the comparison is reversed for two of the three schedules: for 1C+CA and for 1C+CA/CA(5) the larger pool requires fewer evaluations ($5.12\times10^8$ against $7.98\times10^8$, and $5.61\times10^8$ against $7.56\times10^8$), while the two pool sizes are indistinguishable for the CA scheduler ($7.57\times10^8$ against $8.11\times10^8$). Since none of these simulations is seeded, each is a single realization: ten repetitions of the run of panel f, reported in the Supplementary Material, spread by $4.7\%$ in total energy evaluations, so that two of the runs compared here must differ by about $13\%$ before the difference reaches two standard deviations. This is comfortably satisfied by the $44\%$ and $29\%$ separating the two pool sizes under 1C+CA and 1C+CA/CA(5), but not by the $7\%$ between panels e and f. The largest reduction in energy evaluations compared to standard NS, close to a hundredfold, is obtained by the longest-trained protocol of panel a, at the price of $33$ hours of wall-clock time, nearly three times that of the reference. Conversely, the fastest simulation (panel f) completes the full $5\times10^5$-step annealing trajectory in less than $8.5$ hours using $7.57\times10^8$ energy evaluations, a reduction of roughly seventy-fold relative to standard NS (Fig.~\ref{fig:validation_conditioning}) while shortening the overall simulation time by about $30\%$. These improvements are consistent with the performance gains reported by Williams \emph{et al.}~\cite{williams_nested_2021} for gravitational-wave inference using NESSAI~\cite{nessai}, where speedups between 1.40 and 2.32 were obtained.

A substantial part of the cost in terms of energy evaluations is attributable to the dilution stage, namely the $\mathscr{M}=100$ conventional NS iterations performed after each pool is exhausted and before the network is retrained. These iterations account for between $25\%$ and $40\%$ of the energy evaluations of the large-pool simulations (panels a, c and e) and for between $56\%$ and $59\%$ of those of the small-pool ones (panels b, d and f), where retraining is more than twice as frequent; in the fastest simulation of panel f they alone consume $4.44\times10^8$ of the $7.57\times10^8$ total evaluations. Their purpose, as discussed above, is to avoid training the generative model on its own output: each dilution iteration replaces part of the live set with walkers evolved by the constrained random walk of conventional NS, so that the next training set is not composed entirely of synthetic configurations~\cite{shumailov_ai_2024}. In the Supplementary Material we estimate the probability that a stage of $\mathscr{M}$ iterations refreshes every one of the $K$ walkers at least once: for $\mathscr{M}=100$ this probability is $0.76$, while raising it to $0.99$ would require $\mathscr{M}=132$. How much dilution is genuinely necessary, however, remains an open question. The acceptance-rejection step already corrects the generated distribution towards the constrained target, so a residual correlation between successive training sets is expected to degrade the efficiency of the generator rather than the correctness of the sampling, and the appropriate value of $\mathscr{M}$ is therefore a matter of stability over many retraining cycles rather than of bias at any single one. The two limits bracket the attainable performance. Suppressing the dilution stage entirely would leave the $3.13\times10^8$ evaluations spent on pool generation, training and pool draws, a $162$-fold reduction with respect to conventional NS, whereas $\mathscr{M}=132$ would raise the cost to $8.99\times10^8$, a $56$-fold reduction. In wall-clock time the same range is almost invisible, since the dilution iterations account for only about $1.3\%$ of the duration of the run: the speedup with respect to conventional NS moves merely from $1.44$ to $1.41$. The number of walkers refreshed at each dilution iteration, fixed here at $10^3$ by analogy with conventional NS~\cite{partay_nested_2021, unglert_neural-network_2023}, is a second lever on the same cost. As shown in the Supplementary Material, the budget of a dilution stage depends only on the product of that number with $\mathscr{M}$, so that refreshing the whole live set at each iteration would make complete replacement certain at no additional cost, and would allow $\mathscr{M}=1$, raising the reduction of panel f from $67$-fold to $141$-fold. Establishing the smallest amount of dilution that leaves the resulting density of states unaffected is left to future experiments.

These results should be interpreted with some caution. The purpose of the present study is not to provide a highly optimized benchmark of either approach, but rather to assess the potential of generative models for accelerating nested sampling of condensed matter systems. Neither the conventional NS implementation nor the flow-assisted algorithm was systematically optimized beyond the parameter ranges explored here. Nevertheless, the reduction in energy evaluations, combined with a measurable decrease in wall-clock time despite the overhead of training a relatively large neural network, demonstrates that a substantial fraction of the computational effort can be shifted away from expensive energy evaluations. This effort is instead invested in training a generative model, a stage that is largely independent of the underlying potential and can benefit directly from continued advances in machine-learning hardware and software. This is particularly relevant for computational materials science, where realistic interatomic potentials, machine-learning force fields, or first-principles electronic-structure methods are often orders of magnitude more expensive than the simple LJ model considered here. In such cases, the reduction in energy evaluations achieved by the flow-assisted approach is expected to translate into substantially larger practical gains than those observed in the present benchmark.

\subsection{Generative efficiency across the nested-sampling trajectory}
\begin{figure}
    \centering
    \includegraphics[width=\linewidth]{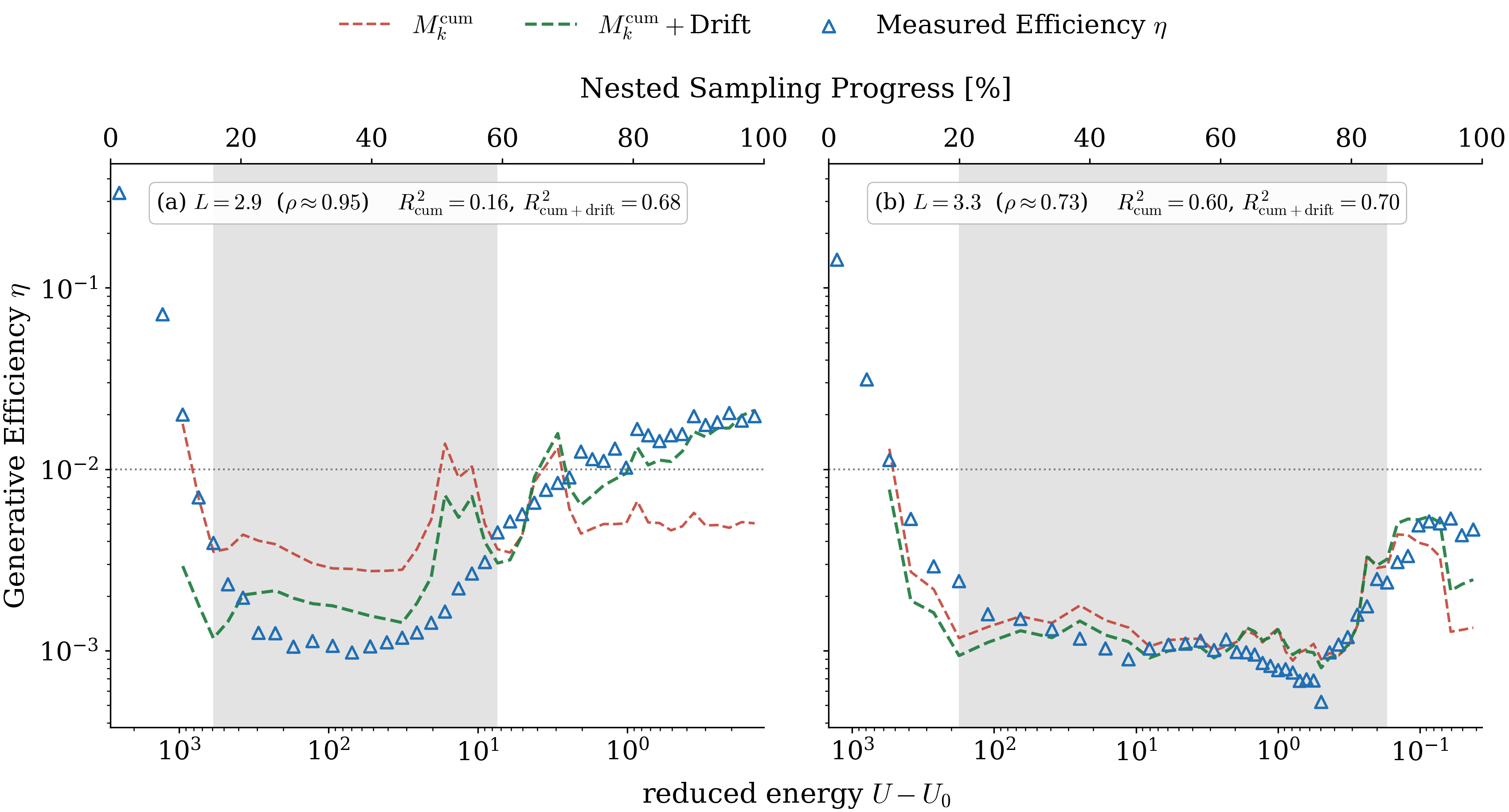}
    \caption{Generative efficiency as a function of the reduced energy $U-U_0$ along the nested-sampling trajectory, with the corresponding nested-sampling progress on the upper axis. The gray shaded regions identify the low-efficiency regime associated with the emergence of strong many-body correlations, defined as the contiguous range in which the generative efficiency stays within a factor of five of its minimum. Dashed lines represent different models fitted to the observed generation attempts (Eq.~\eqref{eq:model} below; see also Appendix~\ref{app:deff}). The model curves begin at the third pool, the first whose cumulation window of $c=3$ live sets is complete.
    The red dashed line is the model of Eq.~\eqref{eq:model} driven by the internal mode complexity alone, evaluated on the pooled last $c=3$ live sets, i.e.\ on the dataset the flow is actually trained on, and therefore denoted $M_k^{\mathrm{cum}}$. The green dashed line adds the drift term $\delta_k$ of Eq.~\eqref{eq:drift}, which measures how far the target distribution moves across that same window and so accounts for the flow lagging behind it. The fit quality of both is given in each panel. The dotted gray line marks $\eta=10^{-2}$, a common reference against which the recovery of the two densities can be compared: at $\rho\approx0.95$ the efficiency climbs back above it in the solid-like regime, whereas at $\rho\approx0.73$ it plateaus well below. \textbf{Left:} $\rho = 0.95$, \textbf{Right:} $\rho=0.73$.
    }
    \label{fig:generator_efficiency}
\end{figure}

Beyond the purely computational aspects discussed above, the present simulations also reveal a number of interesting physical and statistical features associated with the evolution of the learned constrained distributions along the nested-sampling trajectory. In particular, a remarkably structured behavior emerges when monitoring the generative efficiency of the normalizing flow as a function of the pool index and, therefore, indirectly as a function of decreasing energy. Three clearly distinct regimes can be identified throughout the annealing process (see the left panel of Fig.~\ref{fig:generator_efficiency}).

At the beginning of the simulation, where the live set is still characterized by very high energies, the likelihood-constrained distribution remains relatively close to the base distribution, namely the ideal gas. The corresponding configurations resemble weakly correlated fluid-like states with an almost uniform spatial distribution. As a consequence, the flow can learn the transformation between base and target distribution comparatively easily, resulting in high generative efficiencies and correspondingly low rejection rates during pool construction.

As the nested-sampling evolution proceeds toward lower energies, the system enters a regime characterized by dense-fluid or liquid-like configurations. In this region (gray area in Fig.~\ref{fig:generator_efficiency}), the flow efficiency decreases significantly for otherwise identical training protocols. Although the target distributions remain qualitatively disordered, they now contain strong many-body correlations and geometric constraints associated with local packing effects. As a consequence, obtaining an accurate representation of the constrained ensemble becomes increasingly challenging when training from the ideal-gas base distribution. This behavior manifests itself as a marked reduction in the acceptance probability during resampling and therefore in the overall efficiency of pool generation.

Interestingly, upon further annealing, the system eventually enters a low-energy solid-like regime where the efficiency of the generative model increases again. In this phase, the target distribution becomes highly localized around a relatively small number of ordered configurations. Despite the fact that the latent prior and the target distribution are now dramatically different, the training procedure remains effective in producing accurate approximations of the constrained ensemble. In fact, the observed generative efficiencies in the solid regime can exceed those obtained at the highest energies.

A quantitative account of this behavior can be built directly from the live-point ensembles, without reference to the flow (see Appendix~\ref{app:deff} for the full derivation). Each live configuration is mapped to a permutation- and translation-invariant fingerprint by collecting the sorted minimum-image pair distances,
\begin{equation}
  f(x) \;=\; \mathrm{sort}\Bigl(\bigl\{\,\|x_i-x_j\|_{\mathrm{PBC}}\,\bigr\}_{i<j}\Bigr)
  \;\in\;\R^{n_p},
  \qquad n_p=\binom{N}{2},
  \label{eq:fingerprint}
\end{equation}
so that variation in $f$ across the live set at iteration $k$ reflects genuine configurational spread rather than particle relabeling or rigid motion.

The natural first guess is that the difficulty grows with the overall geometric spread of the constrained ensemble, i.e., with the fingerprint covariance $\Sigma_k$ (Eq.~\eqref{eq:fpcov-app} in Appendix~\ref{app:deff}). This explanation fails: normalizing flows can readily accommodate global geometric transformations, such as translations and rescalings, through the learned invertible mapping~\cite{meng_gaussianization_2020,coretti_learning_2025}, so a purely second-moment measure of spread is essentially free for the flow and correspondingly uninformative about the generative burden (Spearman rank correlation $r_s=0.17$ with the measured efficiency; see Appendix~\ref{app:deff}). What poses a much greater challenge for the flow is the multimodality of the target: when the configurations split into several structurally distinct families along a principal fingerprint direction $i$, the corresponding marginal becomes flat-topped or multi-peaked (a platykurtic, negative-excess-kurtosis signature, $\kappa_i<0$) rather than bell-shaped. Summing this signature over the structural directions of the ensemble defines an internal mode complexity
\begin{equation}
  M_k \;=\; \sum_{i:\,\lambda^{(k)}_i>\tau\lambda^{(k)}_{\max}} \max\!\big(0,\,-\kappa_i\big),
  \label{eq:M}
\end{equation}
where $\lambda^{(k)}_i$ are the eigenvalues of $\Sigma_k$ and $\tau$ is a threshold used to retain only the important directions in the sum. This isolates the discrete, combinatorial multi-basin structure characteristic of atomistic landscapes --- precisely the structure a second-moment measure of spread cannot see --- and correctly predicts the generative burden, $r_s=+0.75$.

A second, dynamical term accounts for the flow lagging a moving target. With a cumulation depth $c$ (here $c=3$), the flow generating at threshold $U_k$ is trained on the union of the last $c$ live-point datasets; even when most of that data remains feasible, the underlying distribution has moved, and the stale flow is mismatched in proportion to that motion. This is quantified by the displacement of the fingerprint mean across the training window,
\begin{equation}
  \delta_k \;=\; \big\Vert \bar{f}^{(k)} - \bar{f}^{(k-c+1)}\big\Vert,
  \label{eq:drift}
\end{equation}
which explains the residual left after removing the $M_k$ trend, $r_s=+0.81$. Because the cumulation window is incomplete for the first $c-1$ pools, $\delta_k$ is undefined there and the fits begin at the third pool; the measured efficiencies of all pools are shown. A complementary feasibility test, the fraction of cumulated training mass still below $U_k$, bounds the static staleness contribution alone at a factor $\le 3$, confirming that this is genuine distributional motion and not mere threshold crossing.

Treating $M_k$ as the global term and $\delta_k$ as the local term, an additive-burden ansatz
\begin{equation}
  \log_{10}\eta_k^{-1}
  \;=\; c_0 \;+\; c_1\,\log_{10} M_k \;+\; c_2\,\log_{10}\delta_k ,
  \label{eq:model}
\end{equation}
fitted by least squares over the scale-stable regime, reproduces the measured generative efficiency across the entire trajectory (Fig.~\ref{fig:generator_efficiency}).

The nonmonotonic behavior of the generative efficiency identified in Fig.~\ref{fig:generator_efficiency} is particularly noteworthy because it suggests that the intrinsic difficulty of learning molecular distributions with normalizing flows is not determined solely by the apparent geometric difference between prior and target distributions. Rather, the liquid regime itself appears to represent a particularly challenging class of distributions for current flow architectures. Similar difficulties have been reported previously in the literature on flow-based molecular sampling~\cite{jung_normalizing_2024, coretti_learning_2025, schebek_efficient_2024}, where liquid-like systems are often found to be substantially harder to model than either gas-like or solid-like phases. However, the NS framework employed here provides a particularly systematic perspective on this phenomenon, since the method continuously traverses the entire phase diagram within a single simulation and therefore allows the evolution of the generative efficiency to be monitored across different thermodynamic regimes in a controlled manner.

\subsection{Density dependence and the role of crystalline order}

\begin{figure}[htbp]
    \centering
    \includegraphics[width=\linewidth]{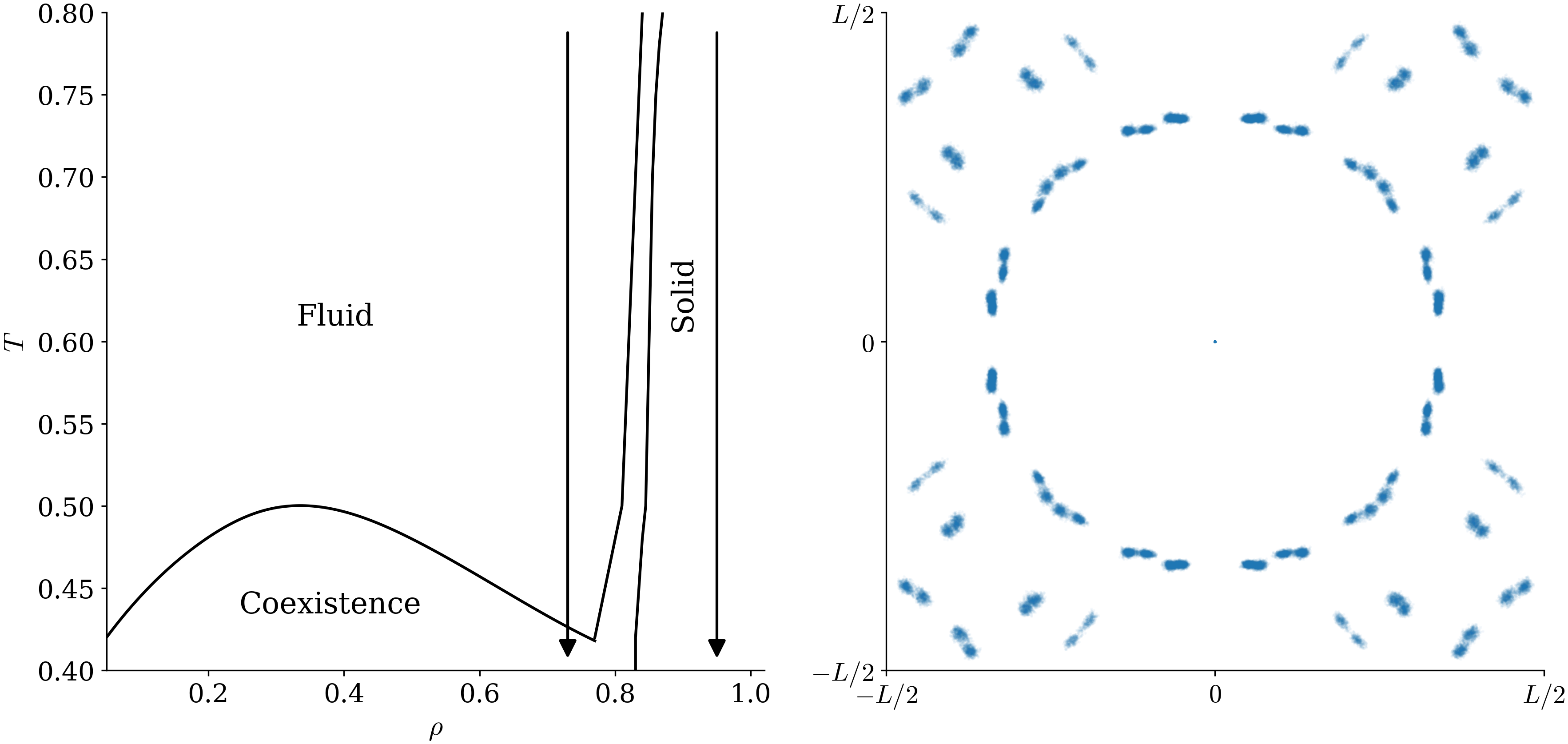}
    \caption{\textbf{Left:} Phase diagram of the two-dimensional Lennard--Jones system, adapted from Ref.~\cite{li_phase_2020}. The two black arrows indicate the trajectories followed by the nested-sampling simulations across the phase diagram. \textbf{Right:} Final configurations obtained after $5\times10^5$ nested-sampling steps for a system at constant density $\rho=0.73$. The configurations are not aligned to a reference, and degeneracies associated with hyperoctahedral symmetry operations are therefore still present within the ensemble (see the Supplementary Material).}
    \label{fig:LJ2D_PD}
\end{figure}

A further indication that the observed three-regime structure is closely connected to the emergence of crystalline order is provided by simulations performed at lower density, specifically at $\rho=0.73$. According to the phase diagram of the two-dimensional LJ system~\cite{li_phase_2020} (left panel of Fig.~\ref{fig:LJ2D_PD}), the system enters an extended fluid--solid coexistence region before reaching the lowest accessible energies. Unlike the higher-density case, however, the low-energy landscape is not dominated by a single well-defined crystalline structure. As a consequence, even at low energies the constrained distributions remain significantly broader and more structurally heterogeneous than those observed at $\rho=0.95$ (right panel of Fig.~\ref{fig:LJ2D_PD}).

Consistent with this picture, the generative efficiency at the lower density follows the same qualitative evolution observed at $\rho=0.95$, albeit shifted to much later stages of the nested-sampling trajectory. As shown in the right panel of Fig.~\ref{fig:generator_efficiency}, the low-efficiency regime (here defined as the contiguous range over which the generative efficiency stays within a factor of five of its minimum) extends approximately from $20\%$ to $85\%$ of the simulation, compared with only $16\%$ to $59\%$ at the higher density. This shift directly reflects the phase behavior of the system, where crystallization is delayed by the broader coexistence region and the constrained distributions retain their highly correlated, structurally heterogeneous character over a much larger fraction of the simulation. Although the generative efficiency eventually increases as the sampled configurations become confined to low-energy basins, the recovery is noticeably weaker than at higher density, reaching a plateau well below the $\sim1\%$ acceptance rate (horizontal dotted line in Fig.~\ref{fig:generator_efficiency}) observed at $\rho=0.95$. This reduced asymptotic efficiency reflects the more complex low-energy landscape associated to $\rho=0.73$, with numerous nearly degenerate minima originating not only from the symmetries of the crystal but also from the incommensurability between the simulation cell and the equilibrium lattice. As a consequence, the flow must learn a substantially broader and more intricate target distribution even in the lowest-energy region explored by nested sampling.

At present, the microscopic origin of the reduced learnability in the liquid regime has yet to be fully characterized. It is possible that the coexistence of strong local correlations with large-scale configurational disorder creates probability manifolds that are intrinsically difficult to represent through smooth invertible transformations. A deeper understanding of this phenomenon will likely require further investigation into both the geometry of molecular probability distributions and the inductive biases of current normalizing-flow architectures.

\section{Conclusion}

In this work, we have presented a nested-sampling algorithm for atomistic systems based on a learned generative model. In our implementation, the likelihood-constrained prior is represented throughout the nested-sampling trajectory by a single conditional normalizing flow, conditioned on the energy bound and trained on a sliding window of recent live sets. The flow replaces the slow MCMC decorrelation step of classical nested sampling with direct, parallel sampling, while the nested-sampling framework supplies the flow with training data, an importance-corrected accept/reject step, and a principled estimate of the evidence. 

The reason this composition works is that nested sampling and normalizing flows are well matched for atomistic systems. First, the proposed framework produces the training data required by the generative model self-consistently via the nested-sampling simulation itself. Unlike many machine-learning approaches to atomistic sampling, which rely on pre-existing datasets or seek fully data-free training strategies to avoid the circularity of learning from the very configurations one wishes to generate~\cite{wirnsberger_normalizing_2022,schebek_efficient_2024,schebek_scalable_2026}, the present algorithm naturally resolves this issue. The live sets generated during the nested-sampling evolution provide a continuous stream of representative constrained configurations, which are immediately recycled to train the conditional normalizing flow. In this way, the algorithm simultaneously learns and samples, progressively adapting the generative model as it traverses the thermodynamic landscape from high- to low-energy states. This on-the-fly training strategy is especially effective because of the particular distributions generated by nested sampling. The constrained prior samples configurations throughout the entire region below the current energy threshold, ensuring that low-energy configurations remain well represented in the training data rather than being suppressed by their small configurational volume, as occurs in Boltzmann sampling at finite temperature. Consequently, the flow is naturally trained on precisely those rare configurations that become increasingly important as the simulation progresses toward the low-energy regime.

Second, since nested sampling anneals through a controlled sequence of flat, constrained targets rather than relying on a structured prior and a fixed temperature, the learning burden placed on the flow remains bounded and recoverable throughout the simulation. This contrasts with fixed-prior Boltzmann generator, for which the learning problem typically becomes substantially more difficult as the system passes through the liquid regime. It is this gradual evolution of the target distribution, rather than flow expressivity alone, that enables a single network to traverse the entire phase diagram in a single run. The broader idea of progressively transforming a difficult learning problem into a sequence of nearby target distributions has recently emerged as an effective strategy in generative modeling, with applications ranging from annealed normalizing flows and variational inference to temperature-conditioned generative models~\cite{fogliani_annealing_2026,wu_annealing_2025,wang_mitigating_2025,moqvist_thermodynamic_2025,qin_flowvat_2025}.

The ability to traverse the entire phase diagram comes at a favorable computational cost. The flow-assisted scheme reproduces the full annealing trajectory at a small fraction of the energy evaluations demanded by a conventional nested sampling simulation, at comparable or shorter wall-clock time. What we demonstrate is therefore a redistribution of computational effort rather than an outright saving: the flow exchanges the many energy evaluations of Markov Chain decorrelation for the cost of training a generative model --- a cost largely independent of the underlying potential. We characterized this exchange across training schedules and pool sizes: energy evaluations fall by one and a half to two orders of magnitude, while the wall-clock time is governed by how much training is given to the flow. When energy evaluations are cheap, the exchange is close to break-even; its advantage grows with the cost of evaluating the potential, so that the approach is most attractive precisely for the costly, high-fidelity models~\cite{batatia_mace_2022} that are increasingly used in modern materials simulation. A different strategy decouples the two costs rather than exchanging them: Slice Monte Carlo integration performs a nested-sampling-like partitioning on a cheap surrogate and spends the expensive evaluations only on a stratified integral over the resulting slices~\cite{krondorfer_slice_2026}. The two compose naturally, since that surrogate partitioning is itself a constrained-sampling problem of the kind addressed here.

The same runs also turn the flow's efficiency into a physical diagnostic. The difficulty of the constrained ensembles varies non-monotonically with energy. It is greatest in the dense, disordered regime and decreases again when the system begins to order. This behavior reflects not the geometric gap between prior and target, but the extent to which the target fragments into distinct configurations. Dense, disordered phases thus stand out as the hard case for current flow models.

Our work therefore suggests a clear path toward global sampling of realistic atomistic systems. The property that makes nested sampling attractive is that its per-iteration cost remains bounded and does not deteriorate with system size. The framework demonstrated here is therefore well suited to scale to larger systems; what remains is to pair it with a flow that scales equally well. Here, the physics is favorable: correlations in liquids are typically short-ranged, so the collective fluctuations that challenge global flow models are largely local. This makes them well suited to architectures built to exploit locality~\cite{schebek_scalable_2026}, such as local-attention or message-passing equivariant models, and semi-local couplings that respect the correlation length rather than coupling all coordinates at once. Although the present work has focused on normalizing flows, the framework itself is not intrinsically tied to this particular class of generative models. Because nested sampling generates its own training data throughout the annealing trajectory, it naturally provides the supervision required by alternative approaches such as diffusion models or flow matching. Normalizing flows nevertheless retain a distinctive advantage in the present setting through their tractable likelihood, which enables the exact importance reweighting and accept/reject correction employed here; generative models without exact density evaluation generally forfeit this capability, or require additional approximation strategies~\cite{wang_energy_2026}. 
Very recently, also a complementary strategy has been proposed to use flow based models as local proposals within MCMC kernels for sampling constrained priors \cite{yallup_neural_2026}.
Combining the prior-free annealing of nested sampling with such locality-aware generative models could provide a practical route to global, flow-based sampling of many-component materials --- and with it, to bringing the full thermodynamics of realistic systems within reach of a single simulation.

\section*{Acknowledgments}
The authors are deeply grateful to Livia B.-P\'artay for initiating the discussion that ultimately led to this work. This research was funded in part by the Austrian Science Fund (FWF) through 10.55776/F8100 and 10.55776/COE5. For open access purposes, the authors have applied a CC BY public copyright license to any author accepted manuscript version arising from this submission.

\section*{Author contributions}
AC, SF and NU conceived the initial idea for the project. AC led the software development and carried out the numerical experiments presented in Sec.~\ref{sec:results}. NU and SF contributed to the implementation and the refinement of the numerical experiments. NU also conducted the numerical experiments presented in Sec.~\ref{sec:landscapes}. NU and AC prepared the initial version of the manuscript. GKHM and CD supervised the project, secured funding, and contributed to the preparation of the manuscript. All authors reviewed and approved the final manuscript.

\section*{Declaration of generative AI use}
Anthropic Claude Opus 5 was used for code debugging and for porting the simulation and analysis code into the public companion repository. Anthropic Claude Opus 5 and OpenAI ChatGPT-5.6 Sol were used for polishing and proofreading the text. No generative model was used to design the method, to produce the numerical results, or to interpret them. All code, numbers and statements in this work were verified by the authors, who take full responsibility for its content.

\section*{Supplementary Material}
The Supplementary Material reports the alignment procedure used for the configuration-space figures, the consumption of the generated pool along the NS trajectory, the run-to-run variability of the reported costs, measured from ten independent repetitions of the run of Fig.~\ref{fig:timings}f, results for a nine-particle system in a rectangular cell of aspect ratio $2\!:\!\sqrt{3}$, an analysis of the dilution stage and of the number of walkers it refreshes, and the complete network architecture, training protocol and computational budget of every run shown here.

\section*{Code and data availability}
The \texttt{nsflows} package, the notebooks and scripts that produce every figure of this
work and of the Supplementary Material, and the simulation data they read are available at
\url{https://github.com/CompPhysVienna/paper_nsflows}. Data provided are sufficient to reproduce the reported results up to statistical noise: besides the inputs of every figure, it includes the initial live sets and the complete parameter record of each run. Runs are not seeded, therefore a repetition reproduces a given result to within the run-to-run variability quantified in the Supplementary Material. The simulation output of the particular runs used here is available from the authors on request.

\appendix

\section{The gravitational-wave benchmark: injection and likelihood}
\label{app:gw}

We obtain the GW posterior in the standard Bayesian parameter-estimation framework~\cite{thrane_introduction_2019,veitch_parameter_2015}. Given the data $d$ and a waveform model, the posterior density over the source parameters $\vartheta$ follows from Bayes' theorem,
\begin{equation}
  p(\vartheta\,|\,d) \;=\; \frac{\mathcal{L}(d\,|\,\vartheta)\,\pi(\vartheta)}{Z},
  \qquad
  Z \;=\; \int \mathcal{L}(d\,|\,\vartheta)\,\pi(\vartheta)\,\dd\vartheta ,
  \label{eq:gwbayes}
\end{equation}
with $\mathcal{L}(d\,|\,\vartheta)$ the likelihood of the data given the parameters, $\pi(\vartheta)$ the prior, and $\mathcal{Z}$ the evidence~\cite{thrane_introduction_2019}. We evaluate this likelihood object with the bilby library~\cite{ashton_bilby_2019,talbot_bilby-devbilby_2025}.

The posterior is not built from real detector data but from a \emph{simulated injection}, the standard controlled-test setup in gravitational-wave parameter estimation~\cite{veitch_parameter_2015}: a signal with known parameters $\vartheta^\star$ is added to detector noise. Because the true source parameters are then known exactly, $\theta^\star$ serves as the reference point at which we evaluate the local diagnostics of App.~\ref{app:diagnostics}. A binary-black-hole waveform is generated with the precessing IMRPhenomPv2 approximant (reference frequency $50$~Hz, minimum frequency $f_{\min}=20$~Hz; segments of duration $T=4$~s sampled at $f_s=2048$~Hz) for a GW150914-like source (component masses $m_1=36\,M_\odot$, $m_2=29\,M_\odot$); the full injected parameter vector $\vartheta^\star$ is listed in Table~\ref{tab:gwparams}. 
The waveform is generated in the frequency domain, with $f$ the Fourier frequency and $\tilde{\cdot}$ denoting the Fourier transform. It is projected onto detector $I$ of the three-detector network (LIGO Hanford H1, LIGO Livingston L1, Virgo V1) through that instrument's antenna-pattern functions $F^I_{+},F^I_{\times}$~\cite{veitch_parameter_2015},
\begin{equation}
  h_I(\vartheta;f) \;=\; F^I_{+}(\alpha,\delta,\psi)\,h_{+}(\vartheta;f)
  \;+\; F^I_{\times}(\alpha,\delta,\psi)\,h_{\times}(\vartheta;f) ,
  \label{eq:gwproj}
\end{equation}
and a noise realization $n_I(f)$ is added, giving the frequency-domain strain
\begin{equation}
  d_I(f) \;=\; h_I(\vartheta^\star;f) + n_I(f).
  \label{eq:gwdata}
\end{equation}

The noise in each detector is modeled as stationary, zero-mean, Gaussian, and uncorrelated between detectors, and is fully characterized by its one-sided power spectral density (PSD) $S_{n,I}(f)$~\cite{veitch_parameter_2015}, which we set to the design sensitivity of each instrument.
Under this noise model the likelihood of the data given parameters $\vartheta$ is the Whittle likelihood~\cite{thrane_introduction_2019,veitch_parameter_2015}
\begin{equation}
\begin{aligned}
  \log\mathcal{L}(d\,|\,\vartheta)
  &\;=\; -\tfrac12 \sum_I
  \big\langle\, d_I - h_I(\vartheta)\ \big|\ d_I - h_I(\vartheta)\,\big\rangle_I \;+\; \Psi \\
  &\;=\; \sum_I\Big[\,\langle d_I\,|\,h_I(\vartheta)\rangle_I
  - \tfrac12\,\langle h_I(\vartheta)\,|\,h_I(\vartheta)\rangle_I\,\Big]
  \;-\;\tfrac12\sum_I\langle d_I\,|\,d_I\rangle_I \;+\; \Psi ,
\end{aligned}
  \label{eq:gwlike}
\end{equation}
summed over the detectors $I$, with the noise-weighted inner product
\begin{equation}
  \langle a\,|\,b\rangle_I \;=\; 4\,\mathrm{Re}\!\int_{f_{\min}}^{f_{\max}}
  \frac{\tilde a^{*}(f)\,\tilde b(f)}{S_{n,I}(f)}\,\dd f ,
  \label{eq:inner}
\end{equation}
integrated from $f_{\min}=20$~Hz up to the Nyquist frequency $f_{\max}=f_s/2=1024$~Hz. Expanding the Gaussian residual on the first line gives the second: the matched-filter term $\langle d_I\,|\,h_I\rangle_I$ and half the optimal signal-to-noise $\langle h_I\,|\,h_I\rangle_I$, the only $\vartheta$-dependent pieces, separated from the $\vartheta$-independent noise inner product $\tfrac12\sum_I\langle d_I|d_I\rangle_I$ and the log-normalization $\Psi=-\tfrac12\sum_I\sum_k\ln\!\big(2\pi\,S_{n,I}(f_k)\big)$. In practice we evaluate the log-likelihood \emph{ratio} relative to the signal-absent (noise) hypothesis $h_I\equiv0$,
\begin{equation}
  \log\Lambda(\vartheta) \;\equiv\;
  \log\mathcal{L}(d\,|\,\vartheta) - \log\mathcal{L}(d\,|\,0)
    \;=\; \sum_I\Big[\,\langle d_I\,|\,h_I(\vartheta)\rangle_I
  - \tfrac12\,\langle h_I(\vartheta)\,|\,h_I(\vartheta)\rangle_I\,\Big],
  \label{eq:gwratio}
\end{equation}
which is exactly the $\vartheta$-dependent first bracket of Eq.~\eqref{eq:gwlike}: forming the ratio cancels both $\Psi$ and $\tfrac12\sum_I\langle d_I|d_I\rangle_I$, leaving the shape of the landscape unchanged~\cite{thrane_introduction_2019}. This matched-filter form is Eq.~(B7) of Ref.~\cite{thrane_introduction_2019}. It is this $\log\Lambda$ (bilby's \texttt{log\_likelihood\_ratio}) that we take as $\log\mathcal{L}$ throughout.

The optimal network signal-to-noise ratio (SNR) of a signal $h$ is $\rho^2=\sum_I\langle h_I\,|\,h_I\rangle_I$, and the matched-filter SNR is $\rho_{\mathrm{mf}}=\rho^{-1}\sum_I\langle d_I\,|\,h_I\rangle_I$~\cite{thrane_introduction_2019}. The GW150914-like injection is a loud signal, and the ``high-SNR (linear-signal) limit'' invoked in App.~\ref{app:diagnostics} --- in which the Fisher matrix approximates the Hessian of $-\log\mathcal{L}$ --- refers to this regime.
The quantity $-\log\Lambda(\vartheta)$ is the GW ``energy'' of the landscape and the direct analogue of the LJ potential $U(x)$; its minimum sits at (or, given the noise realization, very close to) the injection truth $\vartheta^\star$. We sample the 15 parameters of Table~\ref{tab:gwparams}, each rescaled to the unit interval through its prior cumulative distribution function (the probability-integral transform). In these unit-cube coordinates the prior $\pi$ of Eq.~\eqref{eq:gwbayes} is uniform, so the posterior is proportional to the likelihood and $-\log\Lambda(\vartheta)$ coincides, up to an additive constant, with the negative log-posterior. The priors are \texttt{bilby}'s default precessing binary-black-hole set (\texttt{BBHPriorDict})~\cite{ashton_bilby_2019}: uniform in the component masses (sampled in chirp mass $\mathcal{M}$ and mass ratio $q$), uniform in comoving source-frame volume in luminosity distance (\texttt{UniformSourceFrame}), isotropic in sky position and in the spin orientations, and uniform in the dimensionless spin magnitudes; the coalescence time is given a narrow uniform prior about the injected value.

\begin{table}[htbp]
  \centering
  \renewcommand{\arraystretch}{1.15}
  \begin{tabular}{llrp{0.4\linewidth}}
    \toprule
    Symbol & \texttt{bilby} name & Injected value & Description \\
    \midrule
    $\mathcal{M}$    & \texttt{chirp\_mass}          & $28.10$          & chirp mass $(m_1m_2)^{3/5}/(m_1+m_2)^{1/5}$ \\
    $q$              & \texttt{mass\_ratio}          & $0.806$          & mass ratio $m_2/m_1\le1$ \\
    $a_1$            & \texttt{a\_1}                 & $0.4$            & dimensionless spin magnitude, primary \\
    $a_2$            & \texttt{a\_2}                 & $0.3$            & dimensionless spin magnitude, secondary \\
    $\theta_1$       & \texttt{tilt\_1}              & $0.5$            & spin tilt angle, primary \\
    $\theta_2$       & \texttt{tilt\_2}              & $1.0$            & spin tilt angle, secondary \\
    $\phi_{12}$      & \texttt{phi\_12}              & $1.7$            & azimuthal angle between the two spin vectors \\
    $\phi_{JL}$      & \texttt{phi\_jl}              & $0.3$            & azimuthal angle between total and orbital angular momentum \\
    $d_L$            & \texttt{luminosity\_distance} & $2000$           & luminosity distance \\
    $\delta$         & \texttt{dec}                  & $-1.2108$        & declination \\
    $\alpha$         & \texttt{ra}                   & $1.375$          & right ascension \\
    $\theta_{JN}$    & \texttt{theta\_jn}            & $0.4$            & inclination (total angular momentum vs.\ line of sight) \\
    $\psi$           & \texttt{psi}                  & $2.659$          & polarisation angle \\
    $\varphi$        & \texttt{phase}                & $1.3$            & coalescence phase \\
    $t_c$            & \texttt{geocent\_time}        & $1126259642.413$ & geocentric coalescence time \\
    \bottomrule
  \end{tabular}
  \caption{The 15 sampled parameters of the GW150914-like
    binary-black-hole problem (standard \texttt{bilby}/\texttt{nessai}
    setup), with the symbols used in Fig.~\ref{fig:landscapes} and the
    injected true values $\vartheta^\star$. Angles are in radians,
    $\mathcal{M}$ in $M_\odot$, $d_L$ in Mpc, and $t_c$ in seconds (GPS);
    $\mathcal{M}=28.10\,M_\odot$ and $q=0.806$ follow from the injected
    component masses $m_1=36\,M_\odot$, $m_2=29\,M_\odot$. The added
    Gaussian noise uses seed $170817$.}
  \label{tab:gwparams}
\end{table}

\section{Local-geometry and coupling diagnostics}
\label{app:diagnostics}

The diagnostics in the right-hand columns of Fig.~\ref{fig:landscapes}
characterize the curvature and coordinate coupling at the dominant mode.
We collect their definitions here.

At an inherent structure $x^\star$ (a local minimum of the potential),
the Hessian is the matrix of second derivatives of the potential energy,
\begin{equation}
  H_{ij} \;=\; \left.\frac{\partial^2 U}{\partial x_i\,\partial x_j}
  \right|_{x=x^\star},
  \qquad i,j = 1,\dots,Nd ,
  \label{eq:hessian}
\end{equation}
evaluated here by automatic differentiation ($Nd=16$ for LJ-8). The two
translational zero modes are projected out before any eigen-analysis.
$H$ is the local precision (inverse-covariance) of the Boltzmann
distribution in the harmonic approximation, $\pi_\beta\approx
\mathcal{N}(x^\star,(\beta H)^{-1})$.

For the stationary-Gaussian-noise likelihood of App.~\ref{app:gw}, the
Fisher information matrix~\cite{vallisneri_use_2008} is the noise-weighted overlap of the waveform derivatives,
\begin{equation}
  \mathcal{F}_{ij}
  \;=\; \sum_I \left\langle\frac{\partial h}{\partial\vartheta_i}\,\bigg|\,
  \frac{\partial h}{\partial\vartheta_j}\right\rangle_I ,
  \label{eq:fisher}
\end{equation}
with the per-detector inner product of Eq.~\eqref{eq:inner}. At the
injection truth and in the high-SNR (linear-signal) limit, $\mathcal{F}$
coincides with the Hessian of the negative log-likelihood,
\begin{equation}
  \mathcal{F}_{ij} \;=\;
  \left.\frac{\partial^2(-\log\mathcal{L})}{\partial\vartheta_i\,
  \partial\vartheta_j}\right|_{\vartheta=\vartheta^\star},
  \label{eq:fisher-nll}
\end{equation}
which is how we evaluate it (centered finite differences of the
\texttt{bilby} log-likelihood on the unit-cube-rescaled parameters). It
is the GW analogue of the LJ Hessian: the local precision at the mode.

To probe coordinate coupling beyond the matrix structure of
Eqs.~\eqref{eq:hessian}--\eqref{eq:fisher} we draw Laplace samples at the
mode --- $x^{(m)}\sim\mathcal{N}(x^\star,H^{+})$ for LJ and
$\vartheta^{(m)}\sim\mathcal{N}(\vartheta^\star,\mathcal{F}^{-1})$ for GW, with
$H^{+}$ the pseudoinverse on the nonzero-mode subspace --- and compute the
pairwise normalized mutual information, normalized by the smaller of the two
marginal entropies~\cite{vinh_information_2010},
\begin{equation}
  \mathrm{NMI}(X_i,X_j) \;=\;
  \frac{I(X_i;X_j)}{\min\!\big(H(X_i),\,H(X_j)\big)}\;\in[0,1].
  \label{eq:nmi-app}
\end{equation}
The mutual information is the Kullback--Leibler divergence between the
joint distribution of the coordinate pair and the product of its
marginals, equivalently a difference of Shannon entropies,
\begin{equation}
  I(X_i;X_j) \;=\; \sum_{a,b} p_{ab}\,\log\frac{p_{ab}}{p_a\,p_b}
  \;=\; H(X_i)+H(X_j)-H(X_i,X_j),
  \label{eq:mi-app}
\end{equation}
\begin{equation}
  H(X_i) = -\sum_a p_a\log p_a,
  \qquad
  H(X_i,X_j) = -\sum_{a,b} p_{ab}\log p_{ab},
  \label{eq:entropy-app}
\end{equation}
where $p_a$ and $p_{ab}$ are the empirical marginal and joint frequencies
of a plug-in histogram estimator ($B=30$ bins per axis; $\{p_{ab}\}$ the
$B\times B$ joint histogram), with the convention $0\log0=0$. Entropies
are in nats; $I$ is clipped at $0$ to remove the finite-sample negative
bias. $\mathrm{NMI}=0$ iff the coordinates are independent and $1$ for an
exact functional dependence, giving a scale-free measure of shared
information that complements the (Gaussian, second-order) coupling in
$H$ and $\mathcal{F}$.

\section{The internal mode complexity $M_k$}
\label{app:deff}

This appendix gives the full derivation behind the internal mode
complexity $M_k$ [Eq.~\eqref{eq:M}] introduced in Sec.~\ref{sec:results},
and behind the failure of the simpler second-moment measure $\Deff$ used
to motivate it.

Recall the fingerprint $f(x)\in\R^{n_p}$ of
Eq.~\eqref{eq:fingerprint}, collecting the sorted minimum-image pair
distances of all particles; for $N=8$, $n_p=\binom{N}{2}=28$. The sort
removes the particle-label permutation symmetry, the minimum-image
convention removes the continuous box translation, and reflections of
the cell leave the multiset of distances unchanged. The fingerprint
therefore collapses each symmetry orbit to a single point, so that
variation in $f$ across the live set reflects genuine configurational
spread rather than relabeling or rigid motion.

At NS iteration $k$ we snapshot the $K$ live points
$\{x_n^{(k)}\}_{n=1}^{K}$, compute their fingerprints
$f_n^{(k)}=f(x_n^{(k)})$, and form the
empirical fingerprint covariance
\begin{equation}
  \Sigma_k \;=\; \frac{1}{K-1}\sum_{n=1}^{K}
  \bigl(f_n^{(k)}-\bar{f}^{(k)}\bigr)
  \bigl(f_n^{(k)}-\bar{f}^{(k)}\bigr)^{\!\top}
  \;=\; V^{(k)}\,\mathrm{diag}\bigl(\lambda_1^{(k)},\dots,\lambda_{n_p}^{(k)}\bigr)\,V^{(k)\top},
  \label{eq:fpcov-app}
\end{equation}
with $\bar{f}^{(k)}$ the live-set mean and $V^{(k)}$ the orthogonal
matrix whose columns $v_i$ are the eigenvectors of the covariance matrix $\Sigma_k$, ordered according to the
eigenvalues $\lambda_1^{(k)}\ge\dots\ge\lambda_{n_p}^{(k)}$.

\begin{figure}[htbp]
  \centering
  \includegraphics[width=\linewidth]{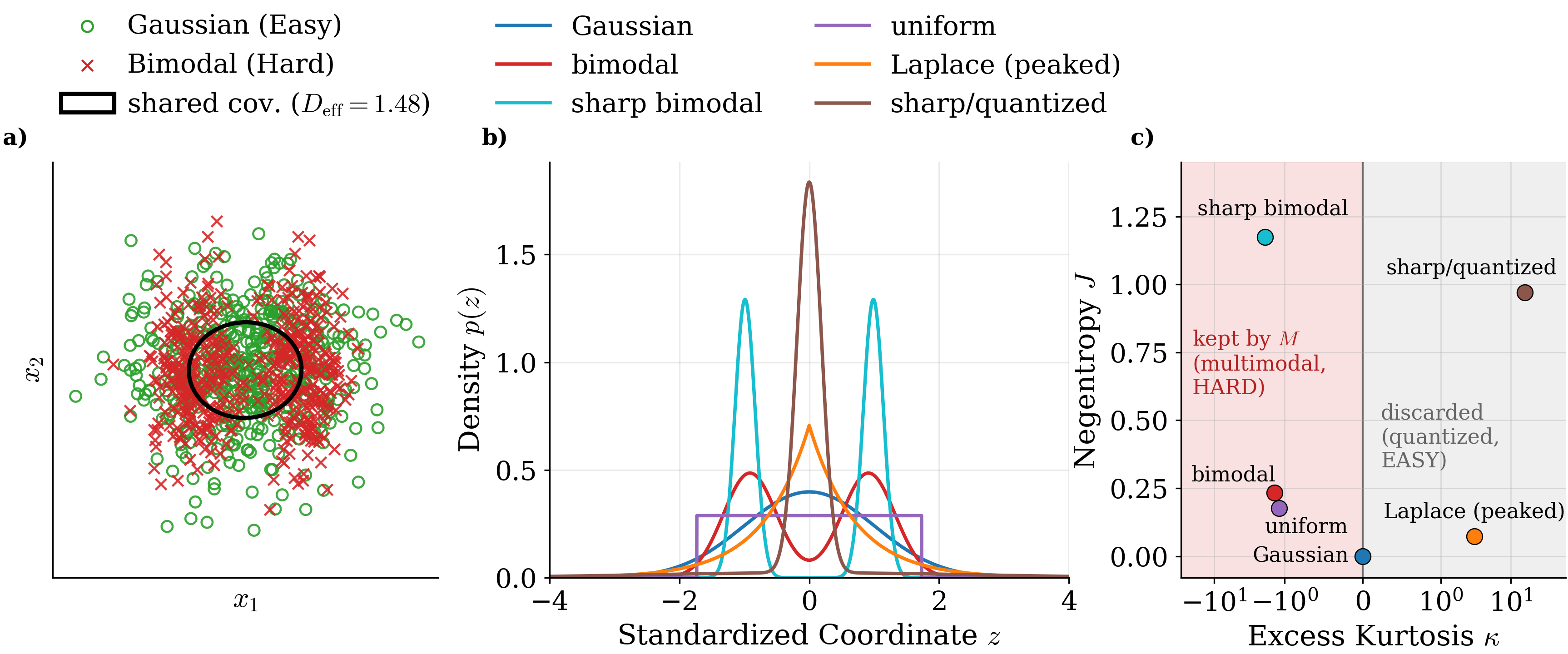}
  \caption{The three concepts of this section on toy distributions.
  \textbf{a)} A Gaussian cloud (green circles, easy) and a bimodal cloud
  (red crosses, hard), built with the \emph{identical} covariance,
  therefore share the \emph{identical} $\Deff=1.48$: their $1\sigma$
  covariance ellipse (black) coincides and is all that $\Deff$ can see,
  yet the Gaussian is affine-trivial for a flow while the bimodal cloud
  forces it to split its base into two modes.
  \textbf{b)} Negentropy $J$ measures the distance from Gaussian (the
  non-affine work) for six unit-variance densities; $J=0$ only for the
  Gaussian. \textbf{c)} The same six densities placed in
  (kurtosis,\,negentropy) space: high $J$ splits by the \emph{sign} of
  $\kappa$ -- multimodality is platykurtic ($\kappa<0$, hard) while a
  \emph{single} sharp/quantized peak is leptokurtic ($\kappa>0$, easy) --
  so the internal complexity $M=\sum_i\max(0,-\kappa_i)$ keeps the red
  region and discards the gray. A \emph{doubly}-peaked sharp density
  (cyan, $\kappa=-1.9$) stays platykurtic, confirming that sharpness
  \emph{within} a multimodal structure does not flip the sign -- only a single sharp peak does.}
  \label{fig:concepts}
\end{figure}

A first, second-moment diagnostic of this ensemble is the eigenvalue
participation ratio of the fingerprint covariance $\Sigma_k$
\begin{equation}
  \Deff \;=\; \frac{\big(\sum_i \lambda_i\big)^2}{\sum_i \lambda_i^2} \;\in\;[1,\,n_p],
  \label{eq:deff}
\end{equation}
the effective number of fingerprint directions of comparable variance (here and below we drop the iteration superscript on $\lambda_i$).

The key observation is that $\Deff$ is a function of the
\emph{covariance alone}, i.e.\ of second moments. 
Even the simplest flow, whose first layer is affine, $x\mapsto Ax + b$, represents an
arbitrary Gaussian \emph{exactly}: if the constrained target $\rho_{X|C}$ were Gaussian the flow could achieve $\rho_{X|C}^\theta=\rho_{X|C}$ and a generative efficiency $\eta_k=1$, with $\eta_k$ being the fraction of flow proposals accepted at iteration $k$, regardless of $\Sigma_k$.
This is the \emph{Gaussianization} view of flows: the residual work
after the affine/whitening step is precisely the non-Gaussianity of the
target \cite{chen_gaussianization_2020,meng_gaussianization_2020}.
Hence the entire second-moment structure, and with it $\Deff$, is
\emph{free} for the flow.
The rational-quadratic spline couplings we actually use~\cite{durkan_neural_2019} contain this affine map as a special case, so the whitening argument is only a lower bound on their capacity and $\Deff$ is free in the same way. $\Deff$ measures precisely the part of the geometry that does not contribute to the burden --- two clouds with the same covariance but different modality share the same $\Deff$ (Fig.~\ref{fig:concepts}a).

Ref.~\cite{meng_gaussianization_2020} defines the irreducible burden as the residual after the best affine map, i.e.\
the \emph{non-Gaussianity} of the target. Whiten the cloud onto its
principal modes, $z_i = \langle f-\bar{f},\, v_i\rangle /
\sqrt{\lambda_i}$ (unit variance). The non-affine cost along mode $i$ is
the negentropy
\begin{equation}
  J(z_i) \;=\; H\!\left[\mathcal{N}(0,1)\right] - H[z_i] \;\ge\; 0,
  \qquad
  J=0 \iff z_i \text{ Gaussian},
  \label{eq:negentropy}
\end{equation}
where $H[\cdot]$ is the differential entropy. Note that $J(z_i)$ is a
\emph{scalar functional} of the mode's marginal law $p_{z_i}$ --- a single
number per mode quantifying its total departure from Gaussianity, not a
function of the coordinate $z$ (the subscript $i$ indexes the mode). 
It equals exactly $\KL \big(p_{z_i}\,\Vert\,\mathcal N(0,1)\big)$ --- the density-ratio variance the flow must generate along that mode, the per-mode version of the $\Var[\log p_{\rm target}-\log p_{\rm prior}]$ burden (with $p_{\rm target}$ the constrained target $\rho_{X|C}$ and $p_{\rm prior}$ the flow's base density), and the standard non-Gaussianity contrast of independent component analysis and projection pursuit \cite{comon_independent_1994,hyvarinen_new_1997}.
Using the classical cumulant (moment) approximation of negentropy
\cite{jones_what_1987} -- a scalar built from the (scalar)
third and fourth moments of $z$,
\begin{equation}
  J(z) \;\approx\; \tfrac{1}{12}\,\E[z^3]^2
              \;+\; \tfrac{1}{48}\,\big(\E[z^4]-3\big)^2 .
  \label{eq:hyvarinen}
\end{equation}
In Fig.~\ref{fig:concepts}b the \emph{curves} are the densities
$p_{z}$; the negentropy $J$ of each --- Eq.~\eqref{eq:negentropy}, the
integrated gap to the Gaussian --- is the vertical axis of panel Fig.~\ref{fig:concepts}c.
It vanishes only for the Gaussian and grows with any departure from it.

The total negentropy $\sum_i J(z_i)$ \emph{fails} as a predictor of the per-iteration generative burden $\log_{10}\eta_k^{-1}$ --- the number of flow proposals needed to fill the pool --- because it
is dominated by the \emph{leptokurtic}
($\E[z^4]-3>0$) contribution of the deep, collapsed basin, where the
sorted pair distances lock onto discrete lattice values and produce
sharply peaked, but easy, marginals. The generation-relevant
structure is the opposite tail: \emph{multimodality}, i.e.\ several
distinct basins occupied along a principal direction, which makes a
marginal \emph{platykurtic} ($\E[z^4]-3<0$; Fig.~\ref{fig:concepts}c) --
the same
negative-excess-kurtosis signature exploited by the bimodality
coefficient
\cite{pfister_good_2013}. 
The negative part of the excess kurtosis $\kappa_i \equiv \E[z_i^4]-3$
quantifies this multimodality; summing it over the significant modes
(threshold $\tau=10^{-3}$ excludes degenerate ones) recovers the internal
mode complexity $M_k$ of Eq.~\eqref{eq:M}. The decisive point behind its
success is that the two halves of the excess
kurtosis carry \emph{opposite-signed} predictive signal: the sharp-peak
(leptokurtic) complement $\sum_i\max(0,\kappa_i)$ runs in the opposite direction because it grows in the deep, easy basins where the burden
is \emph{low}. Their mixture, the full negentropy $\sum_i J(z_i)$,
therefore inherits the wrong trend and fails --- not because
the multimodal signal is absent but because the leptokurtic part, larger
in magnitude, dominates it. That the two contributions predict in opposite
directions, rather than merely with different strength, is what confirms
the split is physical and not a fit artifact.

This split is ultimately a property of the spline flow, not of affine
Gaussianization alone. The monotone spline couplings reshape
\emph{unimodal} non-Gaussianity --- skew, peakedness, the sharp
leptokurtic basins --- by a gentle, well-conditioned deformation that
merely concentrates mass, which is why the full non-Gaussianity
$\sum_i J$, the residual for a \emph{purely} affine flow, overcounts the
burden. A \emph{platykurtic} marginal demands the opposite operation:
generating it from the flow's unimodal base means splitting mass into
separated basins through near-singular, precisely placed maps that
coupling flows represent poorly --- the more so for the combinatorial,
coupled multi-basin structure of atomistic targets. The sign of the
excess kurtosis thus tracks a genuine easy/hard divide of the spline
itself --- concentrate versus split --- so the operative residual for our
flow is the multimodality $M_k$, not $\sum_i J$; for a purely affine flow,
by contrast, the sharp leptokurtic basins would be costly too.

\printbibliography
\end{refsection}

\clearpage
\raggedbottom
\setcounter{section}{0}
\setcounter{figure}{0}
\setcounter{table}{0}
\setcounter{equation}{0}
\renewcommand{\thesection}{S\arabic{section}}
\renewcommand{\thesubsection}{S\arabic{section}.\arabic{subsection}}
\renewcommand{\thefigure}{S\arabic{figure}}
\renewcommand{\thetable}{S\arabic{table}}
\renewcommand{\theequation}{S\arabic{equation}}
\makeatletter
\begingroup
  \renewcommand{\thefootnote}{\fnsymbol{footnote}}
  \setcounter{footnote}{0}
  \noindent\vbox{\hsize\textwidth \linewidth\hsize
    \centering
    {\LARGE\bf Supplementary Material: Generative Nested Sampling of Atomistic
     Thermodynamic Landscapes\par}
    \vskip 1.2em
    {\normalsize\lineskip .5em
      \def\And{\end{tabular}\hfil\linebreak[0]\hfil\begin{tabular}[t]{c}\bf\rule{\z@}{24\p@}\ignorespaces}\def\AND{\end{tabular}\hfil\linebreak[4]\hfil\begin{tabular}[t]{c}\bf\rule{\z@}{24\p@}\ignorespaces}\begin{tabular}[t]{c}\bf\rule{\z@}{24\p@}\textbf{Alessandro Coretti}\footnotemark[1]\footnotemark[2]\\
        Faculty of Physics,\\ University of Vienna,\\ 1090 Vienna, Austria
      \And
        \textbf{Nico Unglert}\footnotemark[1]\footnotemark[3]\\
        Institute of Materials Chemistry,\\ TU Wien,\\ 1060 Vienna, Austria
      \And
        \textbf{Sebastian Falkner}\\
        Institute of Physics,\\ University of Augsburg,\\ 86159 Augsburg, Germany
      \AND
        \textbf{Georg K. H. Madsen}\\
        Institute of Materials Chemistry,\\ TU Wien,\\ 1060 Vienna, Austria
      \And
        \textbf{Christoph Dellago}\\
        Faculty of Physics,\\ University of Vienna,\\ 1090 Vienna, Austria
      \end{tabular}\par}
    \vskip 0.3in \@minus 0.1in
  }
  \footnotetext[1]{These authors contributed equally to this work.}
  \footnotetext[2]{\texttt{alessandro.coretti@univie.ac.at}}
  \footnotetext[3]{\texttt{nico.unglert@tuwien.ac.at}}
\endgroup
\makeatother

\begin{refsection}
\newrefcontext[labelprefix={S}]

Throughout, equations, figures and sections without an ``S'' prefix refer to the main text.
All simulations use $K = 10^4$ walkers and $5\times10^5$ NS iterations; lengths and energies
are in Lennard--Jones reduced units.

\section{Aligning configurations}
\label{sec:alignment}

The configuration-space panels of Figs.~2 and~\ref{fig:hex_configs} superimpose many
configurations after expressing each in internal coordinates: particle~0 is moved to the
origin and all other particles are wrapped into the box with the minimum-image convention.
This removes translations and, because the particles are identical, the resulting image is
also invariant under relabelling. It is not invariant, however, under the point-group
operations of the simulation cell, which map one equilibrium configuration onto another
that is equally probable. Superimposing configurations that are related by such an
operation blurs the image even when every single configuration is perfectly ordered.
Whether this happens depends on which operations of the cell are \emph{not} already
symmetries of the structure being sampled.

\paragraph{Square box.}
The point group of the square box is $D_4$, with eight elements: the identity $E$, the
rotations $C_4$, $C_2$, $C_4^3$ by $\pi/2$, $\pi$, $3\pi/2$, the reflections $\sigma_x$,
$\sigma_y$ across the axes and $\sigma_d$, $\sigma_{d'}$ across the diagonals. To find which
of them are symmetries of the low-energy structure, we take as reference $x_\star$ the first
configuration of the final live set of the standard-NS run and compare it with its eight
images $g\,x_\star$, $g\in D_4$. Two configurations $x$ and $x'$ in internal coordinates are
compared through the matching cost
\begin{equation}
  C(x, x') = \min_{\pi\in S_{N'}} \frac{1}{N'} \sum_{i=1}^{N'}
  \bigl\| x_i - x'_{\pi(i)} \bigr\|_{\mathrm{PBC}}^{2},
  \label{eq:matching_cost}
\end{equation}
the mean squared minimum-image displacement after the optimal relabelling $\pi$, which we
obtain with the Hungarian algorithm~\cite{kuhn_hungarian_1955}. Figure~\ref{fig:alignment}b shows that $C(g\,x_\star,
x_\star)$ vanishes, up to thermal noise, for $g\in\{E, C_2, \sigma_d, \sigma_{d'}\}$ and is
two orders of magnitude larger for $g\in\{C_4, C_4^3, \sigma_x, \sigma_y\}$. The low-energy
crystal, whose close-packed rows run along one of the box diagonals, is therefore invariant
under the subgroup $H=\{E, C_2, \sigma_d, \sigma_{d'}\}$, and $D_4$ splits into the two
cosets $H$ and $C_4H$: the crystal can sit in the box in exactly two distinguishable
orientations, rotated by $\pi/2$ with respect to each other. This remains true if the
reference particle is allowed to change, so the two orientations are not related by a
translation either. Superimposing both produces the twelve-spot pattern of
Fig.~\ref{fig:alignment}c.

The low-energy crystal is a triangular lattice strained to fit the box. With the eight
sites of the box spanned by $b_1 = L(1,1)/4$ and $b_2 = L(-1,3)/8$, the box edges are
themselves lattice vectors, $L\hat{e}_x = 3b_1 - 2b_2$ and $L\hat{e}_y = b_1 + 2b_2$.
The nearest-neighbour bonds $b_1$, $b_2$ and $b_2 - b_1$ have lengths $1.025$, $1.146$ and
$1.146$, against $1.102$ for an unstrained triangular lattice at the same density, and
$b_1$ lies along the box diagonal. These vectors reproduce the lattice sites of the
reference to within the thermal displacements. Figure~\ref{fig:alignment}c overlays this
lattice (dotted) and its image under $C_4$ (solid): every spot of the unaligned live set
lies on a site of one of the two.

\begin{figure}[t]
  \centering
  \includegraphics[width=4.380in]{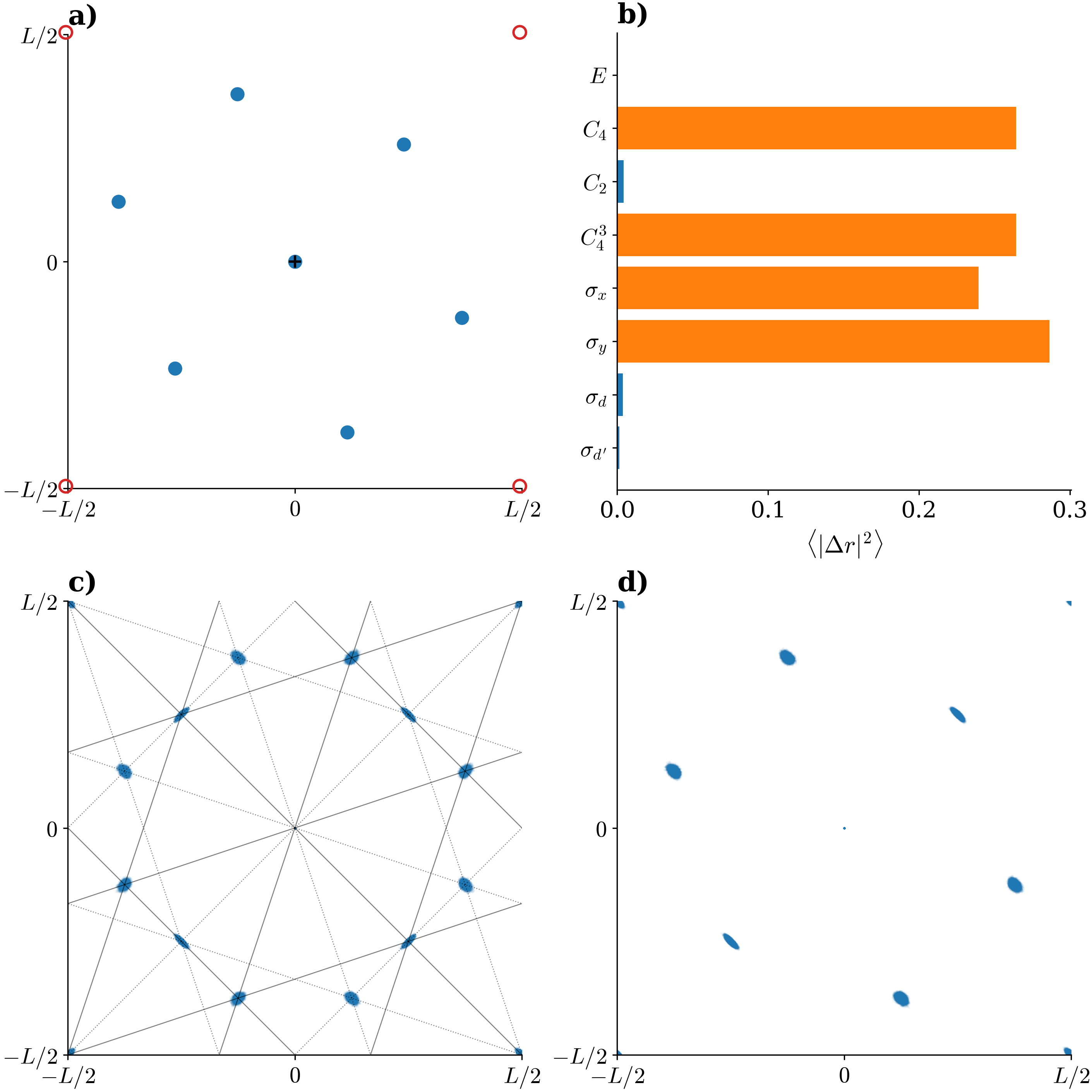}
  \caption{Alignment of low-energy configurations in the square box ($N=8$, $L=2.9$).
  \textbf{a)} Reference configuration in internal coordinates. The outermost particle
  (open circles) lies at the box corner, where its minimum image is ambiguous between the
  four corners, and is excluded from the matching. \textbf{b)} Matching cost,
  Eq.~\eqref{eq:matching_cost}, between the reference and its images under the operations
  of $D_4$. Blue: the subgroup $H$ leaving the crystal invariant. \textbf{c)} Final live set of
  the flow-based run of Fig.~4f, superimposed without alignment. Lines: bonds of the
  strained triangular lattice of the reference (dotted) and of its image under $C_4$
  (solid). \textbf{d)} The same live
  set after alignment: 50.4\% of the configurations have been rotated by $\pi/2$.}
  \label{fig:alignment}
\end{figure}

\paragraph{Alignment procedure.}
Since the two orientations are related by $C_4$, aligning a configuration only requires
deciding whether or not to rotate it by $\pi/2$. For each configuration $x$ we
\begin{enumerate}
  \item remove the particle farthest from the reference particle. In the square box at this
  density this particle sits close to a corner of the box, $|x_i|\approx L/\sqrt{2}$, where
  the minimum-image convention can place it at any of the four corners
  (Fig.~\ref{fig:alignment}a), so it would make the matching unreliable;
  \item compute the matching cost of the remaining $N'=N-1$ particles with the reduced
  reference, for $x$ and for $C_4 x$;
  \item apply to the full configuration the operation with the lower cost.
\end{enumerate}
Checking one representative of each coset is enough, because the other elements of the
coset give the same cost up to thermal noise. Figure~\ref{fig:alignment}d shows the result:
after alignment the final live set collapses onto the seven lattice sites of the
reference. The procedure also shows that both orientations are sampled with equal weight,
as symmetry requires. They account for 50.4\% and 49.6\% of the final live set of the
flow-based run, and for 51.4\% and 48.6\% of that of standard NS. So the flow, trained on
symmetry-augmented data, does not collapse onto one of the two degenerate orientations.

Alignment is used only for visualization. The flow is trained on unaligned live sets, and
each training configuration is augmented with a random choice of reference particle and a
random operation of the hyperoctahedral group (Sec.~3.5). The ensemble at $\rho=0.73$ in
Fig.~6 is shown unaligned, since its low-energy landscape is not dominated by a single
crystal and no single reference exists.

\paragraph{Rectangular box.}
In the $2:\sqrt{3}$ cell of Sec.~\ref{sec:hex}, the point group of the cell is reduced to
$D_2=\{E, C_2, \sigma_x, \sigma_y\}$, and every one of its operations maps the low-energy
crystal onto itself up to a change of the reference particle (matching costs below
$4\times10^{-3}$ for all four elements). Changing the reference particle does not change the
superimposed density, so in this cell all low-energy configurations produce the same image
and no alignment is needed. Accordingly, the final live set is spread evenly over the four
operations, 25\% each, when particle~0 is kept as the reference.

\section{Additional results}
\label{sm:results}

\subsection{Square box: consumption of the pool}
\label{sec:pool}

Each NS iteration draws configurations uniformly from the pool, without replacement, until
one lies below the current bound $\Umax$ (lines 20--23 of Alg.~2). The pool was generated
at $\Utrain$, and $\Umax$ decreases at every nested sampling iteration, so a growing fraction of the pool
becomes infeasible, and the number of draws per iteration grows until the pool is exhausted
and the algorithm returns to standard NS. Figure~\ref{fig:pool_picks} shows this for the
two cosine-annealing runs of Fig.~4, with pools of $2\times10^4$ (Fig.~4f) and $10^5$
(Fig.~4e) configurations.

\begin{figure}[t]
  \centering
  \includegraphics[width=5.480in]{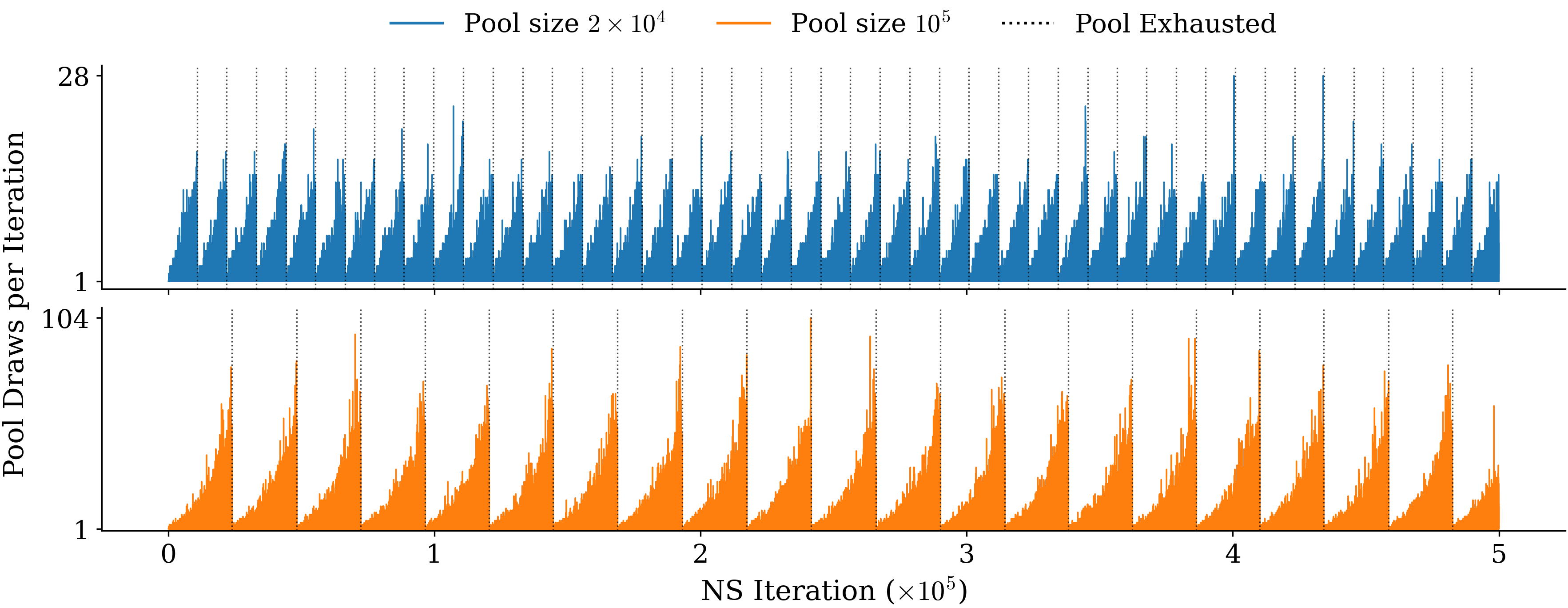}
  \caption{Number of configurations drawn from the pool at each NS iteration before one
  satisfies the current energy bound, for pools of $2\times10^4$ (top, run of Fig.~4f) and
  $10^5$ (bottom, run of Fig.~4e) configurations. Dotted lines mark the iterations at which
  the pool is exhausted and the algorithm switches back to $\mathscr{M}=100$ iterations of
  standard NS before retraining. The upper tick on each vertical axis is the maximum
  number of draws in the run.}
  \label{fig:pool_picks}
\end{figure}

The sawtooth shape follows from the compression of the prior mass. A pool generated at
$\Utrain$ is distributed uniformly in the region $U<\Utrain$. After $t$ further NS
iterations the bound encloses on average a fraction $\alpha_t/\alpha_0 = (K/(K+1))^t
\approx e^{-t/K}$ of that region, so the expected number of draws at iteration $t$ is
$e^{t/K}$.\pagebreak{} Every draw removes a configuration from the pool, whether it is accepted or not.
A pool of $P$ configurations is therefore exhausted after $t^\star$ iterations, with
\mbox{$\sum_{t<t^\star} e^{t/K} \approx K\,(e^{t^\star/K}-1) = P$}, i.e.
\begin{equation}
  t^\star \approx K \ln\!\left(1+\frac{P}{K}\right).
  \label{eq:pool_lifetime}
\end{equation}
With $K=10^4$, Eq.~\eqref{eq:pool_lifetime} predicts $t^\star = 1.10\times10^4$
iterations for $P=2\times10^4$ and $t^\star = 2.40\times10^4$ for $P=10^5$.  The run generates $45$ and $21$ pools respectively; averaged over those that are actually exhausted --- the last pool of a run never is --- the measured lifetimes are $1.10\times10^4$ and $2.40\times10^4$, within $0.5\%$ of the prediction. The mean number of draws per iteration, $P/t^\star$, rises from $1.81$ to $4.16$, and the maximum observed number of draws from $28$ to $104$.

Pool size therefore has diminishing returns. A pool five times larger sustains only about
twice as many NS iterations, because the extra configurations are consumed by an
exponentially growing number of rejected draws as $\Umax$ drifts below $\Utrain$. Only
55\% of a pool of $2\times10^4$ configurations is eventually accepted into the live set,
against 24\% for a pool of $10^5$. This helps explain why the small-pool runs of Fig.~4 need less wall-clock time than
their large-pool counterparts, despite retraining about twice as often. The same logarithmic dependence applies to any flow quality: it is a
property of NS, not of the generator.

\subsection{Run-to-run variability}
\label{sec:replicas}

None of the runs is seeded, and the reductions performed on the GPU are not bitwise
reproducible, so two simulations launched with identical hyperparameters diverge after the
first few operations. Every single-run number quoted in this work therefore carries a
run-to-run uncertainty, and the differences between the protocols of Fig.~4 are only
meaningful when read against it. To measure it we repeated the run of Fig.~4f --- the
cheapest of the eight in wall-clock time, and the one whose cost the main text quotes ---
ten times with identical settings, each starting from the same stored initial live set.
Table~\ref{tab:replicas} collects the result.

\begin{table}[b]
  \centering
  \caption{Ten independent repetitions of the run of Fig.~4f, with identical
  hyperparameters and the same stored initial live set. ``Spread'' is the sample standard
  deviation as a percentage of the mean. The reduction is with respect to the
  $5.05\times10^{10}$ energy evaluations of the standard NS reference. The efficiency
  $\eta$ is the inverse number of generation attempts per pool, as in
  Fig.~\ref{fig:hex_efficiency}a.}
  \label{tab:replicas}
  \small
  \begin{tabular}{@{}l r r r@{}}
    \toprule
    & Mean $\pm$ s.d. & Spread & Range \\
    \midrule
    Generation attempts                  & $15442 \pm 1790$      & $11.6\%$ & $13111$--$19120$ \\
    Energy evaluations ($\times10^{8}$)  & $7.67 \pm 0.36$       & $4.7\%$  & $7.20$--$8.40$ \\
    Reduction vs.\ standard NS           & $66.0 \pm 3.0$        & $4.5\%$  & $60.1$--$70.1$ \\
    Wall-clock time (h)                  & $8.67 \pm 0.52$       & $6.0\%$  & $8.08$--$9.49$ \\
    \quad of which training (h)          & $3.67 \pm 0.30$       & $8.1\%$  & $3.45$--$4.47$ \\
    \quad of which pool generation (h)   & $4.65 \pm 0.52$       & $11.1\%$ & $4.14$--$5.67$ \\
    Minimum $\eta$ ($\times10^{-4}$)     & $9.4 \pm 1.3$         & $13.7\%$ & $6.8$--$11.2$ \\
    Final $\eta$ ($\times10^{-2}$)       & $1.63 \pm 0.29$       & $17.8\%$ & $1.14$--$1.96$ \\
    Final $\Umax$                        & $-14.7932 \pm 0.0016$ & $0.01\%$ & $-14.7960$ to $-14.7908$ \\
    \bottomrule
  \end{tabular}
\end{table}

Three points follow. First, the trajectory is reproducible even though its cost is not.
All ten repetitions generate exactly 45 pools, and their final bounds lie in a range of
$0.005$, a relative spread of $10^{-4}$; what fluctuates is the work needed to get there.

Second, the fluctuation is not uniform along the trajectory. Grouping the generation
attempts by the threshold $\Utrain$ of the pool they were spent on, measured from the
final bound $U_{\mathrm{f}}$, gives $4951\pm353$ attempts ($7.1\%$) for
$\Utrain-U_{\mathrm{f}}>100$, $9637\pm1447$ ($15.0\%$) for
$1<\Utrain-U_{\mathrm{f}}<100$, and $854\pm178$ ($20.8\%$) below that. The dense-liquid
and crystalline stages, where the flow is hardest to train, are also where two identical
runs differ most.

Third, and this is what matters for reading Table~\ref{tab:runs}, the energy budget is a
far steadier quantity than the number of generation attempts, $4.7\%$ against $11.6\%$,
because the dilution stages contribute a fixed $4.44\times10^{8}$ evaluations that do not
fluctuate at all. The difference of two independent runs has $\sqrt{2}$ times the
single-run spread, so two entries of Table~\ref{tab:runs} must differ by about $13\%$ in
total energy evaluations, or by about $33\%$ in generation attempts, before the difference
reaches two standard deviations. On that scale the two large-pool protocols that include a
1C phase are genuinely cheaper than their small-pool counterparts ($5.12$ against
$7.98\times10^{8}$ and $5.61$ against $7.56\times10^{8}$, $44\%$ and $29\%$ apart, more
than four standard deviations each), whereas for CA alone the $7.57$ against $8.11\times10^{8}$ of
the two pool sizes are $7\%$ apart, one standard deviation, and do not establish an
ordering. Differences of that size between neighbouring rows of Table~\ref{tab:runs}
should not be read as real.

These ten repetitions measure the variability of one protocol at one state point. Three
runs with the long 1C+CA schedule of Fig.~4a give a wider spread, about $21\%$ in
generation attempts against the $11.6\%$ of Table~\ref{tab:replicas}, which is a
preliminary indication that the run-to-run variability grows with the amount of training
performed at each stage. We do not pursue it here: the quantitative statements of this
work rest on the run of Fig.~4f, which is the one that was repeated in full.

\subsection{Rectangular box with aspect ratio $2:\sqrt{3}$}
\label{sec:hex}

\paragraph{Setup.}
As discussed in Sec.~3.5, the square box was chosen because its point group $D_4$ provides
eight symmetry operations for data augmentation. To check that the method does not rely on
this choice, we simulated $N=9$ particles at the same density $\rho=0.95$ in a rectangular
cell with $L_y/L_x = 2/\sqrt{3}$, i.e.\ $L_x = 2.864$ and $L_y = 3.307$. Such a rectangular
cell matches the aspect ratio of the triangular lattice, the ground state of the 2D
Lennard--Jones system. Compared with the square-box runs, three things change:
\begin{itemize}
  \item the axis permutations are no longer symmetries of the cell, so training
  configurations are augmented only with the random choice of reference particle and a
  random reflection $\sigma_x$, $\sigma_y$ or $C_2$ (the group $D_2$), instead of the full
  hyperoctahedral group;
  \item the flow maps each Cartesian component to $[-1,1]$ with its own box length, and the
  maximum MC displacement is scaled per axis, $\Delta_\alpha = \Delta\,L_\alpha/\max(L_x,
  L_y)$;
  \item the learnable circular shifts between coupling layers are removed for simplicity, so each of the
  28 blocks consists of four spline coupling layers only (Sec.~\ref{sec:hyper}).
\end{itemize}
All other NS, training and generation hyperparameters are those of the best-performing
square-box run of Fig.~4f (Table~\ref{tab:runs}): a pool of $2\times10^4$ configurations,
cosine-annealing learning-rate schedule with 250 optimization steps per training stage, and
a window of three live sets. As for the square box, the initial live set is the
live set of a standard NS run started from uniformly distributed walkers, after about
$10^3$ iterations, and the flow is switched on after $\mathscr{M}'=100$ further standard
NS iterations. Over the
$5\times10^5$ iterations, $\Umax$ decreases from $2899.3$ to $-16.35$.

\begin{figure}[p]
  \centering
  \includegraphics[width=5.479in]{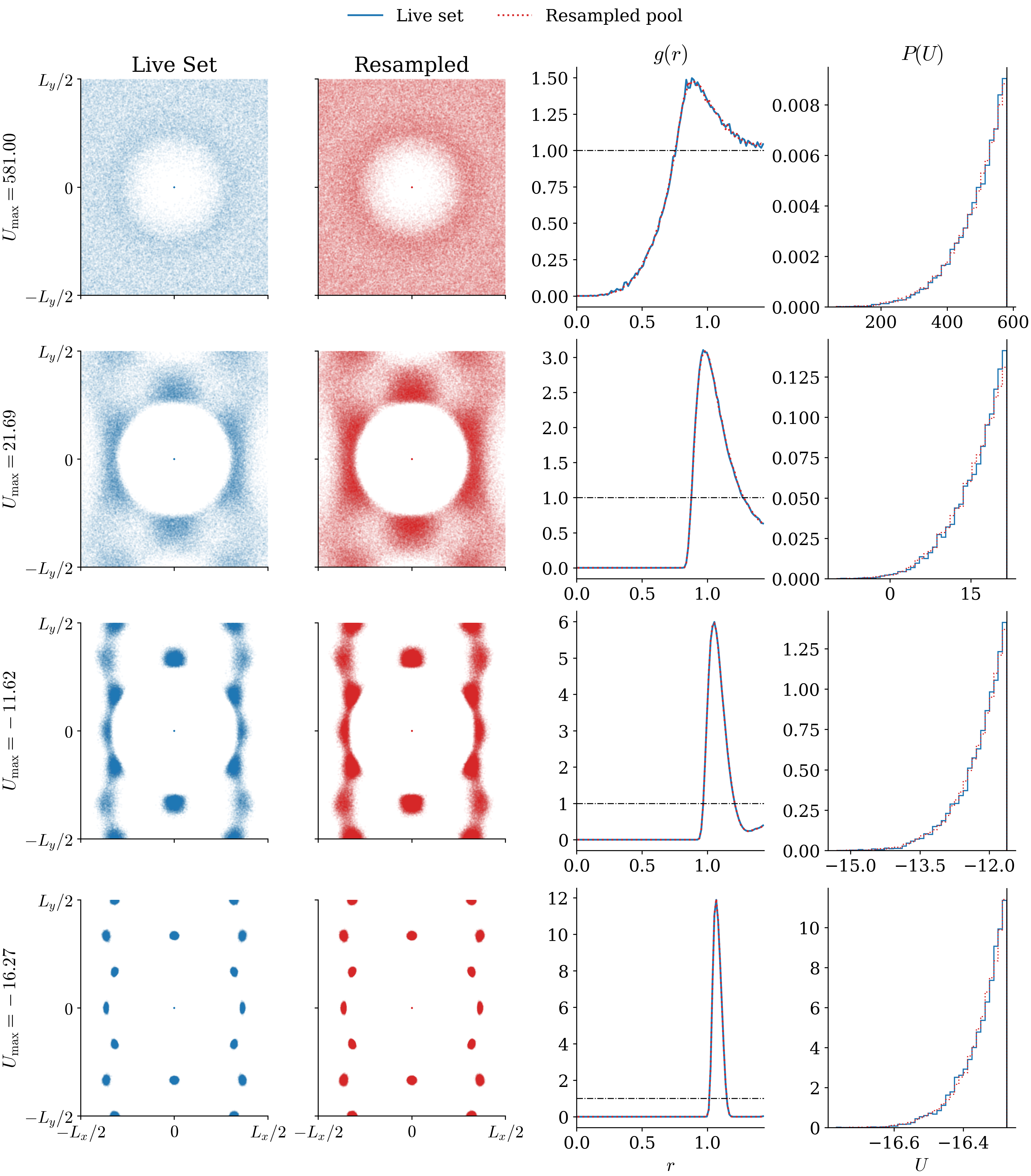}
  \caption{Live sets and resampled pools along the NS trajectory for $N=9$ Lennard--Jones
  disks in a $2:\sqrt{3}$ cell at $\rho=0.95$. Each row corresponds to one training stage,
  labelled by its energy threshold $\Utrain$, decreasing from top to bottom.
  First two columns: live set used for training and pool after rejection resampling,
  superimposed in internal coordinates without alignment (Sec.~\ref{sec:alignment}).
  Third column: radial distribution functions. The dash-dotted line is the ideal-gas
  limit. Fourth column: energy distributions. The vertical line marks $\Utrain$. Compare
  with Fig.~2 for the square box. Unlike in Fig.~2, the configurations generated by the
  flow before resampling, together with their $g(r)$ and $P(U)$, are not shown.}
  \label{fig:hex_configs}
\end{figure}

\paragraph{Structure and validation.}
Figure~\ref{fig:hex_configs} is the rectangular-cell counterpart of Fig.~2. The sequence of
constrained ensembles is qualitatively the same as in the square box: an almost uniform
fluid with an excluded-volume hole around the reference particle at high energy, a dense
liquid with a pronounced first coordination shell at intermediate energies, and a
crystalline solid with sharply localized sites at the lowest energies. As explained in
Sec.~\ref{sec:alignment}, in this cell all low-energy configurations produce the same
image, so the crystal appears in a single orientation without any alignment. At every
stage the resampled pool reproduces the positional density, the radial distribution
function and the energy distribution of the live set used for training, including the
sharp edge of $P(U)$ at $\Utrain$.

\begin{figure}[t]
  \centering
  \includegraphics[width=5.477in]{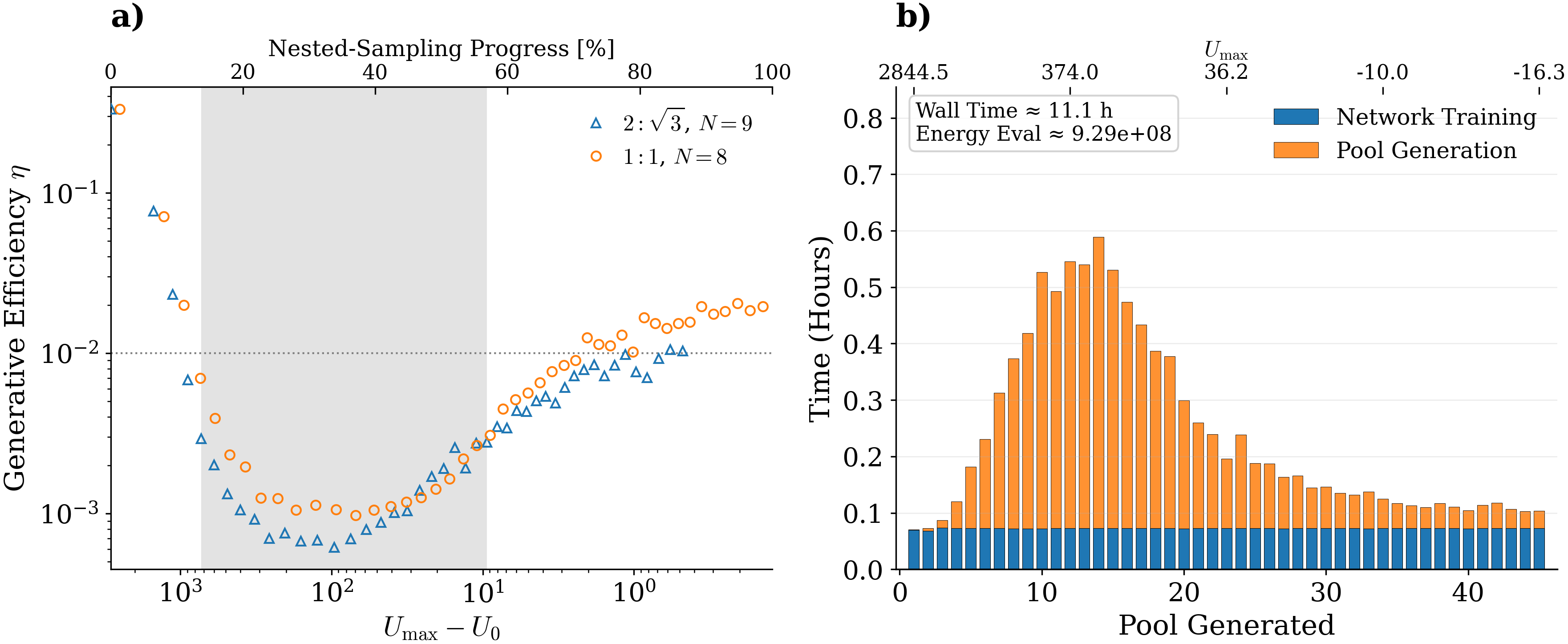}
  \caption{\textbf{a)} Generative efficiency $\eta$, defined as the inverse number of
  generation attempts needed to fill a pool, against the reduced energy $\Utrain - U_0$ of
  the training stage, with nested-sampling progress on the upper axis --- the axes of
  Fig.~5, so that the two figures can be read against each other. $U_0$ is the lowest
  live-point energy of the final live set of each run, as in Fig.~5. Triangles:
  $2:\sqrt{3}$ cell with $N=9$. Circles: square box with $N=8$ (run of Fig.~4f), with the
  same NS and training hyperparameters. The gray band is the low-efficiency regime of the
  rectangular cell, defined as in Fig.~5 as the contiguous range in which $\eta$ stays
  within a factor of five of its minimum; the corresponding range for the square box is
  given in the text. The dotted line marks $\eta=10^{-2}$, the same reference level as in
  Fig.~5.
  \textbf{b)} Wall-clock time spent training the flow (blue) and generating the pool
  (orange) for each pool of the rectangular-cell run, with the energy threshold $\Utrain$
  of the corresponding training stage on the upper axis; cf.\ Fig.~4f. The box reports
  the wall-clock time and the total number of energy evaluations of the whole run,
  including the standard NS stages and training (Table~\ref{tab:runs}).}
  \label{fig:hex_efficiency}
\end{figure}

\paragraph{Generative efficiency and cost.}
The generative efficiency (Fig.~\ref{fig:hex_efficiency}a) shows the same three regimes
as in Fig.~5. It drops from $\eta\approx0.33$ at the start to a minimum of
$6.2\times10^{-4}$ at about 34\% of the trajectory, in the dense-liquid regime. It then
recovers slowly as the system crystallizes and reaches $1.0\times10^{-2}$ in the final
solid-like stages. Applying the criterion of Fig.~5, by which the low-efficiency regime is
the contiguous range in which $\eta$ stays within a factor of five of its minimum, the
rectangular cell is inside it from 14\% to 57\% of the trajectory (gray band in
Fig.~\ref{fig:hex_efficiency}a) and the square box from 16\% to 59\%, the same range it
occupies in Fig.~5, whose abscissa is the same. Read on the energy
axis the two windows nearly coincide, $\Utrain-U_0$ from $7.3\times10^{2}$ to $9.4$ for
the rectangular cell against $5.9\times10^{2}$ to $7.4$ for the square box: where the
low-efficiency regime falls is set by the dense-liquid state itself rather than by the
shape of the cell. The difference between the two cells is one of level. Throughout the
liquid and solid regimes the efficiency of the rectangular cell stays below that of the
square box, whose minimum is $9.8\times10^{-4}$, and its final plateau lies
at $1.0\times10^{-2}$ against $2.0\times10^{-2}$. This is consistent with the larger
number of particles, with the weaker data augmentation, which provides four instead of
eight symmetry-equivalent copies of each training configuration, and with the absence of
the learnable circular shifts between coupling layers. In the square box these shifts move
the position of the periodic boundary of the spline domain from one coupling layer to the
next, so that it does not always fall at the same place in the box. Training takes about the
same time per stage as in the square box, $0.07$\,h against $0.08$\,h
(Fig.~\ref{fig:hex_efficiency}b), and the
additional cost is entirely due to pool generation in the liquid regime. The whole run
takes $11.1$\,h of wall-clock time, of which $3.3$\,h are spent training and $7.5$\,h
generating pools, and requires $9.3\times10^{8}$ energy evaluations in total
(Table~\ref{tab:runs}).

\subsection{Dilution Steps}
\label{sec:dilution}

Whenever the pool is exhausted the algorithm returns to $\mathscr{M}$ iterations of
standard NS before retraining (Alg.~2). Its purpose is to keep the flow from being trained
on its own output: each of these iterations replaces part of the live set with walkers
evolved by the constrained random walk of conventional NS, in which the potential enters
only through the test against the energy bound, so that the next training set is not made
entirely of configurations that a generative model produced. This dilution stage is the
single largest item in
the energy budget of the runs with the smaller pool. Each of its iterations evolves $10^3$
walkers through $10^2$ trial displacements and therefore costs $1.01\times10^5$ energy
evaluations, so a stage of $\mathscr{M}=100$ costs $1.01\times10^7$ evaluations whatever
the size of the pool that preceded it. The last pool of a run is never exhausted, so a run
that generates $N_{\mathrm{pool}}$ pools performs $N_{\mathrm{pool}}-1$ dilution stages.
The run of Fig.~4f generates 45 pools, dilutes 44 times and spends $4.44\times10^8$
evaluations on dilution, $59\%$ of its total (Table~\ref{tab:runs}); the runs with a
$10^5$ pool generate 21 pools, dilute 20 times and spend $2.02\times10^8$. Because this overhead
scales with the number of retraining stages, it places a floor on what can be gained by
reducing the pool size. The value $\mathscr{M}=100$ used throughout this work was chosen
a priori, and this section asks whether it is large enough to refresh the live set, and
what a stricter choice would cost.

\paragraph{Coverage.} A dilution iteration refreshes $n_{\mathrm{prop}}=10^3$ of the
$K=10^4$ live points: the walker at the current bound, which is always replaced, and
$n_{\mathrm{prop}}-1$ others drawn uniformly without replacement. Within an iteration the
selected walkers are distinct, but the draws of successive iterations are independent, so
nothing prevents a walker from being chosen twice, or from never being chosen at all.
Taking one selection to be what removes a walker's synthetic provenance, the number of
iterations needed for the live set to contain no flow-generated configuration at all is a
coupon-collector problem. A given walker is selected with
probability
\begin{equation}
  p = \frac{n_{\mathrm{prop}}-1}{K-1} = 0.0999
\end{equation}
per iteration, so after $\mathscr{M}$ iterations the expected number of walkers never
selected is $K(1-p)^{\mathscr{M}}$, and
\begin{equation}
  P_n(\mathscr{M}) = \Bigl[\,\mathbb{P}\bigl(\mathrm{Bin}(\mathscr{M},p)\ge n\bigr)\Bigr]^{K}
  \label{eq:coupon}
\end{equation}
is the probability that every walker has been selected at least $n$ times, where
$\mathrm{Bin}(\mathscr{M},p)$ denotes a binomial variable counting the successes in
$\mathscr{M}$ independent trials of probability $p$. The characteristic scale is $\ln K/[-\ln(1-p)] = 87.5$ iterations. Equation~\eqref{eq:coupon}
treats the walkers as independent, which is not exact because each iteration selects a
fixed number of them; a direct simulation of the selection gives $0.361$, $0.766$ and
$0.993$ at $\mathscr{M}=87$, $100$ and $132$, against $0.349$, $0.765$ and $0.991$ from
Eq.~\eqref{eq:coupon}.

At $\mathscr{M}=100$ the expected number of walkers left untouched by a dilution stage is
$K(1-p)^{\mathscr{M}} = 0.27$, that is fewer than one walker of the $10^4$, and
$P_1 = 0.76$ (Fig.~\ref{fig:dilution}a). On average, then, the stage does refresh the whole
live set, but it does not do so reliably: about one stage in four leaves at least one
walker behind. The choice is sensible but marginal, sitting just above the
coupon-collector scale, and raising $P_1$ to $0.99$ would require $\mathscr{M}=132$.

\paragraph{Cost.} Only the dilution term depends on $\mathscr{M}$. For the run of Fig.~4f
the remaining $3.13\times10^8$ evaluations, spent on pool generation, training, pool draws
and the $\mathscr{M}'=100$ warm-up iterations, are fixed, so the reduction relative to the
$5.05\times10^{10}$ evaluations of the standard NS reference is
\begin{equation}
  R(\mathscr{M}) = \frac{5.05\times10^{10}}
  {3.13\times10^{8} + N_{\mathrm{dil}}\,\mathscr{M}\,n_{\mathrm{prop}}\,(n_c+1)} ,
  \qquad N_{\mathrm{dil}} = N_{\mathrm{pool}} - 1 = 44 ,
  \label{eq:dilution_cost}
\end{equation}
with the $n_{\mathrm{prop}}=10^3$ walkers moved per iteration defined above and
$n_c=10^2$ trial displacements applied to each of them. This
is the curve of Fig.~\ref{fig:dilution}b. Two features stand out. The gain saturates: even
$\mathscr{M}\to0$ would give only $162\times$, because generation and training do not scale
with $\mathscr{M}$, so no amount of shortening the dilution stage turns this run into a
qualitatively cheaper one. And the region of interest is narrow. Going from the
$\mathscr{M}=100$ used here to the $\mathscr{M}=132$ at which the live set is refreshed
with probability $0.99$ moves the reduction from $67\times$ to $56\times$: the stricter
guarantee costs about $16\%$ of the speed-up, which is a small price for removing the
residual correlation between successive training sets.

\paragraph{Caveats.} Two caveats attach to the criterion itself, and both make it
optimistic. A selection gives a walker $10^2$ trial single-particle displacements at an
acceptance held near $0.5$ by the adaptive step size, that is roughly $50$ accepted moves,
or six per particle for $N=8$; whether this is enough to move the walker away from the
flow-generated configuration it started from, rather than merely to perturb it, is an
assumption rather than a measurement.
If two selections were required, $P_2=0.99$ would need $\mathscr{M}=159$, i.e. $50\times$
(Fig.~\ref{fig:dilution}a). Separately, the walker at the bound is not evolved in place but
replaced by a copy of another live point, so each iteration creates two walkers descended
from a single parent; both are evolved by the random walk and are therefore no longer
synthetic, but they are correlated with each other, and over a stage of $\mathscr{M}=100$ this affects
$1\%$ of the live set. A direct measurement of how far an MCMC stage carries a walker from
its starting configuration would settle the first point and fix $\mathscr{M}$ from below;
we leave it to future work.

\begin{figure}[t]
  \centering
  \includegraphics[width=5.480in]{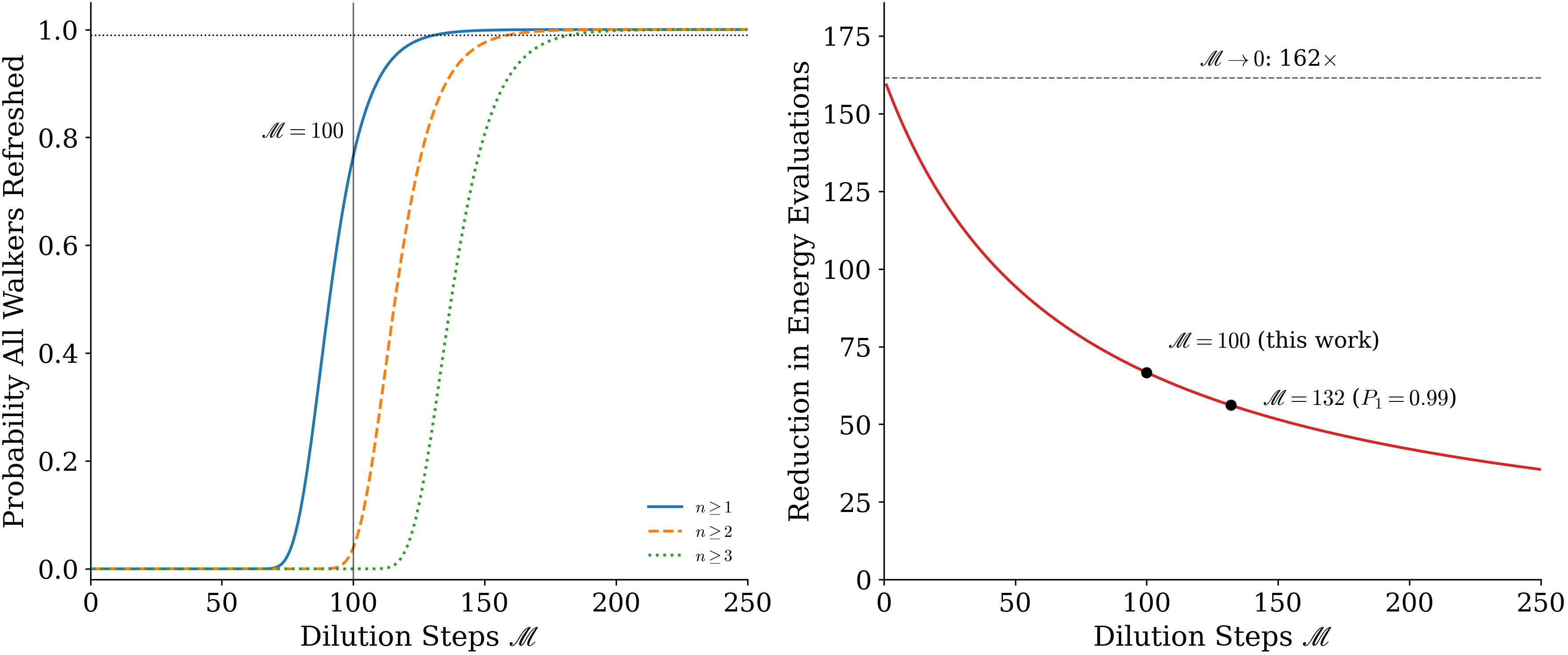}
  \caption{Cost and completeness of the dilution stage. \textbf{a)} Probability that every
  one of the $K=10^4$ walkers has been selected at least $n$ times during a stage of
  $\mathscr{M}$ standard NS iterations, from Eq.~\eqref{eq:coupon}. The vertical line marks
  the $\mathscr{M}=100$ used in this work and the dotted horizontal line is $P_n=0.99$.
  \textbf{b)} Reduction in energy evaluations relative to standard NS for the run of
  Fig.~4f, from Eq.~\eqref{eq:dilution_cost}. The two markers are the $\mathscr{M}=100$
  performed in this work and the $\mathscr{M}=132$ at which every walker is refreshed with
  probability $0.99$; the dashed line is the $\mathscr{M}\to0$ ceiling of $162\times$ set by
  pool generation, training and the warm-up, which do not depend on $\mathscr{M}$.}
  \label{fig:dilution}
\end{figure}

\paragraph{Walkers.} The number of walkers refreshed per iteration,
$n_{\mathrm{prop}}$, is the second parameter of the dilution stage, and it turns out to
matter more than $\mathscr{M}$. Writing the cost of a stage as
\begin{equation}
  \mathscr{M}\,n_{\mathrm{prop}}\,(n_c+1) = \bar{n}\,K\,(n_c+1),
  \qquad \bar{n} \equiv \frac{\mathscr{M}\,n_{\mathrm{prop}}}{K},
  \label{eq:dilution_R}
\end{equation}
shows that the two parameters enter only through their product: the energy budget of a
stage is fixed by $\bar{n}$ alone, the mean number of times each walker is refreshed,
and not by either parameter separately. What
$n_{\mathrm{prop}}$ controls, at fixed budget, is not the cost but the spread. The number
of refreshes a given walker receives is $\mathrm{Bin}(\mathscr{M},p)$, with mean $\bar{n}$
and a tail at zero, and only at $n_{\mathrm{prop}}=K$, where every iteration selects every
walker, does each one receive exactly $\mathscr{M}$ of them. The coupon-collector problem
of the preceding paragraphs is therefore not an intrinsic trade-off but a consequence of
refreshing a subset: at the $\bar{n}=10$ of this work, moving all $K$ walkers and setting
$\mathscr{M}=10$ would have cost exactly what we spent and left nothing to chance
(Fig.~\ref{fig:dilution_nprop}a).

The value $n_{\mathrm{prop}}=10^3$ was chosen by analogy with conventional NS
implementations, where the walk is applied not only to the cloned walker but to a further
$N_p-1$ randomly selected ones, $N_p$ being the number of parallel
tasks~\cite{baldock_determining_2016,baldock_constant-pressure_2017,unglert_neural-network_2023}.
The mean number of walks a configuration there receives before it is removed is
$N_p/K_r$, with $K_r$ the walkers discarded per iteration, which is the analogue of our
$\bar{n}$. The essential difference is that the walk performed at each iteration is
shortened to $\langle L\rangle K_r/N_p$ steps, so that the total walk a configuration
receives before removal stays at $\langle L\rangle$ and the computational work is
independent of $N_p$ altogether~\cite{baldock_constant-pressure_2017}. Load balance then
requires that shortened walk to remain a few times longer than a single step, which caps
$N_p$ at a few tens for the walk lengths in use, and does so independently of $K$: live
sets of $5\times10^2$ to $5\times10^3$ walkers are refreshed a few tens at a
time~\cite{partay_nested_2021}. Our dilution stage instead gives every selected walker a
full walk of $n_c$ trial displacements, so that constraint does not apply, the cost grows
with $n_{\mathrm{prop}}$, and that growth is compensated exactly by the fewer iterations
then required.

Figure~\ref{fig:dilution_nprop}b gives the consequence: the reduction in energy
evaluations obtained when $\mathscr{M}$ is set, for each $n_{\mathrm{prop}}$, to the
smallest value reaching $P_1\geq0.99$. At $n_{\mathrm{prop}}=10^3$ that guarantee costs
$\mathscr{M}=132$ and returns $56\times$; at $n_{\mathrm{prop}}=K$ it costs a single
iteration and returns $141\times$, within a factor $1.14$ of the $162\times$ ceiling. The
two parameters must be moved together to realize this: raising $n_{\mathrm{prop}}$ to $K$
while leaving $\mathscr{M}=100$ would multiply $\bar{n}$ by ten and reduce the gain to
$11\times$.

Whether $\bar{n}=1$ suffices is the question left open above, and this analysis does not settle
it: each walker of this work received about ten refreshes, and a single one may not erase
the memory of a flow-generated starting configuration. What the analysis does show is that
the dilution stage was never tuned. Its parameters were fixed a priori, and the
$67\times$ reduction reported in the main text was obtained with that choice; scheduling
the same stage differently would take it to $141\times$ without touching the flow, the
training protocol or the pool size.

\begin{figure}[t]
  \centering
  \includegraphics[width=5.479in]{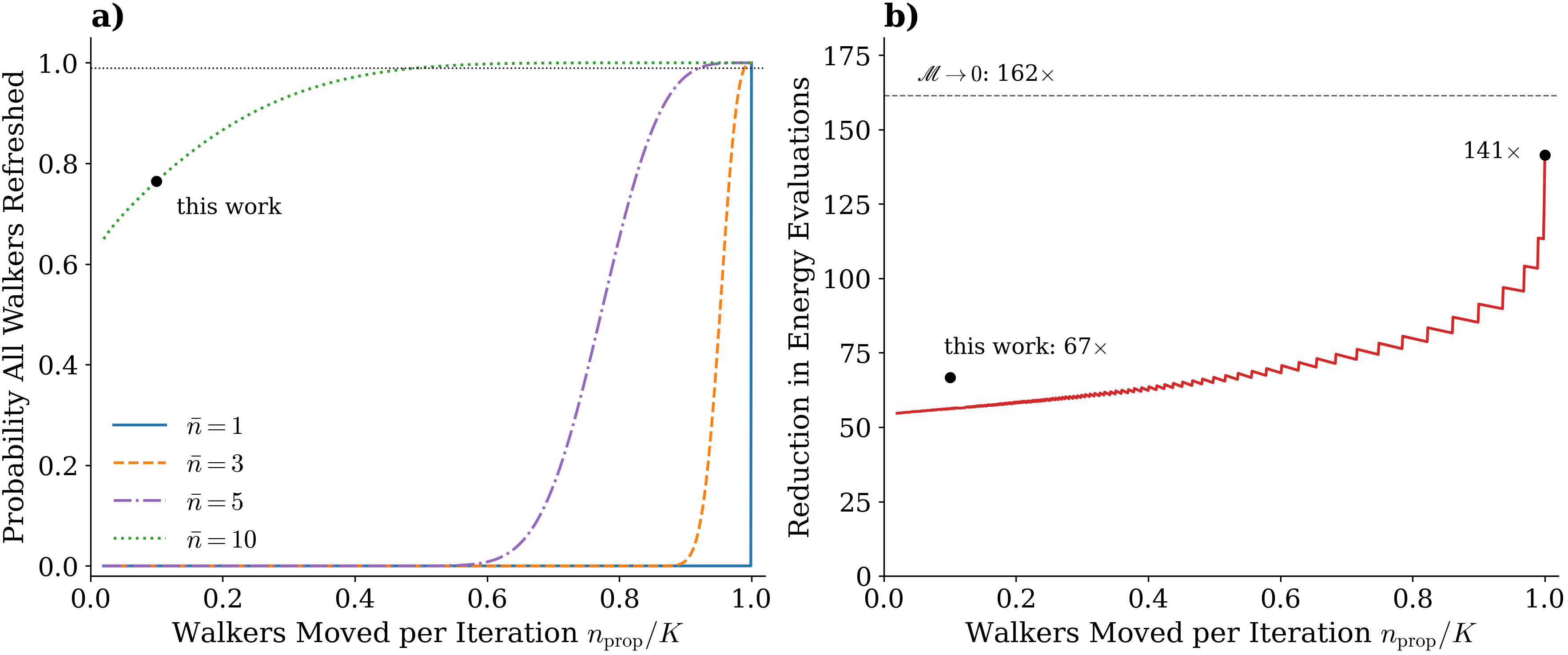}
  \caption{Choosing the number of walkers refreshed per dilution iteration.
  \textbf{a)} Probability that every one of the $K=10^4$ walkers is refreshed during a
  dilution stage, against $n_{\mathrm{prop}}$, at fixed cost: each curve holds
  $\bar{n}=\mathscr{M}n_{\mathrm{prop}}/K$ constant, so every point on a curve spends the same
  number of energy evaluations. The marker is the $n_{\mathrm{prop}}=10^3$,
  $\mathscr{M}=100$ of this work and the dotted line is $P_1=0.99$.
  \textbf{b)} Reduction in energy evaluations for the run of Fig.~4f when $\mathscr{M}$ is
  set, for each $n_{\mathrm{prop}}$, to the smallest integer reaching $P_1\geq0.99$; the
  steps are that rounding. The marker at $n_{\mathrm{prop}}/K=0.1$ is this work, which
  lies above the curve because it does not meet the guarantee ($P_1=0.76$). The dashed
  line is the $\mathscr{M}\to0$ ceiling of Fig.~\ref{fig:dilution}.}
  \label{fig:dilution_nprop}
\end{figure}

\section{Network and training hyperparameters}
\label{sec:hyper}

\paragraph{Flow architecture.}
The base distribution is the ideal gas: $N$ particles uniformly distributed in a box of the
same size as the target system. Both base and target configurations are expressed in
internal coordinates (particle~0 at the origin), and the $N-1$ remaining particles are
mapped to $[-1,1]^2$ by dividing each Cartesian component by half the corresponding box
length. The flow acts on the $2(N-1)$ remaining coordinates. It consists of 28 blocks,
each made of
\begin{equation*}
  \bigl[\,\mathrm{S},\ \mathrm{RQS}_{x|y},\ \mathrm{RQS}_{y|x},\
          \mathrm{S},\ \mathrm{RQS}_{y|x},\ \mathrm{RQS}_{x|y}\,\bigr],
\end{equation*}
where $\mathrm{RQS}_{x|y}$ transforms the $x$ coordinates of all particles conditioned on
their $y$ coordinates and vice versa, and $\mathrm{S}$ is a learnable circular shift of
all particles, $x\mapsto x+s$ wrapped back into $[-1,1)$. In the $2:\sqrt{3}$ cell the
shifts are omitted. Each coupling layer is a circular rational-quadratic spline
\cite{durkan_neural_2019} on $[-1,1]$ with 12 bins. Its derivatives at the two boundaries are
tied, so that the transformation is periodic, and it is initialized to the identity.
The spline parameters of each particle ($3\times12=36$) are predicted by a
permutation-equivariant transformer \cite{vaswani_attention_2017} acting on the particles as
tokens, without positional encoding. Each particle's untransformed coordinate $u\in[-1,1]$
is embedded with $8$ circular Fourier features, $\{\cos(k\pi u), \sin(k\pi u)\}_{k=1}^{8}$.
The normalized energy threshold is appended, and the result is projected linearly to a
width of 128. It then passes through one pre-layer-norm transformer encoder layer (2
attention heads, feed-forward width 512, no dropout) and a final linear layer. Each
coupling layer has $205\,220$ parameters, and the whole flow has
$2.30\times10^{7}$ trainable parameters, independent of $N$.

\paragraph{Conditioning.}
The energy thresholds of the live sets in the training window are rescaled to $[0,1]$ by
min--max normalization over the window. The most recent, lowest threshold, i.e.\ $\Utrain$,
maps to $0$, and this is the value used for pool generation. When the window contains a
single live set, the condition is $0$.

\paragraph{Training.}
The flow is trained by maximum likelihood on the live sets (backward loss only,
$w_{\mathrm{fwd}}=0$, $w_{\mathrm{bwd}}=1$), with Adam, batches of $10^3$ configurations
and gradients clipped to a norm of $100$. The training set is the union of the last three
live sets, i.e.\ up to $3\times10^4$ configurations, of which 10\% are held out for
validation. The number of optimization steps is rounded up to an integer number of epochs,
of 9, 18 and 27 steps per epoch for one, two and three live sets in the window. After every
epoch the flow is evaluated on the validation set, and the parameters with the highest
validation RESS are kept at the end of each training stage. Network parameters are carried
over from one training stage to the next, except where the protocols below re-initialize
them. Two learning-rate schedules are used:
\begin{itemize}
  \item \textbf{1C}: one-cycle schedule, rising from $10^{-5}$ to a peak of
  $2.5\times10^{-4}$ over the first 40\% of the steps and annealing to $10^{-4}$, stepped
  every optimization step;
  \item \textbf{CA}: cosine annealing from $10^{-4}$ to $10^{-6}$, stepped every epoch.
\end{itemize}
In the protocols labelled 1C+CA, a 1C stage is immediately followed by a CA stage at every
training stage, and the CA stage starts from the best parameters of the 1C stage. In those
labelled 1C+CA/CA(5) the same pair is applied only at every fifth training stage, with CA
alone in between; the network is re-initialized at each of those full stages and carried
over through the CA stages between them, five and nine times over the runs of Fig.~4c and
Fig.~4d. The run of Fig.~4a re-initializes before every one of its 21 trainings, and the
CA runs never do. The protocols of the individual
runs are listed in Table~\ref{tab:runs}, and Fig.~\ref{fig:lr_schedules} compares them
within one training stage. The long 1C+CA protocol of Fig.~4a starts each stage from a re-initialized network, and is the only one that keeps improving over the whole stage, reaching a validation RESS of $0.13$; the three shorter protocols continue from the previous stage, starting between $0.008$ and $0.019$ and reaching at most $0.03$. This is consistent with its lower cost of pool
generation in Table~\ref{tab:runs}, obtained at the price of the longest training.

\begin{figure}[t]
  \centering
  \includegraphics[width=5.480in]{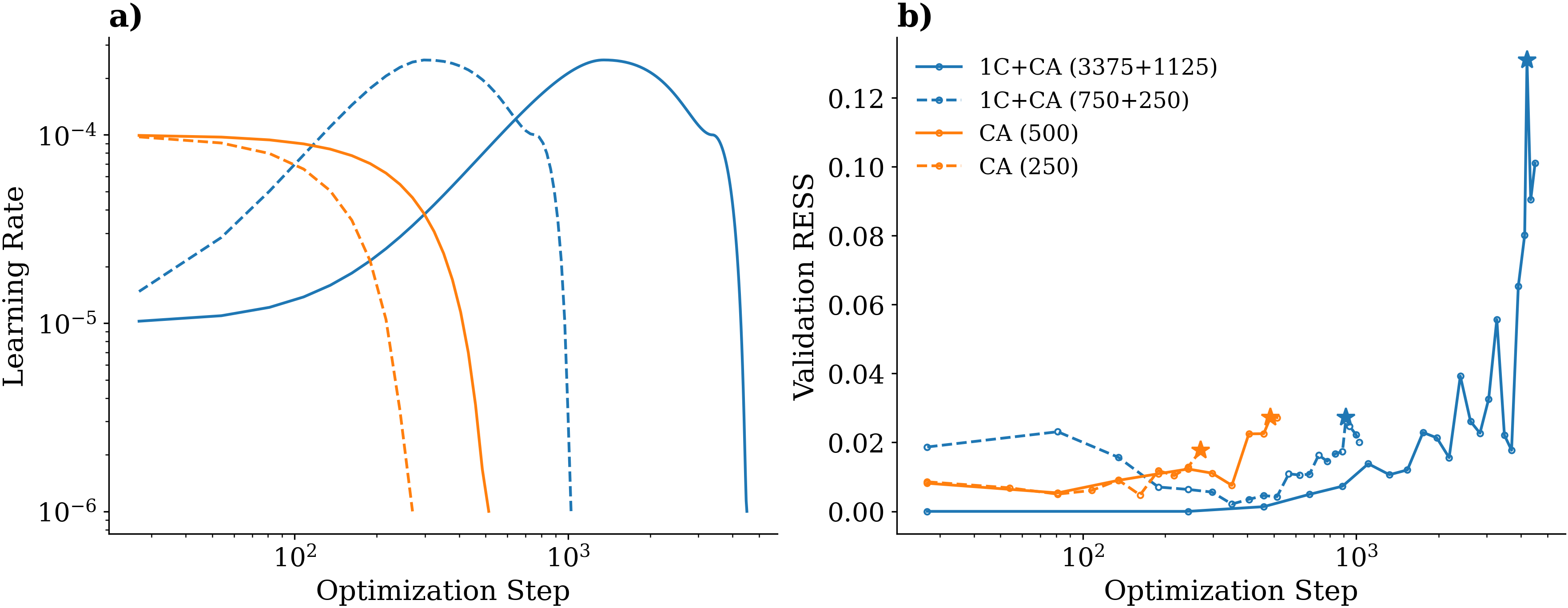}
  \caption{Learning-rate protocols of the runs of Fig.~4, within one training stage whose
  window holds three live sets. Blue: 1C+CA with 3375+1125 steps (Fig.~4a,
  $\Utrain=38.6$, solid) and 750+250 steps (Fig.~4b, $\Utrain=26.1$, dashed). Orange: CA
  with 500 steps (Fig.~4e, $\Utrain=38.1$, solid) and 250 steps (Fig.~4f,
  $\Utrain=25.7$, dashed). For each run the training stage is the one with $\Utrain$
  closest to $30$. \textbf{a)} Learning rate. \textbf{b)} Validation RESS. It is recorded
  after every epoch, but the long protocol runs for 167 epochs against 38, 19 and 10 for
  the shorter ones, so the curves are thinned to every eighth epoch for 1C+CA (3375+1125) and every
  second for 1C+CA (750+250) and CA (500); stars mark the parameters kept at the end of
  the stage, selected over all epochs.}
  \label{fig:lr_schedules}
\end{figure}

\paragraph{Generation.}
Pools are generated in batches of $2\times10^4$ base samples, whatever the target pool
size. Each batch is mapped through the flow at the normalized condition $0$, its energies
are evaluated, and each sample is accepted with probability $w_j/\max_k w_k$, where $w$ is
the importance weight of Eq.~(9) with respect to the indicator target at $\Utrain$ and the
maximum is taken over the batch. Batches are generated until the pool is full, and the
generative efficiency is the inverse of the number of batches.

\begin{table}[p]
  \centering
  \caption{Hyperparameters shared by all flow-based runs.}
  \label{tab:shared}
  \small
  \begin{tabularx}{\linewidth}{@{}l X@{}}
    \toprule
    \multicolumn{2}{@{}l}{\textit{System}} \\
    Particles, cell & $N=8$, square, $L=2.9$ ($\rho=0.951$) or $L=3.3$ ($\rho=0.735$); \\
                    & $N=9$, $L_x\times L_y = 2.864\times3.307$ ($\rho=0.95$) \\
    Potential & LJ, cut and shifted at $R_c = 0.49\,\min(L_x, L_y)$, linearized below
                $R_{\mathrm{cutin}}=0.8$, long-range corrections included \\
    \midrule
    \multicolumn{2}{@{}l}{\textit{Nested sampling}} \\
    Walkers & $K=10^4$ \\
    Iterations & $5\times10^5$ \\
    Standard NS iterations & $\mathscr{M}'=\mathscr{M}=100$ (warm-up and dilution stages) \\
    Walkers moved per standard NS iteration & $10^3$ \\
    MC trial moves per walker & $100$ single-particle displacements \\
    MC maximum displacement & $\Delta=1.2$ initially, halved whenever the acceptance of an
                              iteration drops below $0.5$ \\
    \midrule
    \multicolumn{2}{@{}l}{\textit{Flow}} \\
    Base distribution & ideal gas (uniform in the box), internal coordinates \\
    Blocks & 28 \\
    Coupling layers & circular rational-quadratic splines, 12 bins, on $[-1,1]$ \\
    Conditioner & equivariant transformer: 8 Fourier frequencies, width 128, 1 layer,
                  2 heads, feed-forward width 512, pre-layer-norm, no dropout \\
    Trainable parameters & $2.30\times10^7$ ($205\,220$ per coupling layer) \\
    \midrule
    \multicolumn{2}{@{}l}{\textit{Training}} \\
    Loss & backward (maximum likelihood), $w_{\mathrm{fwd}}=0$, $w_{\mathrm{bwd}}=1$ \\
    Training window & last $h=3$ live sets, 10\% held out for validation \\
    Augmentation & random reference particle; random operation of $D_4$ (square) or
                   $D_2$ (rectangle) \\
    Optimizer & Adam, batch size $10^3$, gradient-norm clipping at 100 \\
    Model selection & best validation RESS within each training stage \\
    Initialization & parameters carried over between stages (online fine-tuning), except
                     in 1C+CA/CA(5), which re-initializes whenever the full 1C+CA pair is
                     run, and in the run of Fig.~4a, which re-initializes before every
                     training \\
    \midrule
    \multicolumn{2}{@{}l}{\textit{Generation}} \\
    Batch & $2\times10^4$ base samples per attempt \\
    Resampling & accept with probability $w_j/\max_k w_k$ within each batch \\
    \bottomrule
  \end{tabularx}
\end{table}

\begin{table}[p]
  \centering
  \caption{Flow-based runs and their computational cost. ``Fig.'' is the figure of the main
  text (or of this SM) the run appears in. ``Schedule'' gives the learning-rate protocol of
  each training stage, with the number of optimization steps in parentheses. $P$: pool
  size; ``Pools'': number of training stages. Energy evaluations are split into pool
  generation, standard NS (warm-up and dilution) stages and training (dataset labels and
  validation); the total also includes the draws from the pool and the initial live set. The standard NS reference is listed for comparison.}
  \label{tab:runs}
  \scriptsize
  \setlength{\tabcolsep}{3pt}
  \begin{tabularx}{\linewidth}{@{}l l l >{\raggedright\arraybackslash}X r r r r r r@{}}
    \toprule
    & & & & & & \multicolumn{4}{c}{Energy evaluations ($\times10^{8}$)} \\
    \cmidrule(l){7-10}
    Fig. & Cell & $P$ & Schedule & Pools & Wall (h) & Pool gen. & Standard NS & Training & Total \\
    \midrule
2, 4a & 1:1 & $10^5$ & 1C+CA (3375+1125) & 21 & 33.0 & 2.87 & 2.12 & 0.111 & 5.12 \\
4b & 1:1 & $2\times10^4$ & 1C+CA (750+250) & 45 & 19.3 & 3.36 & 4.54 & 0.0645 & 7.98 \\
4c & 1:1 & $10^5$ & 1C+CA/CA(5) (3375+1125) & 21 & 17.7 & 3.42 & 2.12 & 0.0512 & 5.61 \\
4d & 1:1 & $2\times10^4$ & 1C+CA/CA(5) (750+250) & 45 & 10.3 & 2.97 & 4.54 & 0.0342 & 7.56 \\
4e & 1:1 & $10^5$ & CA (500) & 21 & 12.4 & 5.95 & 2.12 & 0.0180 & 8.11 \\
3, 4f, 5 (left) & 1:1 & $2\times10^4$ & CA (250) & 45 & 8.4 & 2.99 & 4.54 & 0.0267 & 7.57 \\
5 (right) & 1:1, $\rho\approx0.73$ & $2\times10^4$ & CA (250) & 45 & 13.9 & 6.79 & 4.54 & 0.0267 & 11.4 \\
S3, S4 & $2:\sqrt{3}$ & $2\times10^4$ & CA (250) & 45 & 11.1 & 4.71 & 4.54 & 0.0267 & 9.29 \\
\midrule
3 & 1:1 & -- & standard NS & -- & 11.9 & -- & 505 & -- & 505 \\
    \bottomrule
  \end{tabularx}
\end{table}

\paragraph{Runs and computational budget.}
Table~\ref{tab:runs} lists every flow-based run shown in the main text and in this
Supplementary Material, together with its wall-clock time and number of energy
evaluations. The count includes every evaluation of the potential performed by the code:
\begin{enumerate}
  \item pool generation, one evaluation per flow sample ($2\times10^4$ per attempt);
  \item the standard NS stages, i.e.\ the $\mathscr{M}'=100$ warm-up iterations and the
  $\mathscr{M}=100$ dilution iterations after each pool but the last, which is never
  exhausted. Each iteration evaluates the
  $10^3$ walkers it moves once and then once per trial move, $1.01\times10^5$ evaluations
  in total;
  \item training: the energies of the whole training window when the dataset is built,
  and the validation loss on the held-out base samples after every epoch, which selects
  the saved parameters;
  \item the draws from the pool, whose energy is compared with the current bound, and the
  initial live set. Together these never exceed $2.1\times10^6$ per run and are included
  only in the total.
\end{enumerate}
The same count gives $5.05\times10^{10}$ evaluations for the standard NS reference. Every
run starts from a stored initial live set, obtained with about $10^3$ standard NS
iterations from uniformly distributed walkers; all runs in the square box at $L=2.9$,
including the reference, share the same one. These iterations are not counted.

\printbibliography

@article{partay_nested_2014,
	title = {Nested sampling for materials: {The} case of hard spheres},
	volume = {89},
	shorttitle = {Nested sampling for materials},
	url = {https://link.aps.org/doi/10.1103/PhysRevE.89.022302},
	doi = {10.1103/PhysRevE.89.022302},
	number = {2},
	urldate = {2022-04-11},
	journal = {Phys. Rev. E},
	author = {Pártay, Lívia B. and Bartók, Albert P. and Csányi, Gábor},
	month = feb,
	year = {2014},
	pages = {022302},
}

@article{skilling_nested_2006,
	title = {Nested sampling for general {Bayesian} computation},
	volume = {1},
	issn = {1936-0975, 1931-6690},
	url = {https://projecteuclid.org/journals/bayesian-analysis/volume-1/issue-4/Nested-sampling-for-general-Bayesian-computation/10.1214/06-BA127.full},
	doi = {10.1214/06-BA127},
	number = {4},
	urldate = {2022-02-07},
	journal = {Bayesian Anal.},
	author = {Skilling, John},
	month = dec,
	year = {2006},
	pages = {833--859},
}

@article{skilling_nested_2004,
	title = {Nested {Sampling}},
	volume = {735},
	issn = {0094-243X},
	url = {https://aip.scitation.org/doi/abs/10.1063/1.1835238},
	doi = {10.1063/1.1835238},
	number = {1},
	urldate = {2022-02-11},
	journal = {AIP Conf. Proc.},
	author = {Skilling, John},
	month = nov,
	year = {2004},
	pages = {395--405},
}

@article{martiniani_superposition_2014,
	title = {Superposition {Enhanced} {Nested} {Sampling}},
	volume = {4},
	url = {https://link.aps.org/doi/10.1103/PhysRevX.4.031034},
	doi = {10.1103/PhysRevX.4.031034},
	number = {3},
	urldate = {2022-02-06},
	journal = {Phys. Rev. X},
	author = {Martiniani, Stefano and Stevenson, Jacob D. and Wales, David J. and Frenkel, Daan},
	month = aug,
	year = {2014},
	pages = {031034},
}

@misc{rezende_variational_2016,
	title = {Variational {Inference} with {Normalizing} {Flows}},
	url = {http://arxiv.org/abs/1505.05770},
	doi = {10.48550/arXiv.1505.05770},
	urldate = {2025-09-30},
	publisher = {arXiv},
	author = {Rezende, Danilo Jimenez and Mohamed, Shakir},
	month = jun,
	year = {2016},
	note = {arXiv:1505.05770 [stat]},
}

@article{unglert_neural-network_2023,
	title = {Neural-network force field backed nested sampling: \{{Study}\} of the silicon \$\{p{\textbackslash}text\{-\}{T}\}\$ phase diagram},
	volume = {7},
	shorttitle = {Neural-network force field backed nested sampling},
	url = {https://link.aps.org/doi/10.1103/PhysRevMaterials.7.123804},
	doi = {10.1103/PhysRevMaterials.7.123804},
	number = {12},
	urldate = {2024-01-12},
	journal = {Phys. Rev. Mater.},
	author = {Unglert, Nico and Carrete, Jesús and Pártay, Livia B. and Madsen, Georg K. H.},
	month = dec,
	year = {2023},
	pages = {123804},
}

@article{marchant_exploring_2023,
	title = {Exploring the configuration space of elemental carbon with empirical and machine learned interatomic potentials},
	volume = {9},
	copyright = {2023 The Author(s)},
	issn = {2057-3960},
	url = {https://www.nature.com/articles/s41524-023-01081-w},
	doi = {10.1038/s41524-023-01081-w},
	number = {1},
	urldate = {2023-10-17},
	journal = {npj Comput. Mater.},
	author = {Marchant, George A. and Caro, Miguel A. and Karasulu, Bora and Pártay, Livia B.},
	month = jul,
	year = {2023},
	pages = {1--12},
}

@article{dorrell_thermodynamics_2019,
	title = {Thermodynamics and the potential energy landscape: case study of small water clusters},
	volume = {21},
	issn = {1463-9084},
	shorttitle = {Thermodynamics and the potential energy landscape},
	url = {https://pubs.rsc.org/en/content/articlelanding/2019/cp/c9cp00474b},
	doi = {10.1039/C9CP00474B},
	number = {14},
	urldate = {2022-05-18},
	journal = {Phys. Chem. Chem. Phys.},
	author = {Dorrell, Jordan and Pártay, Livia B.},
	month = apr,
	year = {2019},
	pages = {7305--7312},
}

@article{unglert_replica_2025,
	title = {Replica {Exchange} {Nested} {Sampling}},
	volume = {21},
	issn = {1549-9618},
	url = {https://doi.org/10.1021/acs.jctc.5c00588},
	doi = {10.1021/acs.jctc.5c00588},
	number = {15},
	urldate = {2025-09-09},
	journal = {J. Chem. Theory Comput.},
	author = {Unglert, N. and Pártay, L. B. and Madsen, G. K. H.},
	month = aug,
	year = {2025},
	pages = {7304--7319},
}

@article{baldock_determining_2016,
	title = {Determining pressure-temperature phase diagrams of materials},
	volume = {93},
	url = {https://link.aps.org/doi/10.1103/PhysRevB.93.174108},
	doi = {10.1103/PhysRevB.93.174108},
	number = {17},
	urldate = {2021-10-05},
	journal = {Phys. Rev. B},
	author = {Baldock, Robert J. N. and Pártay, Lívia B. and Bartók, Albert P. and Payne, Michael C. and Csányi, Gábor},
	month = may,
	year = {2016},
	pages = {174108}
}

@article{baldock_constant-pressure_2017,
	title = {Constant-pressure nested sampling with atomistic dynamics},
	volume = {96},
	url = {https://link.aps.org/doi/10.1103/PhysRevE.96.043311},
	doi = {10.1103/PhysRevE.96.043311},
	number = {4},
	urldate = {2022-05-16},
	journal = {Phys. Rev. E},
	author = {Baldock, Robert J. N. and Bernstein, Noam and Salerno, K. Michael and Pártay, Lívia B. and Csányi, Gábor},
	month = oct,
	year = {2017},
	pages = {043311}
}

@article{batatia_mace_2022,
	title = {{MACE}: {Higher} {Order} {Equivariant} {Message} {Passing} {Neural} {Networks} for {Fast} and {Accurate} {Force} {Fields}},
	volume = {35},
	shorttitle = {{MACE}},
	url = {https://proceedings.neurips.cc/paper_files/paper/2022/hash/4a36c3c51af11ed9f34615b81edb5bbc-Abstract-Conference.html},
	language = {en},
	urldate = {2025-10-01},
	journal = {Advances in Neural Information Processing Systems},
	author = {Batatia, Ilyes and Kovacs, David P. and Simm, Gregor and Ortner, Christoph and Csanyi, Gabor},
	month = dec,
	year = {2022},
	pages = {11423--11436},
}

@article{wirnsberger_normalizing_2022,
	title = {Normalizing flows for atomic solids},
	volume = {3},
	issn = {2632-2153},
	url = {https://iopscience.iop.org/article/10.1088/2632-2153/ac6b16},
	doi = {10.1088/2632-2153/ac6b16},
	number = {2},
	urldate = {2023-02-17},
	journal = {Mach. Learn.: Sci. Technol.},
	author = {Wirnsberger, Peter and Papamakarios, George and Ibarz, Borja and Racanière, Sébastien and Ballard, Andrew J and Pritzel, Alexander and Blundell, Charles},
	month = jun,
	year = {2022},
	pages = {025009}
}

@article{papamakarios_normalizing_2021,
	title = {Normalizing {Flows} for {Probabilistic} {Modeling} and {Inference}},
	volume = {22},
	issn = {1533-7928},
	url = {http://jmlr.org/papers/v22/19-1028.html},
	number = {57},
	urldate = {2025-10-13},
	journal = {J. Mach. Learn. Res.},
	author = {Papamakarios, George and Nalisnick, Eric and Rezende, Danilo Jimenez and Mohamed, Shakir and Lakshminarayanan, Balaji},
	year = {2021},
	pages = {1--64},
}

@article{kroupa_resonances_2025,
	title = {Resonances in reflective {Hamiltonian} {Monte} {Carlo}},
	volume = {111},
	url = {https://link.aps.org/doi/10.1103/PhysRevE.111.045308},
	doi = {10.1103/PhysRevE.111.045308},
	number = {4},
	journal = {Physical Review E},
	author = {Kroupa, Namu and Csányi, Gábor and Handley, Will},
	year = {2025},
	pages = {045308},
}

@article{kuhn_hungarian_1955,
  title   = {The {H}ungarian method for the assignment problem},
  author  = {Kuhn, H. W.},
  journal = {Naval Research Logistics Quarterly},
  volume  = {2},
  number  = {1--2},
  pages   = {83--97},
  year    = {1955},
}

@misc{yallup_neural_2026,
	title = {Neural {Transport} {Nested} {Sampling}},
	url = {http://arxiv.org/abs/2609.29413},
	doi = {10.48550/arXiv.2609.29413},
	urldate = {2026-09-28},
	publisher = {arXiv},
	author = {Yallup, David and Handley, Will},
	year = {2026},
	note = {arXiv:2609.29413 [cs.LG]},
}

@misc{krondorfer_slice_2026,
	title = {Slice {Monte} {Carlo} {Integration}},
	url = {http://arxiv.org/abs/2606.25967},
	doi = {10.48550/arXiv.2606.25967},
	language = {en},
	urldate = {2026-09-22},
	publisher = {arXiv},
	author = {Krondorfer, Johannes K. and Binder, Christian W. and Neumann, Matthias and Linden, Wolfgang von der},
	month = jun,
	year = {2026},
	note = {arXiv:2606.25967 [stat.ME]},
}

@article{wang_energy_2026,
title = {Energy-based diffusion generator for efficient sampling of Boltzmann distributions},
journal = {Neural Networks},
volume = {194},
pages = {108126},
year = {2026},
issn = {0893-6080},
doi = {https://doi.org/10.1016/j.neunet.2025.108126},
url = {https://www.sciencedirect.com/science/article/pii/S0893608025010068},
author = {Yan Wang and Ling Guo and Hao Wu and Tao Zhou}
}

@BOOK{Kish1966-hz,
  title     = "Survey Sampling",
  author    = "Kish, Leslie",
  publisher = "John Wiley \& Sons",
  month     =  jan,
  year      =  1966,
  address   = "Nashville, TN"
}

@article{villa_improving_2026,
	title = {Improving {Bayesian} inference in {PTA} data analysis: {Importance} nested sampling with {Normalizing} {Flows}},
	volume = {55},
	issn = {22131337},
	shorttitle = {Improving {Bayesian} inference in {PTA} data analysis},
	url = {https://linkinghub.elsevier.com/retrieve/pii/S221313372600003X},
	doi = {10.1016/j.ascom.2026.101061},
	language = {en},
	urldate = {2026-08-28},
	journal = {Astronomy and Computing},
	author = {Villa, Eleonora and Shaifullah, Golam Mohiuddin and Possenti, Andrea and Carbone, Carmelita},
	month = apr,
	year = {2026},
	pages = {101061},
}

@article{prathaban_accelerated_2025,
	title = {Accelerated nested sampling with posterior repartitioning and \textit{$\beta$} -flows for gravitational waves},
	volume = {541},
	copyright = {https://creativecommons.org/licenses/by/4.0/},
	issn = {0035-8711, 1365-2966},
	url = {https://academic.oup.com/mnras/article/541/1/200/8163830},
	doi = {10.1093/mnras/staf962},
	language = {en},
	number = {1},
	urldate = {2026-08-28},
	journal = {Monthly Notices of the Royal Astronomical Society},
	author = {Prathaban, Metha and Bevins, Harry and Handley, Will},
	month = jun,
	year = {2025},
	pages = {200--213},
}

@misc{lemos_improving_2023,
	title = {Improving {Gradient}-guided {Nested} {Sampling} for {Posterior} {Inference}},
	url = {http://arxiv.org/abs/2312.03911},
	doi = {10.48550/arXiv.2312.03911},
	language = {en},
	urldate = {2025-10-21},
	publisher = {arXiv},
	author = {Lemos, Pablo and Malkin, Nikolay and Handley, Will and Bengio, Yoshua and Hezaveh, Yashar and Perreault-Levasseur, Laurence},
	month = dec,
	year = {2023},
	note = {arXiv:2312.03911 [cs]},
}

@article{baruah_normalizing_2025,
	title = {Normalizing flow-assisted nested sampling on {Type}-{II} {Seesaw} model},
	volume = {85},
	issn = {1434-6052},
	url = {https://link.springer.com/10.1140/epjc/s10052-025-14502-5},
	doi = {10.1140/epjc/s10052-025-14502-5},
	language = {en},
	number = {7},
	urldate = {2026-08-28},
	journal = {The European Physical Journal C},
	author = {Baruah, Rajneil and Mondal, Subhadeep and Patra, Sunando Kumar and Roy, Satyajit},
	month = jul,
	year = {2025},
	pages = {816},
}

@misc{qin_flowvat_2025,
	title = {{FlowVAT}: {Normalizing} {Flow} {Variational} {Inference} with {Affine}-{Invariant} {Tempering}},
	shorttitle = {{FlowVAT}},
	url = {http://arxiv.org/abs/2505.10466},
	doi = {10.48550/arXiv.2505.10466},
	language = {en},
	urldate = {2025-08-16},
	publisher = {arXiv},
	author = {Qin, Juehang and Liang, Shixiao and Tunnell, Christopher},
	month = may,
	year = {2025},
	note = {arXiv:2505.10466 [stat]},
}

@article{moqvist_thermodynamic_2025,
	title = {Thermodynamic {Interpolation}: {A} {Generative} {Approach} to {Molecular} {Thermodynamics} and {Kinetics}},
	volume = {21},
	copyright = {https://creativecommons.org/licenses/by/4.0/},
	issn = {1549-9618, 1549-9626},
	shorttitle = {Thermodynamic {Interpolation}},
	url = {https://pubs.acs.org/doi/10.1021/acs.jctc.4c01557},
	doi = {10.1021/acs.jctc.4c01557},
	language = {en},
	number = {5},
	urldate = {2026-08-28},
	journal = {Journal of Chemical Theory and Computation},
	author = {Moqvist, Selma and Chen, Weilong and Schreiner, Mathias and Nüske, Feliks and Olsson, Simon},
	month = mar,
	year = {2025},
	pages = {2535--2545},
}

@misc{wu_annealing_2025,
	title = {Annealing {Flow} {Generative} {Models} {Towards} {Sampling} {High}-{Dimensional} and {Multi}-{Modal} {Distributions}},
	url = {http://arxiv.org/abs/2409.20547},
	doi = {10.48550/arXiv.2409.20547},
	language = {en},
	urldate = {2026-08-28},
	publisher = {arXiv},
	author = {Wu, Dongze and Xie, Yao},
	month = may,
	year = {2025},
	note = {arXiv:2409.20547 [stat.ML]},
}

@misc{fogliani_annealing_2026,
	title = {Annealing in variational inference mitigates mode collapse: {A} theoretical study on {Gaussian} mixtures},
	shorttitle = {Annealing in variational inference mitigates mode collapse},
	url = {http://arxiv.org/abs/2602.12923},
	doi = {10.48550/arXiv.2602.12923},
	language = {en},
	urldate = {2026-08-28},
	publisher = {arXiv},
	author = {Fogliani, Luigi and Loureiro, Bruno and Gabrié, Marylou},
	month = feb,
	year = {2026},
	note = {arXiv:2602.12923 [stat.ML]},
}

@inproceedings{chen_gaussianization_2020,
 author = {Chen, Scott and Gopinath, Ramesh},
 booktitle = {Advances in Neural Information Processing Systems},
 editor = {T. Leen and T. Dietterich and V. Tresp},
 pages = {},
 publisher = {MIT Press},
 title = {Gaussianization},
 url = {https://proceedings.neurips.cc/paper_files/paper/2000/file/3c947bc2f7ff007b86a9428b74654de5-Paper.pdf},
 volume = {13},
 year = {2000}
}

@software{nessai,
  author       = {Michael J. Williams},
  title        = {nessai: Nested Sampling with Artificial Intelligence},
  month        = feb,
  year         = 2021,
  publisher    = {Zenodo},
  version      = {latest},
  doi          = {10.5281/zenodo.4550693},
  url          = {https://doi.org/10.5281/zenodo.4550693}
}

@article{turisini_leonardo_2024,
	title = {{LEONARDO}: {A} {Pan}-{European} {Pre}-{Exascale} {Supercomputer} for {HPC} and {AI} applications},
	volume = {9},
	copyright = {http://creativecommons.org/licenses/by/4.0},
	issn = {2364-091X},
	shorttitle = {{LEONARDO}},
	url = {https://jlsrf.org/index.php/lsf/article/view/186},
	doi = {10.17815/jlsrf-8-186},
	language = {en},
	number = {1},
	urldate = {2025-10-27},
	journal = {Journal of large-scale research facilities JLSRF},
	author = {Turisini, Matteo and Cestari, Mirko and Amati, Giorgio},
	month = jan,
	year = {2024},
}

@book{chopin_introduction_2020,
    address = {Cham},
    series = {Springer {Series} in {Statistics}},
    title = {An {Introduction} to {Sequential} {Monte} {Carlo}},
    isbn = {978-3-030-47845-2},
    url = {https://doi.org/10.1007/978-3-030-47845-2},
    doi = {10.1007/978-3-030-47845-2},
    publisher = {Springer},
    author = {Chopin, Nicolas and Papaspiliopoulos, Omiros},
    year = {2020},
}

@article{john_comparison_2025,
	title = {A comparison of probabilistic generative frameworks for molecular simulations},
	volume = {162},
	issn = {0021-9606, 1089-7690},
	url = {https://pubs.aip.org/jcp/article/162/11/114121/3340279/A-comparison-of-probabilistic-generative},
	doi = {10.1063/5.0249683},
	language = {en},
	number = {11},
	urldate = {2026-07-01},
	journal = {The Journal of Chemical Physics},
	author = {John, Richard and Herron, Lukas and Tiwary, Pratyush},
	month = mar,
	year = {2025},
	pages = {114121},
}

@article{coretti_boltzmann_2024,
  author       = {Coretti, Alessandro and Falkner, Sebastian and Weinreich, Jan and Dellago, Christoph and von Lilienfeld, O. Anatole},
  title        = {Boltzmann Generators and the New Frontier of Computational Sampling in Many-Body Systems},
  journal      = {KIM Review},
  volume       = {2},
  articleno    = {03},
  year         = {2024},
  doi          = {10.25950/bfa99422},
  url          = {https://doi.org/10.25950/bfa99422}
}

@misc{durkan_neural_2019,
	title = {Neural {Spline} {Flows}},
	url = {http://arxiv.org/abs/1906.04032},
	doi = {10.48550/arXiv.1906.04032},
	language = {en},
	urldate = {2025-08-17},
	publisher = {arXiv},
	author = {Durkan, Conor and Bekasov, Artur and Murray, Iain and Papamakarios, George},
	month = dec,
	year = {2019},
	note = {arXiv:1906.04032 [stat]},
}

@misc{vaswani_attention_2017,
      title={Attention Is All You Need}, 
      author={Ashish Vaswani and Noam Shazeer and Niki Parmar and Jakob Uszkoreit and Llion Jones and Aidan N. Gomez and Lukasz Kaiser and Illia Polosukhin},
      year={2023},
      eprint={1706.03762},
      archivePrefix={arXiv},
      primaryClass={cs.CL},
      url={https://arxiv.org/abs/1706.03762}, 
}

@article{shumailov_ai_2024,
	title = {{AI} models collapse when trained on recursively generated data},
	volume = {631},
	issn = {0028-0836, 1476-4687},
	url = {https://www.nature.com/articles/s41586-024-07566-y},
	doi = {10.1038/s41586-024-07566-y},
	language = {en},
	number = {8022},
	urldate = {2026-06-30},
	journal = {Nature},
	author = {Shumailov, Ilia and Shumaylov, Zakhar and Zhao, Yiren and Papernot, Nicolas and Anderson, Ross and Gal, Yarin},
	month = jul,
	year = {2024},
	pages = {755--759},
}

@article{de_santis_sampling_2026,
	title = {Sampling the liquid–gas critical point with {Boltzmann} generators},
	volume = {164},
	issn = {0021-9606, 1089-7690},
	url = {https://pubs.aip.org/jcp/article/164/9/094108/3381947/Sampling-the-liquid-gas-critical-point-with},
	doi = {10.1063/5.0314239},
	language = {en},
	number = {9},
	urldate = {2026-03-04},
	journal = {The Journal of Chemical Physics},
	author = {De Santis, Luigi and Russo, John and Ninarello, Andrea},
	month = mar,
	year = {2026},
	pages = {094108},
}

@article{kohler_equivariant_2020,
	%title = {Equivariant {Flows}: {Exact} {Likelihood} {Generative} {Learning} for {Symmetric} {Densities}},
	%abstract = {Normalizing ﬂows are exact-likelihood generative neural networks which approximately transform samples from a simple prior distribution to samples of the probability distribution of interest. Recent work showed that such generative models can be utilized in statistical mechanics to sample equilibrium states of many-body systems in physics and chemistry. To scale and generalize these results, it is essential that the natural symmetries in the probability density – in physics deﬁned by the invariances of the target potential – are built into the ﬂow. We provide a theoretical sufﬁcient criterion showing that the distribution generated by equivariant normalizing ﬂows is invariant with respect to these symmetries by design. Furthermore, we propose building blocks for ﬂows which preserve symmetries which are usually found in physical/chemical many-body particle systems. Using benchmark systems motivated from molecular physics, we demonstrate that those symmetry preserving ﬂows can provide better generalization capabilities and sampling efﬁciency.},
	%language = {en},
	%author = {Kohler, Jonas and Klein, Leon and Noe, Frank},
	%year = {2020},
	%file = {PDF:/Users/acoretti/Zotero/storage/S2KNWB2E/Kohler et al. - 2020 - Equivariant Flows Exact Likelihood Generative Learning for Symmetric Densities.pdf:application/pdf},
%}

@article{asghar_efficient_2024,
	title = {Efficient rare event sampling with unsupervised normalizing flows},
	volume = {6},
	issn = {2522-5839},
	url = {https://www.nature.com/articles/s42256-024-00918-3},
	doi = {10.1038/s42256-024-00918-3},
	language = {en},
	number = {11},
	urldate = {2026-06-30},
	journal = {Nature Machine Intelligence},
	author = {Asghar, Solomon and Pei, Qing-Xiang and Volpe, Giorgio and Ni, Ran},
	month = nov,
	year = {2024},
	pages = {1370--1381},
}

@article{gabrie_adaptive_2021,
author = {Marylou Gabrié  and Grant M. Rotskoff  and Eric Vanden-Eijnden },
title = {Adaptive Monte Carlo augmented with normalizing flows},
journal = {Proceedings of the National Academy of Sciences},
volume = {119},
number = {10},
pages = {e2109420119},
year = {2022},
doi = {10.1073/pnas.2109420119},
URL = {https://www.pnas.org/doi/abs/10.1073/pnas.2109420119},
eprint = {https://www.pnas.org/doi/pdf/10.1073/pnas.2109420119}
}

@article{wirnsberger_targeted_2020,
	title = {Targeted free energy estimation via learned mappings},
	volume = {153},
	issn = {0021-9606, 1089-7690},
	url = {http://arxiv.org/abs/2002.04913},
	doi = {10.1063/5.0018903},
	language = {en},
	number = {14},
	urldate = {2021-10-22},
	journal = {The Journal of Chemical Physics},
	author = {Wirnsberger, Peter and Ballard, Andrew J. and Papamakarios, George and Abercrombie, Stuart and Racanière, Sébastien and Pritzel, Alexander and Rezende, Danilo Jimenez and Blundell, Charles},
	month = oct,
	year = {2020},
	note = {Number: 14
arXiv: 2002.04913},
	pages = {144112},
}

@article{noe_boltzmann_2019,
	title = {Boltzmann generators: {Sampling} equilibrium states of many-body systems with deep learning},
	volume = {365},
	issn = {0036-8075, 1095-9203},
	shorttitle = {Boltzmann generators},
	url = {https://www.sciencemag.org/lookup/doi/10.1126/science.aaw1147},
	doi = {10.1126/science.aaw1147},
	language = {en},
	number = {6457},
	urldate = {2020-07-30},
	journal = {Science},
	author = {Noé, Frank and Olsson, Simon and Köhler, Jonas and Wu, Hao},
	month = sep,
	year = {2019},
	pages = {eaaw1147},
}

@article{williams_nested_2021,
	title = {Nested sampling with normalizing flows for gravitational-wave inference},
	volume = {103},
	issn = {2470-0010, 2470-0029},
	url = {https://link.aps.org/doi/10.1103/PhysRevD.103.103006},
	doi = {10.1103/PhysRevD.103.103006},
	language = {en},
	number = {10},
	urldate = {2023-02-15},
	journal = {Physical Review D},
	author = {Williams, Michael J. and Veitch, John and Messenger, Chris},
	month = may,
	year = {2021},
	pages = {103006},
}

@article{partay_efficient_2010,
	title = {Efficient {Sampling} of {Atomic} {Configurational} {Spaces}},
	volume = {114},
	issn = {1520-6106, 1520-5207},
	url = {https://pubs.acs.org/doi/10.1021/jp1012973},
	doi = {10.1021/jp1012973},
	language = {en},
	number = {32},
	urldate = {2023-02-15},
	journal = {The Journal of Physical Chemistry B},
	author = {Pártay, Lívia B. and Bartók, Albert P. and Csányi, Gábor},
	month = aug,
	year = {2010},
	pages = {10502--10512},
}

@article{partay_nested_2021,
	title = {Nested sampling for materials},
	volume = {94},
	issn = {1434-6028, 1434-6036},
	url = {https://link.springer.com/10.1140/epjb/s10051-021-00172-1},
	doi = {10.1140/epjb/s10051-021-00172-1},
	language = {en},
	number = {8},
	urldate = {2023-02-15},
	journal = {The European Physical Journal B},
	author = {Pártay, Livia B. and Csányi, Gábor and Bernstein, Noam},
	month = aug,
	year = {2021},
	pages = {159},
}

@article{li_phase_2020,
  title = {Phase behavior of Lennard-Jones particles in two dimensions},
  author = {Li, Yan-Wei and Ciamarra, Massimo Pica},
  journal = {Phys. Rev. E},
  volume = {102},
  issue = {6},
  pages = {062101},
  numpages = {7},
  year = {2020},
  month = {12},
  publisher = {American Physical Society},
  doi = {10.1103/PhysRevE.102.062101},
  url = {https://link.aps.org/doi/10.1103/PhysRevE.102.062101}
}

@article{coretti_learning_2025,
	title = {Learning mappings between equilibrium states of liquid systems using normalizing flows},
	volume = {162},
	issn = {0021-9606},
	url = {https://doi.org/10.1063/5.0253034},
	doi = {10.1063/5.0253034},
	number = {18},
	urldate = {2025-06-07},
	journal = {The Journal of Chemical Physics},
	author = {Coretti, Alessandro and Falkner, Sebastian and Geissler, Phillip L. and Dellago, Christoph},
	month = may,
	year = {2025},
	pages = {184102},
}

@misc{midgley_flow_2023,
      title={Flow Annealed Importance Sampling Bootstrap}, 
      author={Laurence Illing Midgley and Vincent Stimper and Gregor N. C. Simm and Bernhard Schölkopf and José Miguel Hernández-Lobato},
      year={2023},
      eprint={2208.01893},
      archivePrefix={arXiv},
      primaryClass={cs.LG},
      url={https://arxiv.org/abs/2208.01893}, 
}

@article{falkner_conditioning_2023,
doi = {10.1088/2632-2153/acf55c},
url = {https://doi.org/10.1088/2632-2153/acf55c},
year = {2023},
month = {09},
publisher = {IOP Publishing},
volume = {4},
number = {3},
pages = {035050},
author = {Falkner, Sebastian and Coretti, Alessandro and Romano, Salvatore and Geissler, Phillip L and Dellago, Christoph},
title = {Conditioning Boltzmann generators for rare event sampling},
journal = {Machine Learning: Science and Technology}
}

@article{jung_normalizing_2024,
doi = {10.1088/2632-2153/ad6ca0},
url = {https://doi.org/10.1088/2632-2153/ad6ca0},
year = {2024},
month = {08},
publisher = {IOP Publishing},
volume = {5},
number = {3},
pages = {035053},
author = {Jung, Gerhard and Biroli, Giulio and Berthier, Ludovic},
title = {Normalizing flows as an enhanced sampling method for atomistic supercooled liquids},
journal = {Machine Learning: Science and Technology}
}

@article{wang_mitigating_2025,
title = {Mitigating mode collapse in normalizing flows by annealing with an adaptive schedule: Application to parameter estimation},
journal = {Journal of Computational Physics},
volume = {564},
pages = {115115},
year = {2026},
issn = {0021-9991},
doi = {https://doi.org/10.1016/j.jcp.2026.115115},
url = {https://www.sciencedirect.com/science/article/pii/S0021999126004675},
author = {Yihang Wang and Chris Chi and Aaron R. Dinner}
}

@article{williams_importance_2023,
	title = {Importance nested sampling with normalising flows},
	volume = {4},
	issn = {2632-2153},
	url = {https://iopscience.iop.org/article/10.1088/2632-2153/acd5aa},
	doi = {10.1088/2632-2153/acd5aa},
	language = {en},
	number = {3},
	urldate = {2025-09-03},
	journal = {Machine Learning: Science and Technology},
	author = {Williams, Michael J and Veitch, John and Messenger, Chris},
	month = sep,
	year = {2023},
	pages = {035011},
}

@article{schebek_efficient_2024,
	title = {Efficient mapping of phase diagrams with conditional {Boltzmann} {Generators}},
	volume = {5},
	issn = {2632-2153},
	url = {https://iopscience.iop.org/article/10.1088/2632-2153/ad849d},
	doi = {10.1088/2632-2153/ad849d},
	language = {en},
	number = {4},
	urldate = {2025-08-21},
	journal = {Machine Learning: Science and Technology},
	author = {Schebek, Maximilian and Invernizzi, Michele and Noé, Frank and Rogal, Jutta},
	month = dec,
	year = {2024},
	pages = {045045},
}

@article{veitch_parameter_2015,
	title = {Parameter estimation for compact binaries with ground-based gravitational-wave observations using the {LALInference} software library},
	volume = {91},
	url = {https://link.aps.org/doi/10.1103/PhysRevD.91.042003},
	doi = {10.1103/PhysRevD.91.042003},
	number = {4},
	urldate = {2026-05-26},
	journal = {Physical Review D},
	publisher = {American Physical Society},
	author = {Veitch, J. and Raymond, V. and Farr, B. and Farr, W. and Graff, P. and Vitale, S. and Aylott, B. and Blackburn, K. and Christensen, N. and Coughlin, M. and Del Pozzo, W. and Feroz, F. and Gair, J. and Haster, C.-J. and Kalogera, V. and Littenberg, T. and Mandel, I. and O’Shaughnessy, R. and Pitkin, M. and Rodriguez, C. and Röver, C. and Sidery, T. and Smith, R. and Van Der Sluys, M. and Vecchio, A. and Vousden, W. and Wade, L.},
	month = feb,
	year = {2015},
	pages = {042003},
}

@article{ashton_nested_2022,
	title = {Nested sampling for physical scientists},
	volume = {2},
	copyright = {2022 Springer Nature Limited},
	issn = {2662-8449},
	url = {https://www.nature.com/articles/s43586-022-00121-x},
	doi = {10.1038/s43586-022-00121-x},
	language = {en},
	number = {1},
	urldate = {2026-02-10},
	journal = {Nature Reviews Methods Primers},
	publisher = {Nature Publishing Group},
	author = {Ashton, Greg and Bernstein, Noam and Buchner, Johannes and Chen, Xi and Csányi, Gábor and Fowlie, Andrew and Feroz, Farhan and Griffiths, Matthew and Handley, Will and Habeck, Michael and Higson, Edward and Hobson, Michael and Lasenby, Anthony and Parkinson, David and Pártay, Livia B. and Pitkin, Matthew and Schneider, Doris and Speagle, Joshua S. and South, Leah and Veitch, John and Wacker, Philipp and Wales, David J. and Yallup, David},
	month = may,
	year = {2022},
	pages = {39},
}

@article{ashton_bilby_2019,
	title = {Bilby: {A} {User}-friendly {Bayesian} {Inference} {Library} for {Gravitational}-wave {Astronomy}},
	volume = {241},
	issn = {0067-0049},
	shorttitle = {Bilby},
	url = {https://doi.org/10.3847/1538-4365/ab06fc},
	doi = {10.3847/1538-4365/ab06fc},
	language = {en},
	number = {2},
	urldate = {2026-05-27},
	journal = {The Astrophysical Journal Supplement Series},
	publisher = {The American Astronomical Society},
	author = {Ashton, Gregory and Hübner, Moritz and Lasky, Paul D. and Talbot, Colm and Ackley, Kendall and Biscoveanu, Sylvia and Chu, Qi and Divakarla, Atul and Easter, Paul J. and Goncharov, Boris and Vivanco, Francisco Hernandez and Harms, Jan and Lower, Marcus E. and Meadors, Grant D. and Melchor, Denyz and Payne, Ethan and Pitkin, Matthew D. and Powell, Jade and Sarin, Nikhil and Smith, Rory J. E. and Thrane, Eric},
	month = apr,
	year = {2019},
	pages = {27},
}

@misc{talbot_bilby-devbilby_2025,
	title = {bilby-dev/bilby: v.2.7.1},
	url = {https://doi.org/10.5281/zenodo.17533961},
	doi = {10.5281/zenodo.17533961},
	publisher = {Zenodo},
	author = {Talbot, Colm and Ashton, Gregory and Hübner, Moritz and Pitkin, Matt and {plasky} and Williams, Michael J. and {asb5468} and Vijaykumar, Aditya and Smith, Rory and {SMorisaki} and Veitch, John and Sarin, Nikhil and Macleod, Duncan and Williams, Daniel and {JasperMartins} and {MarcArene} and Berry, C. P. L. and Raymond, Vivien and {Ceciliogq} and Markin, Ivan and Keitel, David and {AlexandreGoettel} and Pompili, Lorenzo and Wright, Mick and {oliviawilk} and {noahewolfe} and {jacobgolomb} and Wu, Shichao and Udall, Rhiannon and Pürrer, Michael},
	month = nov,
	year = {2025},
}

@article{gw150914,
  title = {Observation of Gravitational Waves from a Binary Black Hole Merger},
  author = {Abbott, B. P. and Abbott, R. and Abbott, T. D. and Abernathy, M. R. and Acernese, F. and Ackley, K. and Adams, C. and Adams, T. and Addesso, P. and Adhikari, R. X. and Adya, V. B. and Affeldt, C. and Agathos, M. and Agatsuma, K. and Aggarwal, N. and Aguiar, O. D. and Aiello, L. and Ain, A. and Ajith, P. and Allen, B. and Allocca, A. and Altin, P. A. and Anderson, S. B. and Anderson, W. G. and Arai, K. and Arain, M. A. and Araya, M. C. and Arceneaux, C. C. and Areeda, J. S. and Arnaud, N. and Arun, K. G. and Ascenzi, S. and Ashton, G. and Ast, M. and Aston, S. M. and Astone, P. and Aufmuth, P. and Aulbert, C. and Babak, S. and Bacon, P. and Bader, M. K. M. and Baker, P. T. and Baldaccini, F. and Ballardin, G. and Ballmer, S. W. and Barayoga, J. C. and Barclay, S. E. and Barish, B. C. and Barker, D. and Barone, F. and Barr, B. and Barsotti, L. and Barsuglia, M. and Barta, D. and Bartlett, J. and Barton, M. A. and Bartos, I. and Bassiri, R. and Basti, A. and Batch, J. C. and Baune, C. and Bavigadda, V. and Bazzan, M. and Behnke, B. and Bejger, M. and Belczynski, C. and Bell, A. S. and Bell, C. J. and Berger, B. K. and Bergman, J. and Bergmann, G. and Berry, C. P. L. and Bersanetti, D. and Bertolini, A. and Betzwieser, J. and Bhagwat, S. and Bhandare, R. and Bilenko, I. A. and Billingsley, G. and Birch, J. and Birney, I. A. and Birnholtz, O. and Biscans, S. and Bisht, A. and Bitossi, M. and Biwer, C. and Bizouard, M. A. and Blackburn, J. K. and Blair, C. D. and Blair, D. G. and Blair, R. M. and Bloemen, S. and Bock, O. and Bodiya, T. P. and Boer, M. and Bogaert, G. and Bogan, C. and Bohe, A. and Bojtos, P. and Bond, C. and Bondu, F. and Bonnand, R. and Boom, B. A. and Bork, R. and Boschi, V. and Bose, S. and Bouffanais, Y. and Bozzi, A. and Bradaschia, C. and Brady, P. R. and Braginsky, V. B. and Branchesi, M. and Brau, J. E. and Briant, T. and Brillet, A. and Brinkmann, M. and Brisson, V. and Brockill, P. and Brooks, A. F. and Brown, D. A. and Brown, D. D. and Brown, N. M. and Buchanan, C. C. and Buikema, A. and Bulik, T. and Bulten, H. J. and Buonanno, A. and Buskulic, D. and Buy, C. and Byer, R. L. and Cabero, M. and Cadonati, L. and Cagnoli, G. and Cahillane, C. and Bustillo, J. Calder\'on and Callister, T. and Calloni, E. and Camp, J. B. and Cannon, K. C. and Cao, J. and Capano, C. D. and Capocasa, E. and Carbognani, F. and Caride, S. and Diaz, J. Casanueva and Casentini, C. and Caudill, S. and Cavagli\`a, M. and Cavalier, F. and Cavalieri, R. and Cella, G. and Cepeda, C. B. and Baiardi, L. Cerboni and Cerretani, G. and Cesarini, E. and Chakraborty, R. and Chalermsongsak, T. and Chamberlin, S. J. and Chan, M. and Chao, S. and Charlton, P. and Chassande-Mottin, E. and Chen, H. Y. and Chen, Y. and Cheng, C. and Chincarini, A. and Chiummo, A. and Cho, H. S. and Cho, M. and Chow, J. H. and Christensen, N. and Chu, Q. and Chua, S. and Chung, S. and Ciani, G. and Clara, F. and Clark, J. A. and Cleva, F. and Coccia, E. and Cohadon, P.-F. and Colla, A. and Collette, C. G. and Cominsky, L. and Constancio, M. and Conte, A. and Conti, L. and Cook, D. and Corbitt, T. R. and Cornish, N. and Corsi, A. and Cortese, S. and Costa, C. A. and Coughlin, M. W. and Coughlin, S. B. and Coulon, J.-P. and Countryman, S. T. and Couvares, P. and Cowan, E. E. and Coward, D. M. and Cowart, M. J. and Coyne, D. C. and Coyne, R. and Craig, K. and Creighton, J. D. E. and Creighton, T. D. and Cripe, J. and Crowder, S. G. and Cruise, A. M. and Cumming, A. and Cunningham, L. and Cuoco, E. and Canton, T. Dal and Danilishin, S. L. and D'Antonio, S. and Danzmann, K. and Darman, N. S. and Da Silva Costa, C. F. and Dattilo, V. and Dave, I. and Daveloza, H. P. and Davier, M. and Davies, G. S. and Daw, E. J. and Day, R. and De, S. and DeBra, D. and Debreczeni, G. and Degallaix, J. and De Laurentis, M. and Del\'eglise, S. and Del Pozzo, W. and Denker, T. and Dent, T. and Dereli, H. and Dergachev, V. and DeRosa, R. T. and De Rosa, R. and DeSalvo, R. and Dhurandhar, S. and D\'{\i}az, M. C. and Di Fiore, L. and Di Giovanni, M. and Di Lieto, A. and Di Pace, S. and Di Palma, I. and Di Virgilio, A. and Dojcinoski, G. and Dolique, V. and Donovan, F. and Dooley, K. L. and Doravari, S. and Douglas, R. and Downes, T. P. and Drago, M. and Drever, R. W. P. and Driggers, J. C. and Du, Z. and Ducrot, M. and Dwyer, S. E. and Edo, T. B. and Edwards, M. C. and Effler, A. and Eggenstein, H.-B. and Ehrens, P. and Eichholz, J. and Eikenberry, S. S. and Engels, W. and Essick, R. C. and Etzel, T. and Evans, M. and Evans, T. M. and Everett, R. and Factourovich, M. and Fafone, V. and Fair, H. and Fairhurst, S. and Fan, X. and Fang, Q. and Farinon, S. and Farr, B. and Farr, W. M. and Favata, M. and Fays, M. and Fehrmann, H. and Fejer, M. M. and Feldbaum, D. and Ferrante, I. and Ferreira, E. C. and Ferrini, F. and Fidecaro, F. and Finn, L. S. and Fiori, I. and Fiorucci, D. and Fisher, R. P. and Flaminio, R. and Fletcher, M. and Fong, H. and Fournier, J.-D. and Franco, S. and Frasca, S. and Frasconi, F. and Frede, M. and Frei, Z. and Freise, A. and Frey, R. and Frey, V. and Fricke, T. T. and Fritschel, P. and Frolov, V. V. and Fulda, P. and Fyffe, M. and Gabbard, H. A. G. and Gair, J. R. and Gammaitoni, L. and Gaonkar, S. G. and Garufi, F. and Gatto, A. and Gaur, G. and Gehrels, N. and Gemme, G. and Gendre, B. and Genin, E. and Gennai, A. and George, J. and Gergely, L. and Germain, V. and Ghosh, Abhirup and Ghosh, Archisman and Ghosh, S. and Giaime, J. A. and Giardina, K. D. and Giazotto, A. and Gill, K. and Glaefke, A. and Gleason, J. R. and Goetz, E. and Goetz, R. and Gondan, L. and Gonz\'alez, G. and Castro, J. M. Gonzalez and Gopakumar, A. and Gordon, N. A. and Gorodetsky, M. L. and Gossan, S. E. and Gosselin, M. and Gouaty, R. and Graef, C. and Graff, P. B. and Granata, M. and Grant, A. and Gras, S. and Gray, C. and Greco, G. and Green, A. C. and Greenhalgh, R. J. S. and Groot, P. and Grote, H. and Grunewald, S. and Guidi, G. M. and Guo, X. and Gupta, A. and Gupta, M. K. and Gushwa, K. E. and Gustafson, E. K. and Gustafson, R. and Hacker, J. J. and Hall, B. R. and Hall, E. D. and Hammond, G. and Haney, M. and Hanke, M. M. and Hanks, J. and Hanna, C. and Hannam, M. D. and Hanson, J. and Hardwick, T. and Harms, J. and Harry, G. M. and Harry, I. W. and Hart, M. J. and Hartman, M. T. and Haster, C.-J. and Haughian, K. and Healy, J. and Heefner, J. and Heidmann, A. and Heintze, M. C. and Heinzel, G. and Heitmann, H. and Hello, P. and Hemming, G. and Hendry, M. and Heng, I. S. and Hennig, J. and Heptonstall, A. W. and Heurs, M. and Hild, S. and Hoak, D. and Hodge, K. A. and Hofman, D. and Hollitt, S. E. and Holt, K. and Holz, D. E. and Hopkins, P. and Hosken, D. J. and Hough, J. and Houston, E. A. and Howell, E. J. and Hu, Y. M. and Huang, S. and Huerta, E. A. and Huet, D. and Hughey, B. and Husa, S. and Huttner, S. H. and Huynh-Dinh, T. and Idrisy, A. and Indik, N. and Ingram, D. R. and Inta, R. and Isa, H. N. and Isac, J.-M. and Isi, M. and Islas, G. and Isogai, T. and Iyer, B. R. and Izumi, K. and Jacobson, M. B. and Jacqmin, T. and Jang, H. and Jani, K. and Jaranowski, P. and Jawahar, S. and Jim\'enez-Forteza, F. and Johnson, W. W. and Johnson-McDaniel, N. K. and Jones, D. I. and Jones, R. and Jonker, R. J. G. and Ju, L. and Haris, K. and Kalaghatgi, C. V. and Kalogera, V. and Kandhasamy, S. and Kang, G. and Kanner, J. B. and Karki, S. and Kasprzack, M. and Katsavounidis, E. and Katzman, W. and Kaufer, S. and Kaur, T. and Kawabe, K. and Kawazoe, F. and K\'ef\'elian, F. and Kehl, M. S. and Keitel, D. and Kelley, D. B. and Kells, W. and Kennedy, R. and Keppel, D. G. and Key, J. S. and Khalaidovski, A. and Khalili, F. Y. and Khan, I. and Khan, S. and Khan, Z. and Khazanov, E. A. and Kijbunchoo, N. and Kim, C. and Kim, J. and Kim, K. and Kim, Nam-Gyu and Kim, Namjun and Kim, Y.-M. and King, E. J. and King, P. J. and Kinzel, D. L. and Kissel, J. S. and Kleybolte, L. and Klimenko, S. and Koehlenbeck, S. M. and Kokeyama, K. and Koley, S. and Kondrashov, V. and Kontos, A. and Koranda, S. and Korobko, M. and Korth, W. Z. and Kowalska, I. and Kozak, D. B. and Kringel, V. and Krishnan, B. and Kr\'olak, A. and Krueger, C. and Kuehn, G. and Kumar, P. and Kumar, R. and Kuo, L. and Kutynia, A. and Kwee, P. and Lackey, B. D. and Landry, M. and Lange, J. and Lantz, B. and Lasky, P. D. and Lazzarini, A. and Lazzaro, C. and Leaci, P. and Leavey, S. and Lebigot, E. O. and Lee, C. H. and Lee, H. K. and Lee, H. M. and Lee, K. and Lenon, A. and Leonardi, M. and Leong, J. R. and Leroy, N. and Letendre, N. and Levin, Y. and Levine, B. M. and Li, T. G. F. and Libson, A. and Littenberg, T. B. and Lockerbie, N. A. and Logue, J. and Lombardi, A. L. and London, L. T. and Lord, J. E. and Lorenzini, M. and Loriette, V. and Lormand, M. and Losurdo, G. and Lough, J. D. and Lousto, C. O. and Lovelace, G. and L\"uck, H. and Lundgren, A. P. and Luo, J. and Lynch, R. and Ma, Y. and MacDonald, T. and Machenschalk, B. and MacInnis, M. and Macleod, D. M. and Maga\~na-Sandoval, F. and Magee, R. M. and Mageswaran, M. and Majorana, E. and Maksimovic, I. and Malvezzi, V. and Man, N. and Mandel, I. and Mandic, V. and Mangano, V. and Mansell, G. L. and Manske, M. and Mantovani, M. and Marchesoni, F. and Marion, F. and M\'arka, S. and M\'arka, Z. and Markosyan, A. S. and Maros, E. and Martelli, F. and Martellini, L. and Martin, I. W. and Martin, R. M. and Martynov, D. V. and Marx, J. N. and Mason, K. and Masserot, A. and Massinger, T. J. and Masso-Reid, M. and Matichard, F. and Matone, L. and Mavalvala, N. and Mazumder, N. and Mazzolo, G. and McCarthy, R. and McClelland, D. E. and McCormick, S. and McGuire, S. C. and McIntyre, G. and McIver, J. and McManus, D. J. and McWilliams, S. T. and Meacher, D. and Meadors, G. D. and Meidam, J. and Melatos, A. and Mendell, G. and Mendoza-Gandara, D. and Mercer, R. A. and Merilh, E. and Merzougui, M. and Meshkov, S. and Messenger, C. and Messick, C. and Meyers, P. M. and Mezzani, F. and Miao, H. and Michel, C. and Middleton, H. and Mikhailov, E. E. and Milano, L. and Miller, J. and Millhouse, M. and Minenkov, Y. and Ming, J. and Mirshekari, S. and Mishra, C. and Mitra, S. and Mitrofanov, V. P. and Mitselmakher, G. and Mittleman, R. and Moggi, A. and Mohan, M. and Mohapatra, S. R. P. and Montani, M. and Moore, B. C. and Moore, C. J. and Moraru, D. and Moreno, G. and Morriss, S. R. and Mossavi, K. and Mours, B. and Mow-Lowry, C. M. and Mueller, C. L. and Mueller, G. and Muir, A. W. and Mukherjee, Arunava and Mukherjee, D. and Mukherjee, S. and Mukund, N. and Mullavey, A. and Munch, J. and Murphy, D. J. and Murray, P. G. and Mytidis, A. and Nardecchia, I. and Naticchioni, L. and Nayak, R. K. and Necula, V. and Nedkova, K. and Nelemans, G. and Neri, M. and Neunzert, A. and Newton, G. and Nguyen, T. T. and Nielsen, A. B. and Nissanke, S. and Nitz, A. and Nocera, F. and Nolting, D. and Normandin, M. E. N. and Nuttall, L. K. and Oberling, J. and Ochsner, E. and O'Dell, J. and Oelker, E. and Ogin, G. H. and Oh, J. J. and Oh, S. H. and Ohme, F. and Oliver, M. and Oppermann, P. and Oram, Richard J. and O'Reilly, B. and O'Shaughnessy, R. and Ott, C. D. and Ottaway, D. J. and Ottens, R. S. and Overmier, H. and Owen, B. J. and Pai, A. and Pai, S. A. and Palamos, J. R. and Palashov, O. and Palomba, C. and Pal-Singh, A. and Pan, H. and Pan, Y. and Pankow, C. and Pannarale, F. and Pant, B. C. and Paoletti, F. and Paoli, A. and Papa, M. A. and Paris, H. R. and Parker, W. and Pascucci, D. and Pasqualetti, A. and Passaquieti, R. and Passuello, D. and Patricelli, B. and Patrick, Z. and Pearlstone, B. L. and Pedraza, M. and Pedurand, R. and Pekowsky, L. and Pele, A. and Penn, S. and Perreca, A. and Pfeiffer, H. P. and Phelps, M. and Piccinni, O. and Pichot, M. and Pickenpack, M. and Piergiovanni, F. and Pierro, V. and Pillant, G. and Pinard, L. and Pinto, I. M. and Pitkin, M. and Poeld, J. H. and Poggiani, R. and Popolizio, P. and Post, A. and Powell, J. and Prasad, J. and Predoi, V. and Premachandra, S. S. and Prestegard, T. and Price, L. R. and Prijatelj, M. and Principe, M. and Privitera, S. and Prix, R. and Prodi, G. A. and Prokhorov, L. and Puncken, O. and Punturo, M. and Puppo, P. and P\"urrer, M. and Qi, H. and Qin, J. and Quetschke, V. and Quintero, E. A. and Quitzow-James, R. and Raab, F. J. and Rabeling, D. S. and Radkins, H. and Raffai, P. and Raja, S. and Rakhmanov, M. and Ramet, C. R. and Rapagnani, P. and Raymond, V. and Razzano, M. and Re, V. and Read, J. and Reed, C. M. and Regimbau, T. and Rei, L. and Reid, S. and Reitze, D. H. and Rew, H. and Reyes, S. D. and Ricci, F. and Riles, K. and Robertson, N. A. and Robie, R. and Robinet, F. and Rocchi, A. and Rolland, L. and Rollins, J. G. and Roma, V. J. and Romano, J. D. and Romano, R. and Romanov, G. and Romie, J. H. and Rosi\ifmmode \acute{n}\else \'{n}\fi{}ska, D. and Rowan, S. and R\"udiger, A. and Ruggi, P. and Ryan, K. and Sachdev, S. and Sadecki, T. and Sadeghian, L. and Salconi, L. and Saleem, M. and Salemi, F. and Samajdar, A. and Sammut, L. and Sampson, L. M. and Sanchez, E. J. and Sandberg, V. and Sandeen, B. and Sanders, G. H. and Sanders, J. R. and Sassolas, B. and Sathyaprakash, B. S. and Saulson, P. R. and Sauter, O. and Savage, R. L. and Sawadsky, A. and Schale, P. and Schilling, R. and Schmidt, J. and Schmidt, P. and Schnabel, R. and Schofield, R. M. S. and Sch\"onbeck, A. and Schreiber, E. and Schuette, D. and Schutz, B. F. and Scott, J. and Scott, S. M. and Sellers, D. and Sengupta, A. S. and Sentenac, D. and Sequino, V. and Sergeev, A. and Serna, G. and Setyawati, Y. and Sevigny, A. and Shaddock, D. A. and Shaffer, T. and Shah, S. and Shahriar, M. S. and Shaltev, M. and Shao, Z. and Shapiro, B. and Shawhan, P. and Sheperd, A. and Shoemaker, D. H. and Shoemaker, D. M. and Siellez, K. and Siemens, X. and Sigg, D. and Silva, A. D. and Simakov, D. and Singer, A. and Singer, L. P. and Singh, A. and Singh, R. and Singhal, A. and Sintes, A. M. and Slagmolen, B. J. J. and Smith, J. R. and Smith, M. R. and Smith, N. D. and Smith, R. J. E. and Son, E. J. and Sorazu, B. and Sorrentino, F. and Souradeep, T. and Srivastava, A. K. and Staley, A. and Steinke, M. and Steinlechner, J. and Steinlechner, S. and Steinmeyer, D. and Stephens, B. C. and Stevenson, S. P. and Stone, R. and Strain, K. A. and Straniero, N. and Stratta, G. and Strauss, N. A. and Strigin, S. and Sturani, R. and Stuver, A. L. and Summerscales, T. Z. and Sun, L. and Sutton, P. J. and Swinkels, B. L. and Szczepa\ifmmode \acute{n}\else \'{n}\fi{}czyk, M. J. and Tacca, M. and Talukder, D. and Tanner, D. B. and T\'apai, M. and Tarabrin, S. P. and Taracchini, A. and Taylor, R. and Theeg, T. and Thirugnanasambandam, M. P. and Thomas, E. G. and Thomas, M. and Thomas, P. and Thorne, K. A. and Thorne, K. S. and Thrane, E. and Tiwari, S. and Tiwari, V. and Tokmakov, K. V. and Tomlinson, C. and Tonelli, M. and Torres, C. V. and Torrie, C. I. and T\"oyr\"a, D. and Travasso, F. and Traylor, G. and Trifir\`o, D. and Tringali, M. C. and Trozzo, L. and Tse, M. and Turconi, M. and Tuyenbayev, D. and Ugolini, D. and Unnikrishnan, C. S. and Urban, A. L. and Usman, S. A. and Vahlbruch, H. and Vajente, G. and Valdes, G. and Vallisneri, M. and van Bakel, N. and van Beuzekom, M. and van den Brand, J. F. J. and Van Den Broeck, C. and Vander-Hyde, D. C. and van der Schaaf, L. and van Heijningen, J. V. and van Veggel, A. A. and Vardaro, M. and Vass, S. and Vas\'uth, M. and Vaulin, R. and Vecchio, A. and Vedovato, G. and Veitch, J. and Veitch, P. J. and Venkateswara, K. and Verkindt, D. and Vetrano, F. and Vicer\'e, A. and Vinciguerra, S. and Vine, D. J. and Vinet, J.-Y. and Vitale, S. and Vo, T. and Vocca, H. and Vorvick, C. and Voss, D. and Vousden, W. D. and Vyatchanin, S. P. and Wade, A. R. and Wade, L. E. and Wade, M. and Waldman, S. J. and Walker, M. and Wallace, L. and Walsh, S. and Wang, G. and Wang, H. and Wang, M. and Wang, X. and Wang, Y. and Ward, H. and Ward, R. L. and Warner, J. and Was, M. and Weaver, B. and Wei, L.-W. and Weinert, M. and Weinstein, A. J. and Weiss, R. and Welborn, T. and Wen, L. and We\ss{}els, P. and Westphal, T. and Wette, K. and Whelan, J. T. and Whitcomb, S. E. and White, D. J. and Whiting, B. F. and Wiesner, K. and Wilkinson, C. and Willems, P. A. and Williams, L. and Williams, R. D. and Williamson, A. R. and Willis, J. L. and Willke, B. and Wimmer, M. H. and Winkelmann, L. and Winkler, W. and Wipf, C. C. and Wiseman, A. G. and Wittel, H. and Woan, G. and Worden, J. and Wright, J. L. and Wu, G. and Yablon, J. and Yakushin, I. and Yam, W. and Yamamoto, H. and Yancey, C. C. and Yap, M. J. and Yu, H. and Yvert, M. and Zadro\ifmmode \dot{z}\else \.{z}\fi{}ny, A. and Zangrando, L. and Zanolin, M. and Zendri, J.-P. and Zevin, M. and Zhang, F. and Zhang, L. and Zhang, M. and Zhang, Y. and Zhao, C. and Zhou, M. and Zhou, Z. and Zhu, X. J. and Zucker, M. E. and Zuraw, S. E. and Zweizig, J.},
  collaboration = {LIGO Scientific Collaboration and Virgo Collaboration},
  journal = {Phys. Rev. Lett.},
  volume = {116},
  issue = {6},
  pages = {061102},
  numpages = {16},
  year = {2016},
  month = {2},
  publisher = {American Physical Society},
  doi = {10.1103/PhysRevLett.116.061102},
  url = {https://link.aps.org/doi/10.1103/PhysRevLett.116.061102}
}

@article{thrane_introduction_2019,
	title = {An introduction to {Bayesian} inference in gravitational-wave astronomy: {Parameter} estimation, model selection, and hierarchical models},
	volume = {36},
	issn = {1323-3580, 1448-6083},
	shorttitle = {An introduction to {Bayesian} inference in gravitational-wave astronomy},
	url = {https://www.cambridge.org/core/journals/publications-of-the-astronomical-society-of-australia/article/an-introduction-to-bayesian-inference-in-gravitationalwave-astronomy-parameter-estimation-model-selection-and-hierarchical-models/D459F61D8C37F0BEF86D60F42A418304},
	doi = {10.1017/pasa.2019.2},
	urldate = {2026-05-27},
	journal = {Publications of the Astronomical Society of Australia},
	author = {Thrane, Eric and Talbot, Colm},
	month = jan,
	year = {2019},
	pages = {e010},
}

@article{jones_what_1987,
	title = {What is {Projection} {Pursuit}?},
	volume = {150},
	issn = {0035-9238},
	url = {https://www.jstor.org/stable/2981662},
	doi = {10.2307/2981662},
	number = {1},
	urldate = {2026-06-03},
	journal = {Journal of the Royal Statistical Society. Series A (General)},
	publisher = {[Royal Statistical Society, Oxford University Press]},
	author = {Jones, M. C. and Sibson, Robin},
	year = {1987},
	pages = {1--37},
}

@article{hyvarinen_new_1997,
 author = {Hyv\"{a}rinen, Aapo},
 booktitle = {Advances in Neural Information Processing Systems},
 editor = {M. Jordan and M. Kearns and S. Solla},
 pages = {},
 publisher = {MIT Press},
 title = {New Approximations of Differential Entropy for Independent Component Analysis and Projection Pursuit},
 url = {https://proceedings.neurips.cc/paper_files/paper/1997/file/6d9c547cf146054a5a720606a7694467-Paper.pdf},
 volume = {10},
 year = {1997}
}

@article{pfister_good_2013,
	title = {Good things peak in pairs: a note on the bimodality coefficient},
	volume = {4},
	issn = {1664-1078},
	shorttitle = {Good things peak in pairs},
	url = {https://www.frontiersin.org/journals/psychology/articles/10.3389/fpsyg.2013.00700/full},
	doi = {10.3389/fpsyg.2013.00700},
	urldate = {2026-06-03},
	journal = {Frontiers in Psychology},
	publisher = {Frontiers},
	author = {Pfister, Roland and Schwarz, Katharina A. and Janczyk, Markus and Dale, Rick and Freeman, Jon},
	year = {2013},
}

@article{comon_independent_1994,
	series = {Higher {Order} {Statistics}},
	title = {Independent component analysis, {A} new concept?},
	volume = {36},
	issn = {0165-1684},
	url = {https://www.sciencedirect.com/science/article/pii/0165168494900299},
	doi = {10.1016/0165-1684(94)90029-9},
	number = {3},
	urldate = {2026-06-03},
	journal = {Signal Processing},
	author = {Comon, Pierre},
	year = {1994},
	pages = {287--314},
}

@inproceedings{meng_gaussianization_2020,
	title = {Gaussianization {Flows}},
	issn = {2640-3498},
	url = {https://proceedings.mlr.press/v108/meng20b.html},
	language = {en},
	urldate = {2026-06-03},
	booktitle = {Proceedings of the {Twenty} {Third} {International} {Conference} on {Artificial} {Intelligence} and {Statistics}},
	publisher = {PMLR},
	author = {Meng, Chenlin and Song, Yang and Song, Jiaming and Ermon, Stefano},
	month = jun,
	year = {2020},
	pages = {4336--4345},
}

@article{schebek_scalable_2026,
	title = {Scalable {Boltzmann} generators for equilibrium sampling of large-scale materials},
	volume = {17},
	copyright = {2026 The Author(s)},
	issn = {2041-1723},
	url = {https://www.nature.com/articles/s41467-026-73900-9},
	doi = {10.1038/s41467-026-73900-9},
	language = {en},
	number = {1},
	urldate = {2026-06-09},
	journal = {Nature Communications},
	publisher = {Nature Publishing Group},
	author = {Schebek, Maximilian and Noé, Frank and Rogal, Jutta},
	month = jun,
	year = {2026},
	pages = {5010},
}

@article{unglert_active_2026,
	title = {Active learning potentials for first-principles phase diagrams using replica-exchange nested sampling},
	volume = {12},
	copyright = {2026 The Author(s)},
	issn = {2057-3960},
	url = {https://www.nature.com/articles/s41524-026-01989-z},
	doi = {10.1038/s41524-026-01989-z},
	number = {1},
	urldate = {2026-03-15},
	journal = {npj Comput. Mater.},
	publisher = {Nature Publishing Group},
	author = {Unglert, Nico and Ketter, Michael and Madsen, Georg K. H.},
	month = feb,
	year = {2026},
	pages = {107},
}

@article{vallisneri_use_2008,
	title = {Use and abuse of the {Fisher} information matrix in the assessment of gravitational-wave parameter-estimation prospects},
	volume = {77},
	url = {https://link.aps.org/doi/10.1103/PhysRevD.77.042001},
	doi = {10.1103/PhysRevD.77.042001},
	number = {4},
	journal = {Physical Review D},
	publisher = {American Physical Society},
	author = {Vallisneri, Michele},
	year = {2008},
	pages = {042001},
}

@article{vinh_information_2010,
	title = {Information {Theoretic} {Measures} for {Clusterings} {Comparison}: {Variants}, {Properties}, {Normalization} and {Correction} for {Chance}},
	volume = {11},
	issn = {1533-7928},
	shorttitle = {Information {Theoretic} {Measures} for {Clusterings} {Comparison}},
	url = {http://jmlr.org/papers/v11/vinh10a.html},
	number = {95},
	urldate = {2026-07-22},
	journal = {Journal of Machine Learning Research},
	author = {Vinh, Nguyen Xuan and Epps, Julien and Bailey, James},
	year = {2010},
	pages = {2837--2854},
}
\end{refsection}

\end{document}